\documentclass[12pt,a4paper]{article}
  \usepackage{a4wide}
  \usepackage{latexsym}
  \usepackage{epsf}
  \usepackage{amssymb}
  \usepackage{graphicx}
  \usepackage{amsmath, cite}
  \usepackage{slashed,epsfig}
\usepackage{amsfonts}
  \usepackage{bbm}
  \usepackage{bbm}
  \usepackage{amsmath,amssymb,amsthm}
  \usepackage{pst-all}
  \usepackage{here}
\usepackage{latexsym,amssymb}
\usepackage{mcite}

\usepackage{a4wide}
  \usepackage{latexsym}
  \usepackage{epsf}
  \usepackage{amssymb}
  \usepackage{graphicx}
  \usepackage{amsmath, cite}
  \usepackage{slashed,epsfig}
\usepackage{amsfonts}
  \usepackage{bbm}
  \usepackage{bbm}
  \usepackage{amsmath,amssymb,amsthm}
  \usepackage{pst-all}
  \usepackage{here}
\usepackage{latexsym,amssymb}

 \makeatletter
\@addtoreset{equation}{section} \makeatother

\def\bfone{\relax{\rm 1\kern-.35em 1}}

 \makeatletter
\@addtoreset{equation}{section} \makeatother

\renewcommand{\d}{\textrm{d}}
\newcommand{\Real}{\textrm{I\!R}}

\newcommand{\e}{\textrm{e}}
\renewcommand{\d}{\textrm{d}}

\newcommand{\E}{\mathop{\rm E}}
\newcommand{\SU}{\mathop{\rm SU}}
\newcommand{\SO}{\mathop{\rm SO}}

\newcommand{\ISO}{\mathop{\rm ISO}}
\newcommand{\U}{\mathop{\rm {}U}}

\newcommand{\Sp}{\mathop{\rm {}Sp}}

\newcommand{\SL}{\mathop{\rm SL}}
\newcommand{\GL}{\mathop{\rm GL}}

\DeclareFontFamily{U}{rsf}{} \DeclareFontShape{U}{rsf}{m}{n}{
  <5> <6> rsfs5 <7> <8> <9> rsfs7 <10-> rsfs10}{}
\DeclareMathAlphabet\Scr{U}{rsf}{m}{n}

\begin{document}

\begin{flushright}
{\small
UG-08-05\\
KUL-TF-08/12\\
MAD-TH-08-07\\
UG-FT-229/08,\quad CAFPE-99/08}

\normalsize
\end{flushright}

\begin{center}

{\LARGE \bf{Generating Geodesic Flows and
\\\vspace{0.3cm} Supergravity Solutions }}\\

\vspace{1cm} {\large E. Bergshoeff$^{\dag}$, W.
Chemissany$^{\dag}$, A. Ploegh$^{\dag}$\\
M. Trigiante$^{\ddag}$ and
T. Van Riet$^*$} \\[3mm]\vspace{2mm}
$\dag${\small\slshape
   Centre for Theoretical Physics, University of Groningen,\\
    Nijenborgh 4, 9747 AG Groningen, The Netherlands}\\
{\upshape\ttfamily  e.a.bergshoeff, w.chemissany, a.r.ploegh @rug.nl}\\[3mm]
$\ddag${\small\slshape Dipartimento di Fisica Politecnico di
Torino,\\
C.so Duca degli Abruzzi, 24, I-10129 Torino, Italy}\\
{\upshape\ttfamily   mario.trigiante@polito.it }\\[3mm]

$*${\small\slshape Institute for Theoretical Physics, K.U. Leuven,\\
Celestijnenlaan 200D, B-3001 Leuven, Belgium  and \\ Department of
Physics, University of Wisconsin, \\Madison, WI 53706, USA, \\
and Departamento de F\'isica Te\'orica y del Cosmos\\ and Centro
Andaluz de F\'isica de Part\'iculas Elementales\\ Universidad de
Granada, 18071 Granada, Spain.\\}
{\upshape\ttfamily   Thomasvr@itf.fys.kuleuven.ac.be}\\[15mm]

{\bf Abstract}
\end{center}

\begin{quotation}
\small

We consider the geodesic motion on the symmetric moduli spaces
that arise after timelike and spacelike reductions of supergravity
theories. The geodesics correpond to timelike respectively
spacelike $p$-brane solutions when they are lifted over a
$p$-dimensional flat space. In particular, we consider the problem
of constructing \emph{the minimal generating solution}: A geodesic
with the minimal number of free parameters such that all other
geodesics are generated through isometries. We give an intrinsic
characterization of this solution in a wide class of orbits for
various supergravities in different dimensions. We apply our
method to three cases: (i) Einstein vacuum solutions, (ii) extreme
and non-extreme $D=4$ black holes in $\mathcal{N}=8$ supergravity
and their relation to $\mathcal{N}=2$ STU black holes and (iii)
Euclidean wormholes in $D\geq 3$. In case (iii) we present an easy
and general criterium for the existence of regular wormholes for a
given scalar coset.
\end{quotation}

\newpage

\pagestyle{plain} \tableofcontents

\section{Introduction}

Over the years much effort has been put into the investigation of
(non-)BPS solutions to (matter-coupled) supergravity theories. The
relevance of these solutions relies on the fact that they provide
crucial information about the underlying string theories and their
dualities. In particular, we focus on supergravity solutions that
have the structure of a $p$-brane. In general two different kinds of
$p$-brane solutions are considered: timelike $p$-branes that are
related to the string theory D-branes \cite{Polchinski:1995mt} (or
M-branes) or spacelike $p$-branes (known as S-branes) who are
conjectured to describe time-dependent phenomena in string theory
\cite{Gutperle:2002ai}. Timelike $p$-branes have a Lorentzian
worldvolume and are stationary solutions whereas spacelike
$p$-branes have a Euclidean worldvolume and are explicitly
time-dependent.

In view of the above it is important to find new solutions. One way
to do this is to develop new solution-generating techniques. These
techniques are often based on reducing the $p$-brane solution over
the brane worldvolume to obtain a corresponding ($-1$)-brane
solution. It turns out that the dynamics of these ($-1$)-brane
solutions is described by a geodesic motion on the moduli space that
follows from this
reduction\cite{Gal'tsov:1998yu,Breitenlohner:1987dg}. This has led
to the study of the geodesic solutions and, more general, the study
of the integrability of the geodesic equations on symmetric spaces.
Most of the focus has been on the geodesic curves  that correspond
to time-dependent supergravity solutions \cite{Fre:2003ep,
Fre:2003tg, Fre:2005bs, Fre':2005sr, Kleinschmidt:2005gz,
Fre':2005si, Fre:2006eu, Chemissany:2007fg, Fre':2007hd,
Fre:2008zd}.

We consider the problem of defining, in an intrinsic,
model-independent way, the most general geodesic that corresponds
both to time-dependent and stationary supergravity solutions. In
order to achieve this we use the isometry group of the moduli
space to construct the geodesic with the minimal number of free
parameters such that all other geodesics can now be obtained by an
isometry rotation of this particular solution. We call this
solution the {\sl minimal generating solution}. This method is
closely related to the compensator-algorithm developed in
\cite{Fre:2003ep}.

In our approach there is an important difference between the
Riemannian and pseudo-Riemannian moduli spaces. The generating
geodesic in the Riemannian case was shown to be carried by the
dilatons only \cite{Chemissany:2007fg}. The pseudo-Riemannian case
turns out to be richer. The aim of this paper is to extend the
discussion to the pseudo-Riemannian case. One of the main results,
derived in this paper, is the derivation of a theorem, see
\eqref{theorem}, valid for a wide class of orbits, defined by a
diagonalizable generator $Q$ of the geodesic, that characterizes the
geodesic generating solution in terms of the group-theoretical
properties of the corresponding moduli space. Our theorem applies to
all supergravities with symmetric scalar manifolds. This includes
all theories with more then $8$ supercharges and applies to an
interesting subset of theories with 8 and less supercharges. We show
that the generating solution can be found in a suitable sub-manifold
of the original scalar manifold defining a consistent truncation of
the theory. We also make general comments which apply to all orbits,
including thus the cases in which $Q$ is not diagonalizable.

To illustrate our methods we consider three different classes of
solutions. We first focus on a class of vacuum Einstein solutions.
The application of our theorem to this case reproduces some
well-known and some less known solutions. We next consider
stationary black hole solutions in four-dimensional supergravity.
For that we reduce the four-dimensional black hole solutions, via
a timelike reduction, to three dimensions, where they become
instantons \cite{Breitenlohner:1987dg, Gutperle:2000ve}. This
procedure has been used earlier to better understand black hole
solutions with symmetric scalar cosets \cite{Breitenlohner:1987dg,
Gunaydin:2005mx, Pioline:2006ni, Gaiotto:2007ag, Li:2008ar,
Berkooz:2008rj, Gunaydin:2007bg, Neitzke:2007ke}. It is of
interest to consider the class of black holes that satisfy the
attractor mechanism \cite{Ferrara:1995ih, Strominger:1996kf,
Ferrara:1996dd, Ferrara:1996um, Ferrara:1997tw, Gunaydin:2005mx}.
These black holes play an important role in the microstate
counting of the entropy \cite{Strominger:1996sh,
Maldacena:1997de}. Previously, it was believed that only the set
of extreme BPS black holes could be attractors. Later, it was
realized that non-BPS extreme black holes could also exhibit
attractor behavior \cite{Ortin:1997yn, Goldstein:2005hq,
Kallosh:2005ax} (for recent reviews on extreme black holes in
supergravity see also \cite{Ferrara:2008hw,Andrianopoli:2006ub}) .
We shall observe that, although extreme black holes with
$AdS_{2}\times S^{2}$ horizon are characterized by a nilpotent
(and thus non-diagonalizable) $Q$, the truncated theory defined by
our theorem already comprises all the nilpotent orbits of $Q$
which are relevant for this kind of solutions. We discuss the
application of our technique to the construction of more examples
of such non-BPS black holes in various supergravity models,
focusing, as an example, on the extreme solutions.

Applying our theorem we can write down general instanton solutions
and uplift this back to a general black hole solution in four
dimensions. In particular, we consider black hole solutions of
$D=4$, $\mathcal{N}=8$ supergravity. Our methods enable us to easily
reproduce the known dilatonic extreme black hole solutions
corresponding to this case. Embedding this extreme generating
solution in the ${\cal {N}}=2$ STU model allows us to discuss its
supersymmetry properties. In \cite{Pioline:2006ni} a factorization
property of the corresponding charge matrix has been introduced to
characterize extreme BPS black holes. This property has been
exploited in \cite{Gaiotto:2007ag, Li:2008ar}, and it is given a
simple group-theoretical interpretation. We show in this paper that
this property can be generalized in the ${\cal {N}}=2$ models to
distinguish between two kinds of extreme non-BPS solutions: those
with vanishing central charge at the horizon from the others. This
is illustrated in the simple dilatonic solution and we discuss how
to construct from it, using the three-dimensional isometries, a
generic full-charge $D=4$ black hole.

We also study the generating non-extreme solutions and thereby
demonstrate that no technical complications arise in finding
non-extreme solutions in comparison with extreme solutions in this
approach \footnote{This is along the lines of \cite{Miller:2006ay,
Andrianopoli:2007gt, Janssen:2007rc, Cardoso:2008gm} where BPS
type equations were constructed for non-extreme solutions (and
extreme non-BPS), thereby showing that the technical benefits of
BPS solutions can sometimes be carried over to non-BPS and
non-extreme solutions.}.

As a third application we consider wormhole solutions of Euclidean
supergravity. Recently, it has been shown that there is a simple
bound that needs to be satisfied in order to obtain a regular
wormhole solution \cite{ArkaniHamed:2007js}. Furthermore, examples
of such regular wormhole solutions could be obtained by allowing
Euclidean theories that do not follow from the reduction of a
higher-dimensional Minkowskian supergravity. In our analysis we
restrict to Euclidean supergravities that do follow from the
reduction of a higher-dimensional Minkowskian supergravity.  This
has the advantage that the Euclidean theory has a well-defined
superalgebra.  For this class of supergravities we find, using our
techniques, that there do not exist regular wormhole solutions and
that at most wormhole solutions exist that saturate the bound.

This paper is structured as follows. First, in section 2 we map the
D$p$- and S$p$-brane solutions  to D(-1)- and S(-1)-brane solutions
through dimensional reduction over the brane worldvolume. We show
how brane solutions can be described as the geodesic motion on the
moduli space.  In section 3 we derive the theorem which allows us to
construct the generating solution for diagonalizable $Q$,  for both
split and non-split symmetric spaces, as a solution of a truncation
of the original theory. Next we apply our method to construct three
classes of solutions: Einstein vacuum solutions in section
\ref{PHYSICSI}, $\mathcal{N}=8, D=4$ non-extreme black holes in
section \ref{PHYSICSII} and to Euclidean wormholes in $D\ge 3$ in
section \ref{PHYSICSIII}. In section \ref{PHYSICSII} we also
consider extreme $\mathcal{N}=8, D=4$ black hole solutions from the
same truncated theory in $D=3,$ giving a simple mathematical
characterization of several properties of the general solution.
Finally, in section \ref{CONCLUSION} we present our conclusions.
There are five appendices. In appendix \ref{appendix A} we give our
conventions, in appendix \ref{APPENDIXB} we present the explicit
form of a few Einstein vacuum solutions and in appendix
\ref{APPENDIXC} we present a Wick rotation that allows us to connect
the geodesic motion on Riemannian and pseudo-Riemannian coset
spaces. In appendix \ref{typeIIred} we present the toroidal
reduction of Type II theories. Finally, in appendix \ref{APPENDIXE}
we review some geometric properties of the STU model.

\section{Branes as Geodesics on Moduli Space}\label{section2}
\subsection{From $p$-branes to (-1)-branes}
Many supergravity solutions have the structure of a $p$-brane. The
solutions are charged electrically under a ($p+1$)-form gauge
potential $A_{p+1}$ or magnetically under a ($d-p-3$)-form gauge
potential $A_{d-p-3}$, where $d$ is the space-time dimension of the
supergravity theory. Another characteristic of brane solutions is
that the brane geometry has a flat ($p+1$)-dimensional worldvolume.
The metrics are given by\footnote{In this paper a S$p$-brane has a
$(p+1)$-dimensional Euclidean worldvolume just like a D$p$-brane has
a $(p+1)$-dimensional Lorentzian worldvolume.}
\begin{align}
& \text{timelike brane:} \qquad\,\d s^2_d=\e^{2A(r)}\eta_{\mu\nu}\d
x^{\mu}\d x^{\nu} + \e^{2B(r)}(\d r^2 + r^2\d\Omega^2_{d-p-2}),\nonumber\\
& \text{spacelike brane:}\quad\,\,\,\,\, \d
s^2_d=\e^{2A(t)}\delta_{\mu\nu}\d x^{\mu}\d x^{\nu} + \e^{2B(t)}(-\d
t^2 + t^2\d\Sigma^2_{d-p-2})\label{metric}\,,
\end{align}
where $A,B$ are arbitrary functions and $\delta,\eta$ are
respectively the Euclidean and the Lorentzian metric. The volume
elements $\d \Omega^2, \d\Sigma^2$ are, respectively, the metric on
the unit sphere and hyperboloid. There also exist less symmetric
solutions that break the worldvolume symmetries ($\ISO(p,1)$ and
$\ISO(p+1)$) and the transversal symmetries ($\SO(d-p-1)$ and
$\SO(d-p-2,1)$).

In this paper we develop a technique whose application for instance
allows us to classify and construct a \emph{wide} class of solutions
of 10 and 11-dimensional supergravity that generalize the Ansatz
(\ref{metric}) obeying the following two conditions
\begin{enumerate}
\item The transversal symmetries are unbroken.
\item The worldvolume symmetries ($\ISO(p,1)$ or $\ISO(p+1)$) can be broken down to
the translations along the worldvolume, thus the $\Real^{p+1}$
subgroup remains.
\end{enumerate}
For the second condition to be valid the matter-fields that carry
the solution must also be translation invariant. This implies that
one can effectively dimensionally reduce the solution over its
worldvolume. This maps a $p$-brane solution to a $(-1)$-brane
solution in $D=d-p-1$ dimensions whose equations of motion can be
derived from the following action
\begin{equation}
S=\int \d^Dx \sqrt{|g|} \Bigl\{\mathcal{R} -
\tfrac{1}{2}G_{ij}(\Phi)\partial\Phi^i\partial\Phi^j
\Bigr\}\,,\label{SIGMA-ACTION}
\end{equation}
where $G_{ij}$ is the metric on the moduli space that appears after
dimensional reduction. For timelike branes time is included in the
reduction and the corresponding moduli spaces are
pseudo-Riemannian, in contrast to moduli spaces that appear after
spacelike reductions.

We now consider a metric Ansatz for the ($-1$)-brane solution which
covers a more general slicing of the transverse space than the ones
indicated in eq. \eqref{metric}
\begin{equation}\label{Ansatz}
\d s^2_{D}=\epsilon f^2(r)\d r^2 + g^2(r)g_{ab}^{D-1}\d x^a\d
x^b\,,\qquad \Phi^i=\Phi^i(r)\,.
\end{equation}
Here the indices  $a,b$ run from $1,\ldots, D-1$. For  $\epsilon=-1$
the coordinate $r$ corresponds to time ($r\equiv t$) and $g_{ab}$ is
the metric of a ($D-1$)-dimensional Euclidean maximally symmetric
space (a sphere, flat space or hyperboloid). For $\epsilon=+1$
(\ref{Ansatz}) describes an instanton geometry with $r$ the
direction of the tunnelling process. It is convenient to
re-parameterize the coordinate $r$ to $h(r)$ via
\begin{equation}\label{harmonic}
\d h(r)=g^{1-D}f\d r\,.
\end{equation}
In terms of the new coordinate $h$ the equations of motion for the
scalars are derived from the one-dimensional action
\begin{equation}
S=\int G_{ij}\partial_h\Phi^i\partial_h\Phi^j\d h\,,
\end{equation}
where the metric has decoupled and can be solved independently (see
below). This action demonstrates that the solutions describe a
geodesic motion on the moduli space with $h(r)$ as an affine
parameter. Note that equation (\ref{harmonic}) is the integrated
version of the harmonic equation for $h(r)$ on the ($-1$)-brane
geometry, see eq. \eqref{Ansatz}. In terms of the affine parameter
the velocity $||v||$ is a constant
$||v||^2=G_{ij}\partial_h\Phi^i\partial_h\Phi^j$. The Einstein
equation for $(-1)$-branes is given by
\begin{equation}\label{Einstein}
\mathcal{R}_{rr}=\tfrac{1}{2}G_{ij}\partial_{r}\Phi^i\partial_{r}\Phi^j=\tfrac{1}{2}||v||^2(\partial_r
h(r))^2\,,\qquad \mathcal{R}_{ab}=0\,.
\end{equation}
Note that indeed the scalar fields play no longer a role in the
Einstein equations, their presence is only due to the affine
velocity $||v||$.

Combining the scalar field equations and the Einstein equations we
deduce the following first-order equation
\begin{equation}\label{EinsteinII}
\dot{g}^2=\frac{||v||^2}{2(D-2)(D-1)}f^2g^{4-2D}+\epsilon kf^2\,,
\end{equation}
where a dot denotes differentiation with respect to $r$. A solution
exists when the right-hand side remains positive. There is no
equation of motion for $f$ since it corresponds to the
re-parametrization freedom of $r$.

In the case of timelike branes the correspondence between geodesics
and branes is probably best known in terms of four-dimensional black
holes (0-branes) as three-dimensional instantons
\cite{Breitenlohner:1987dg,Gutperle:2000ve}. For spacelike branes we
refer to \cite{Fre:2003ep, Kleinschmidt:2005gz} for a description in
terms of a geodesic motion.

As an example of a geodesic motion on the moduli space, consider the
supersymmetric IIB instanton \cite{Gibbons:1995vg}. That solution
corresponds to the lightlike geodesics on $\SL(2,\Real)/\SO(1,1)$
(the Euclidean axion-dilaton system) whereas the non-supersymmetric
IIB instantons correspond to spacelike and timelike geodesics on
$\SL(2,\Real)/\SO(1,1)$ \cite{Bergshoeff:2004fq}.

\subsection{($-1$)-brane Geometries}\label{geometries}
\subsubsection*{Spacelike ($-1$)-brane}
We first consider the spacelike ($-1$)-branes. For this case the
target space is Riemannian and all geodesics have strictly
positive affine velocity squared $||v||^2>0$. The solution to the
Einstein equations (\ref{EinsteinII}) gives the following
$D$-dimensional metric
\begin{equation}\label{metricII}
\d s^2_D = -\frac{\d t^2}{a\,t^{-2(D-2)}-k} + t^2\d
\Sigma_k^2\,,\qquad a=\tfrac{||v||^2}{2(D-1)(D-2)}\,,
\end{equation}
while the scalar fields trace out geodesic curves with the harmonic
function $h(t)$ as affine parameter. The harmonic function $h$ is
given by
\begin{equation}\label{harmonicII}
h(t) = \frac{1}{\sqrt{a}(2-D)}\ln
\Bigl|\sqrt{a}t^{2-D}+\sqrt{at^{2(2-D)}-k} \Bigr| +  b\,.
\end{equation}
We take $b=0$ in what follows.

\subsubsection*{Timelike (-1)-brane}
For timelike branes the geometry of the ($-1$)-brane (a.k.a.
\emph{instanton}) entirely depends on the character of the geodesic
curve (spacelike, lightlike or timelike). Some of these solutions
have appeared in the literature before \cite{Arkani-Hamed:2007js,
Bergshoeff:2004fq, Breitenlohner:1987dg, Gutperle:2002km, Bergshoeff:2005zf}.\\

\begin{itemize}
\item $||v||^2>0$

In the table below we present the solution for $f$ in the gauge
$f=g$ and the harmonic function $h$. Note that for all three values
of $k$ the solutions have metric singularities.
\begin{table}[ht]\renewcommand{\arraystretch}{2.2}
\begin{center}
\hspace{-1cm}
\begin{tabular}{|c|l|}
\hline \rule[-1mm]{0mm}{6mm}
               & $f(r)= (\tfrac{||v||^2}{2(D-1)(D-2)})^{\tfrac{1}{2D-4}}\cos^{\tfrac{1}{D-2}}[(D-2)r]$ \\
 $k=-1$        & $h(r)=\sqrt{\tfrac{8(D-1)}{(D-2)||v||^2}}\,\text{arctanh}[\tan(\tfrac{D-2}{2}r)]+b$  \\ \hline
               & $f(r)=\bigl(\sqrt{\tfrac{(D-2)||v||^2}{2(D-1)}}\,r\bigr)^{\tfrac{1}{D-2}}$\\
 $k=0$         & $h(r)=\sqrt{\tfrac{2(D-1)}{(D-2)||v||^2}} \log r +b$  \\  \hline
               & $f(r)=(\tfrac{||v||^2}{2(D-1)(D-2)})^{\tfrac{1}{2D-4}}\sinh^{\tfrac{1}{D-2}}[(D-2)r]$  \\
 $k=+1$        & $h(r)=\sqrt{\tfrac{2(D-1)}{||v||^2(D-2)}}\,\log[\text{tanh}(\tfrac{D-2}{2}r)]+b$  \\

\hline
\end{tabular}
\caption{\it The Euclidean geometries with $||v^2||>0$ in the gauge
$f=g$. The real number $b$ is an integration constant.}
\end{center}
\end{table}

\item $||v||^2=0$

We take the Euclidean ``FLRW gauge" for which $f=1$. It is clear
from (\ref{EinsteinII}) that for $k=-1$ we do not find a solution
and that for $k=0$ we find flat space in Cartesian coordinates
($g=1$) and for $k=+1$ we find flat space in spherical coordinates
($g=r$). This makes sense since a lightlike geodesic motion comes
with zero ``energy-momentum''\footnote{ The fact that the $k=-1$
solution does not exist reflects that there does not exist a
hyperbolic slicing of the Euclidean plane.}. The harmonic function
is
\begin{equation}\label{harmonic lightlike conformal}
\begin{aligned}
k&=0 \qquad h(r)=  c\, r +b \,,\\
k&=1 \qquad h(r)= \frac{c}{r^{D-2}} + b  \,,
\end{aligned}
\end{equation}
where $c$ is a constant. In Euclidean IIB supergravity the
axion-dilaton parameterize $\SL(2,\Real)/\SO(1,1)$ and for
$||v||^2=0$ and $k=1$ we have the
standard half-supersymmetric D-instanton \cite{Gibbons:1995vg}.\\

\item  $||v||^2<0$

For $k=0$ and $k=-1$ we clearly have no solutions since the
right-hand side of (\ref{EinsteinII}) is always negative. For $k=+1$
a solution does exist, and in the conformal gauge ($g=fr$) it is
given by
\begin{equation}
f(r)=\Bigl(1-\tfrac{||v||^2}{8(D-1)(D-2)}r^{-2(D-2)}\Bigr)^{\frac{1}{D-2}}\,,
\end{equation}
where indeed only $||v||^2<0$ is valid. This geometry is smooth
everywhere and describes a wormhole, since there is  a
$\mathbb{Z}_2$-symmetry that acts as follows
\begin{equation}
r^{D-2}\rightarrow \tfrac{-||v||^2}{8(D-1)(D-2)}r^{-(D-2)}\,,
\end{equation}
and interchanges the two asymptotic regions. The harmonic function
is given by
\begin{equation}\label{harmonic function timlike conformal}
h(r)=\sqrt{-\tfrac{8(D-1)}{(D-2)||v||^2}}\,
\text{arctan}\Bigl(\sqrt{\tfrac{-||v||^2}{8(D-1)(D-2)}}\,\,r^{-(D-2)}\Bigr)+b\,.
\end{equation}
\end{itemize}

\subsection{Geodesic Curves}

In this paper we need  the geodesic curves on the moduli spaces of
several supergravity theories. For the case of maximal supergravity
we summarize the moduli spaces in table \ref{table: Euclidean
supergravity} \cite{Hull:1998br,Cremmer:1998em}. The symmetric
moduli spaces for other theories are presented in the tables in
sections \ref{HALF} and \ref{QUARTER}.

\begin{table}\renewcommand{\arraystretch}{1.6}
\begin{center}
\begin{tabular}{|c|c|c|}\hline
& $G/H$& $G/H^*$\\ \hline $D=10$ & {\rm SO}(1,1) & {\rm SO}(1,1)
\\ \hline  $D=9$ &
$\frac{\GL(2,\Real)}{{\rm SO}(2)}$ & $\frac{\GL(2,\Real)}{{\rm SO}(1,1)}$ \\
\hline $D=8$ & $\frac{\SL(3,\Real)}{{\rm SO}(3)}\times\frac{
\SL(2,\Real)}{{\rm SO}(2)}$ & $\frac{\SL(3,\Real)}{{\rm
SO}(2,1)}\times\frac{ \SL(2,\Real)}{ {\rm SO}(1,1)} $ \\ \hline
$D=7$ & $\frac{SL(5,\Real)}{{\rm SO}(5)}$ & $\frac{\SL(5,\Real)}{{\rm SO}(3,2)}$ \\
\hline $D=6$ & $\frac{{\rm SO}(5,5)}{{\rm S}[{\rm O}(5)\times {\rm
O}(5)]}$ & $\frac{{\rm SO}(5,5)}{{\rm SO}(5, \mathbb{C})}$
 \\ \hline
$D=5$ & $\frac{{\rm E}_{6(+6)}}{U\!Sp(8)}$ & $\frac{{\rm
E}_{6(+6)}}{U\!Sp(4,4)}$
 \\ \hline
$D=4$ & $\frac{{\rm E}_{7(+7)}}{\SU(8)}$  &  $\frac{E_{7(+7)}}{\SU^*(8)}$ \\
\hline $D=3$ & $\frac{{\rm E}_{8(+8)}}{\SO(16)}$ & $\frac{{\rm
E}_{8(+8)}}{\SO^*(16)}$
\\\hline
\end{tabular}
\caption{\it The scalar cosets for maximal supergravities in
Minkowskian ($G/H$) and Euclidean ($G/H^*$) signatures.  }
\label{table: Euclidean supergravity}
\end{center}
\end{table}

The cosets $G/H$ (or products thereof) in the left column are called
maximally non-compact since $G$ is the maximal non-compact real
slice of a semi-simple algebra and $H$ is the maximal compact
subgroup. Since $H$ is compact the metric is strictly positive
definite and the coset is Riemannian. The cosets $G/H^*$ in the
right column only differ in the isotropy group $H^*$ which is some
non-compact version of $H$ and, as a consequence, $G/H^*$ is
pseudo-Riemannian.

Our approach to understanding all the geodesic curves is by
constructing \emph{the generating solution}. By definition, a
generating solution is a geodesic with the minimal number of
arbitrary integration constants such that the action of the isometry
group $G$ generates all other geodesics from the generating
solution. It was explained in \cite{Chemissany:2007fg}\footnote{See
 the appendix of \cite{Rosseel:2006fs} for earlier remarks.} that
for maximally non-compact cosets $G/H$, the generating solution can
be taken to be the straight line through the origin carried by the
dilaton fields
\begin{equation}
\phi^I(t)=v^I\,t\,, \qquad \chi^{\alpha}=0\,, \quad I=1,\ldots, r
\,.
\end{equation}
This solution contains only $r$ arbitrary integration constants
$v^I$, with $r$ the rank of $G$. This theorem applies to all the
cosets in the left column of table \ref{table: Euclidean
supergravity}.

Since the straight line solution is the generating solution,
$G$-transformations on this solution generate all the other
geodesic curves. The number of independent constants in $G$ is the
dimension of $G$ which is $r+2\text{dim}H$. In total this gives us
$2r+2\text{dim}H$ arbitrary (integration) constants as expected
since there are $r+\text{dim}H$ scalars (coordinates) for which we
have to specify the initial place and velocity. However this
counting exercise is no proof since it might be that the action of
$G$ does not create independent integration constants or if the
solutions lie in disconnected areas. The latter is the case for
the cosets in the right column of table \ref{table: Euclidean
supergravity}. There the straight line solution is not generating
since the affine velocity is positive
\begin{equation}
||v||^2=\sum (v^I)^2>0\,.
\end{equation}
The affine velocity is invariant under $G$-transformations and by
transforming the straight line we only generate spacelike geodesics.
However,  cosets with non-compact isotropy $H^*$ have metrics with
indefinite signature and therefore also allow $||v||^2\leq 0$.

\section{The Math: Coset Spaces and Normal Forms}\label{MATH}

\subsection{The Generating Geodesic Curve and the Normal Form}
Consider a coset space $G/H$. In this section $H$ can be compact or
non-compact. We define the coset representative $L$ as an element of
$G,$ on which the isometry group $G$ acts on the left $L\rightarrow
g L$ and the local isotropy group $H$ acts from the right:
$L\rightarrow L h .$

In this paper we only consider symmetric spaces. The condition that
the scalar manifold is symmetric is defined as follows. Denote
$\frak{g}$ and $\frak{H}$ for respectively the Lie algebras $G$ and
$H$. Consider a generic decomposition
\begin{equation}
\frak{g}=\frak{H}\oplus \mathcal{K}\,,
\end{equation}
where $\mathcal{K}$ is the complement of $\frak{H}$ in $\frak{g}$.
If there exist a $\mathcal{K}$ such that
\begin{equation}\label{definition: symmetric space}
[\mathfrak{H},\mathfrak{H}]\subset\mathfrak{H}\,,\quad
[\mathcal{K},
\mathfrak{H}]\subset\mathcal{K}\,,\quad[\mathcal{K},\mathcal{K}]\subset\mathfrak{H}\,,
\end{equation}
we call $G/H$ a symmetric space. The above condition is equivalent
to the existence of a so-called Cartan involution $\theta$ which
has the following action on the Lie algebra
\begin{equation}
\theta(\frak{H})=\frak{H}\,,\quad
\theta(\mathcal{K})=-\mathcal{K}\,.
\end{equation}
By definition $\theta$ is an involutive automorphism, which means
that it squares to one, without being trivial anywhere and that it
preserves the Lie bracket
\begin{equation}
\theta^2=1\,,\quad \theta([A,B])=[\theta(A),\theta(B)]\,.
\end{equation}
Let us go back to the manifold $G/H$ and explain how the geodesics
are fully determined in terms of the Lie algebra. For that we
consider the symmetric coset matrix
\begin{equation}\label{symmetric definition of M}
\mathcal{M} = L L^{\sharp}\,.
\end{equation}
Here $\sharp$ is the generalized transpose, defined through the
Cartan involution $\theta$
\begin{equation}
L^{\sharp}=\text{exp}[-\theta(\log L)]=\theta(L^{-1})\,.
\end{equation}
The matrix $\mathcal{M}$ is by construction invariant under local
$H$-transformations that work from the right on $L$ and transforms
as follows under the whole of $G$ (from the left on $L$)
\begin{equation}
\mathcal{M}\rightarrow g \mathcal{M} g^{\sharp}\,.
\end{equation}
Up to a representation-dependent factor the metric on $G/H$ is given
by
\begin{equation}
\textrm{d}s^2 =G_{ij}\d\Phi^i\d \Phi^j= \tfrac{1}{2}\text{Tr}[
\textrm{d}\mathcal{M}\textrm{d}\mathcal{M}^{-1}]\,.
\end{equation}
Clearly, the metric is invariant under a local action of $H$ from
the right on $L$ and under a rigid action from $G$ from the left on
$L$. The latter implies that $G$ is the isometry group of $G/H$ as
it should be. The action for the geodesic curves on $G/H$ now is
\begin{equation}
S\propto\int \text{Tr}[ \mathcal{M}' (\mathcal{M}^{-1})']\,,
\end{equation}
where a prime $'$ denotes differentiation with respect to the affine
parameter $h$. The equations of motion are
\begin{equation}\label{scalar fields equations of motion}
[\mathcal{M}^{-1}\mathcal{M}']'=0\,.
\end{equation}
This implies that $\mathcal{M}^{-1}\mathcal{M}'=Q$ with $Q$ a
constant matrix, which can be seen as the matrix of Noether charges.
The affine velocity squared of the geodesic curve is $||v||^2=
\tfrac{1}{2}\text{Tr}[Q^2]$. Since $\mathcal{M}^{-1}\mathcal{M}'=Q$
the problem is integrable and a general solution is given by
\begin{equation}\label{general solution for M}
\mathcal{M}(h)=\mathcal{M}(0)\e^{Qh}\,.
\end{equation}

Since the action of $G$ on $G/H$ is transitive we can restrict to
the origin of $G/H$ and then $\mathcal{M}(0) = \mathbbm{1}$.
Since $\mathcal{M}$ $\in$ $G$ we have that $Q$ $\in$ $\frak{g}$.
But the requirement $\mathcal{M} = \mathcal{M}^{\sharp}$ gives a
further restriction on $Q$
\begin{equation}\label{constraint}
\theta(Q)=-Q\qquad \Longleftrightarrow\qquad Q \in \mathcal{K}\,.
\end{equation}
Under the adjoint of $G$, $Q$ transforms as
\begin{equation}
Q \rightarrow \Omega\,Q\,\Omega^{-1}\,, \quad \Omega \in G\,.
\end{equation}
While the Casimirs Tr\,$Q^{n}$ are invariant, the constraint
(\ref{constraint}) is not invariant under the total isometry group
but only under the smaller isotropy group $H$.

As an example, let us consider $\SL(p+q,\Real)/\SO(p,q)$. For this
case an explicit realization of $\theta$ is given by\footnote{We
will see later in section 4 that a torus reduction yields a slightly
different matrix $\hat{\mathcal{M}}$ given by
$\hat{\mathcal{M}}=L\eta L^T\,,$ that is $\hat{\mathcal{M}} =
\mathcal{M}\eta$.  They both satisfy similar equations of motion: $
\mathcal{M}^{-1}\textrm{d}\mathcal{M}=Q$ and
$\hat{\mathcal{M}}^{-1}\textrm{d}\hat{\mathcal{M}}=Q^T$.}

\begin{equation}\label{SL(p+q,Real)SO(p,q)
example} \theta(Q)=-\eta Q^T\eta\,,\quad \mathcal{M}=L\eta
L^T\eta\,, \quad \eta=(-\mathbbm{1}_p,\mathbbm{1}_q)\,.
\end{equation}
We find from (\ref{SL(p+q,Real)SO(p,q) example}) that $Q$ is defined
by
\begin{equation}\label{identity 2}
\eta Q = Q^T\eta\,,\qquad \text{Tr}\,Q = 0\,.
\end{equation}

Since the matrix $Q$ determines all geodesics through the origin,
and by transitivity all geodesics on $G/H$ we look for the normal
form $Q_N$ of $Q$ under $(H\subset G)$-transformations. We restrict
to $H$ since only these transformation keep us at the origin. As a
result the geodesics determined by the ``integration constants" in
$Q_N$ generate \emph{all} geodesics through a rigid
$G$-transformation\footnote{The method we use here differs from  the
so-called compensator algorithm developed in \cite{Fre':2005sr}, to
generate geodesic solutions. Our method makes use of the isometry
group $G$ while the compensator algorithm uses the local isotropy
$H$.}.

The problem of constructing normal forms of matrices with given
symmetry properties has been considered by mathematicians some time
ago \cite{Djokovic:1981bh}. Now we will consider an explicit
instructive example.

\subsection{An Example: the Normal Form of $\frak{gl}(p+q)/\frak{so}(p,q)$}
\label{sectionsln} Consider $Q \in \frak{gl}(p+q)/\frak{so}(p,q)$
and its corresponding Jordan form obtained by going to a suitable
basis (empty entries are understood to be filled with zeros)
\begin{eqnarray}
Q_J&=&\left(\begin{matrix} A(\lambda_1)&&\cr
                      &\ddots & \cr
                         &  & A(\lambda_\ell)
\end{matrix}\right)\,,
\end{eqnarray}
where $A(\lambda_i)$, $k=1,\dots, \ell$ is the indecomposable block
corresponding to the eigenvalue $\lambda_i$
\begin{eqnarray}
A(\lambda_i)&=&\left(\begin{matrix}\lambda_i& 1& & \cr
                                    & \ddots & \ddots&\cr
                                      & & &1\cr
                                     & & &\lambda_i
\end{matrix}\right)=\lambda_i\,\bfone_{\mu_i}+J_{\mu_i}\,.
\end{eqnarray}
If $\mu_i=\mu(\lambda_i)$ is the degeneracy of the root $\lambda_i$
of the \emph{minimal} polynomial $m(z)$ corresponding to $Q$ and
$J_{\mu}$ is the $\mu\times \mu $ nilpotent matrix of the form
\begin{eqnarray}
J_{\mu}&=&\left(\begin{matrix}0 & 1& & \cr
                                    & \ddots & \ddots&\cr
                                      & & &1\cr
                                     & & &0
\end{matrix}\right)\,,
\end{eqnarray}
$A(\lambda_i)$ is then a $\mu_i\times \mu_i$ matrix and is
diagonalizable only if $\mu_i=1$. We wish to transform the matrix
$Q_J$ to a real normal form $Q_N$ with the required symmetry
properties
\begin{eqnarray}
Q_N^T\,\eta&=&\eta\,Q_N\,,\label{qeta}
\end{eqnarray}
where $\eta^T=\eta$ and has $p$  eigenvalues $-1$ and $q$
eigenvalues $+1$. To this end we need to work on the blocks
corresponding to complex eigenvalues $\lambda =
\lambda_1+i\,\lambda_2$. Since the original $Q$ is a real matrix,
for each block $A(\lambda)$ there will be a conjugate one
$A(\bar{\lambda})=\overline{A(\lambda)}$. Let $\mu=\mu(\lambda)$ and
consider the following $(2\mu)\times (2\mu)$ matrix
\begin{eqnarray}
\hat{\mathcal{A}}(\lambda,\bar{\lambda})&=&\left(\begin{matrix}A(\lambda)
& {\bf 0}\cr  {\bf 0}&A(\bar{\lambda})
\end{matrix}
\right)\,.
\end{eqnarray}
Using the following unitary transformation $\mathcal{U}$
\begin{eqnarray}
\mathcal{U}&=&\frac{1}{\sqrt{2}}\,\begin{pmatrix}
     1   &    i   &       &       &       &       &       \\
        &        & 1 & i &        &        &         \\
        &        &       &  \ddots & \ddots &       &       \\
        &       &        &       &       &1& i \\
  1   &    -i   &       &       &       &       &       \\
        &        & 1 & -i &        &        &         \\
        &        &       &  \ddots & \ddots &       &       \\
        &       &        &       &       &1& -i
\end{pmatrix}
\end{eqnarray}
we can define the matrix below
\begin{eqnarray}
\mathcal{A}(\lambda,\bar{\lambda})&=&\mathcal{U}^\dagger\,\hat{\mathcal{A}}(\lambda,\bar{\lambda})\,\mathcal{U}=\left(\begin{matrix}\lambda_1
& -\lambda_2\cr \lambda_2 & \lambda_1
\end{matrix}\right)\otimes
\bfone_\mu+\bfone_2\otimes J_\mu\,,
\end{eqnarray}
which is the $\eta$-irreducible block for complex eigenvalues. For
real eigenvalues $\lambda=\bar{\lambda}$ the $\eta$-irreducible
block is $\mu\times \mu$ and coincides with $A(\lambda)$. By
applying the transformation $\mathcal{U}$ on each couple of blocks
$A(\lambda),\,A(\bar{\lambda})$, for complex $\lambda$, leaving the
blocks $A(\lambda)$ unchanged for real $\lambda$, we obtain from
$Q_J$ the following normal form
\begin{eqnarray}
Q_N&=&\left(\begin{matrix} A(\lambda_1) & & & & & \cr
                             &\ddots & & & & \cr
                              & & A(\lambda_k)& & & \cr
&&&\mathcal{A}(\lambda_{k+1},\bar{\lambda}_{k+1})& &\cr
                       &&&&\ddots & \cr
                      &&&& & \mathcal{A}(\lambda_s,\bar{\lambda}_s)
\end{matrix}\right)\,,
\end{eqnarray}
where we have ordered the eigenvalues so that the first $k$ are
real. Each $\eta$-irreducible block $A(\lambda_i)$ has dimension
$N_i\times N_i$, $N_i=\mu_i$, while
$\mathcal{A}(\lambda_i,\bar{\lambda}_i)$ is a $N_i\times N_i$ matrix
with $N_i=2\,\mu_i$. We therefore have
\begin{eqnarray}
\sum_{i=1}^s N_i&=&\sum_{i=1}^k \mu_i+2\sum_{i=k+1}^s \mu_i\,=n\,.
\end{eqnarray}
One can easily verify that eq. (\ref{qeta}) is satisfied with
\begin{eqnarray}
\eta &=&\left(\begin{matrix} \epsilon_1\,\eta^{(\mu_1)} & & & & &
\cr
                             &\ddots & & & & \cr
                              & &\epsilon_k\,\eta^{(\mu_k)}& & & \cr
&&&\eta^{(2\mu_{k+1})}& &\cr
                       &&&&\ddots & \cr
                      &&&& & \eta^{(2\mu_{s})}
\end{matrix}\right)\,,\label{etut}
\end{eqnarray}
where each diagonal block $\eta^{(N)}$ is a  $N\times N$ matrix
defined as follows
\begin{eqnarray}
\eta^{(N)}&=&\left(\begin{matrix} &  & & & 1\cr &  & & \cdot & \cr &
& \cdot & & \cr &  \cdot & & & \cr 1&   & & &
\end{matrix}\right)\,,
\end{eqnarray}
and $\epsilon_i=\pm 1$. The signs $\epsilon_i$ characterize $Q$ and
will be explained in the following construction of the
pseudo-orthogonal matrix $T\in \SO(p,q)$ which brings $Q$ to its
normal form $Q_N=T^{-1}Q T$. In order for the right hand side of eq.
(\ref{etut}) to describe $\eta$, denoting by $s_i$ the signature of
the $i^{th}$ block, the following conditions should be satisfied
\begin{eqnarray}
\sum_{i=1}^k
s_i&=&q-p\,\,\,;\,\,\,\,\,s_i=\frac{\epsilon_i}{2}\,(1-(-)^{\mu_i})\,,\label{andone}\\
p&=&\sum_{i=1}^k\frac{1}{2}\,(\mu_i-s_i)+\sum_{i=k+1}^s
\mu_i\label{andtwo}\,.
\end{eqnarray}
Let us explicitly construct the transformation $T$. Consider a real
eigenvalue $\lambda$, $\mu=\mu(\lambda)$ and let $v_i^\lambda$,
$i=1,\dots, \mu$, denote the corresponding generalised eigenvectors
\begin{eqnarray}
Q\,v^{\lambda}_i&=&\lambda
\,v^{\lambda}_i+v^{\lambda}_{i-1}\,.\label{onr}
\end{eqnarray}
If $v,\,w$ are two generic vectors we shall use the notation
$(v,\,w)\equiv v^T\eta w$. By definition of $Q$ we then have that
$(v,\,Q w)=(Qv,\,w)$. Using this property and (\ref{onr}) we find
that
\begin{eqnarray}
(v^\lambda_i,\,v^\lambda_{j-1})&=&(v^\lambda_{i-1},\,v^\lambda_{j})\,,\label{sameasbefore}
\end{eqnarray}
which in turn implies that
$(v^\lambda_k,\,v^\lambda_{1})=\dots=(v^\lambda_k,\,v^\lambda_{\mu-k})=0$.
We can write the matrix $(v^\lambda_i,\,v^\lambda_{j})$ in the
following form
\begin{eqnarray}
(v^\lambda_i,\,v^\lambda_{j})&=&\begin{pmatrix}
          &   &    &    &            &   v^{(1)} \\
        &      &      &     &  \cdot  &  v^{(2)} \\
        &      &      &  \cdot   &  \cdot &       \\
         &      &  \cdot     & \cdot      &      &   \vdots   \\
         & \cdot & \cdot &      &      &      \\
      v^{(1)} &  v^{(2)}&    & \ldots &    & v^{(\mu)} \\
   \end{pmatrix}\,,
\end{eqnarray}
where $v^{(i)}=(v_i,v_{\mu})$. The quantity $v^{(1)}$ is different
from zero. Otherwise the above matrix would be singular, which
cannot be since it corresponds to the bilinear form $\eta$ on the
invariant subspace. We can construct a matrix $R_i{}^j$ which
reduces the above matrix to $\epsilon\,\eta^{(\mu)}$, where
$\epsilon={\rm sign}(v^{(1)})$. It has the following form
\begin{eqnarray}
R_i{}^j&=&\begin{pmatrix}
        a_\mu  &   & \ldots    &    &        a_2    &  a_1 \\
        &      &      &   \cdot  &  \cdot  &      \\
       \vdots  &      &  \cdot     &  \cdot   &   &       \\
         &   \cdot   &  \cdot     &      &      &     \\
       a_2   & \cdot &  &      &      &      \\
      a_1 &     &    & &    &  \\
   \end{pmatrix}\,,
\end{eqnarray}
where the coefficients $a_i$ are determined recursively
\begin{eqnarray}
a_1&=&\frac{1}{\sqrt{|v^{(1)}|}}\,\,\,;\,\,\,\,a_i=-\frac{1}{2\,a_1\,v^{(1)}}\,\left(\sum_{{\tiny
\begin{matrix}\ell\le i\cr j,k<i\cr\ell+j+k=i+2
\end{matrix}}}a_j\,a_k\,v^{(\ell)}\right)\,.
\end{eqnarray}
Now define a new basis of vectors
\begin{eqnarray}
\tilde{v}^\lambda_{\mu-i+1}&=&R_i{}^j\,v^\lambda_{j}\,.
\end{eqnarray}
They satisfy
\begin{eqnarray}
(\tilde{v}^\lambda_{i},\,\tilde{v}^\lambda_{i})&=&\epsilon\,\eta^{(\mu)}_{ij}\,\,\,,\,\,\,Q\,\tilde{v}^{\lambda}_i=\lambda
\,\tilde{v}^{\lambda}_i+\tilde{v}^{\lambda}_{i-1}\,.
\end{eqnarray}
Consider now a complex eigenvalue $\lambda=\lambda_1+i\,\lambda_2$.
We can define a basis of $2\mu$ real vectors $({\bf
v}^\lambda_{I})=({\bf v}^\lambda_{\alpha,i})$, where $\alpha=0,1$,
$i=1,\dots,\mu$ and $I=(\alpha,i)=((0,1),(1,1),\dots, (1,\mu))$, so
that
\begin{eqnarray}
Q\,{\bf v}^\lambda_{\alpha,i}&=&A_\alpha{}^\beta\,{\bf
v}^\lambda_{\beta,i}+{\bf v}^\lambda_{\alpha,i-1}\,,\qquad
\text{where}\qquad A_\alpha{}^\beta=\left(\begin{matrix}\lambda_1 &
-\lambda_2\cr \lambda_2 & \lambda_1
\end{matrix}\right).\label{eqc}
\end{eqnarray}
Eq. (\ref{eqc}) is solved by vectors of the form ${\bf
v}^\lambda_{\alpha,i}=w_\alpha\otimes v^\lambda_{i}$, where
$w_\alpha^T\eta^{(2)}\,w_\beta=\pm\eta^{(2)}_{\alpha\beta}$. Using
the symmetry properties of $Q$ and $A_\alpha{}^\beta$, one can
easily show that the components $v^\lambda_{i}$ satisfy eq.
(\ref{sameasbefore}). Therefore, using the same matrix $R_i{}^j$ we
can define a new set of vectors
\begin{eqnarray}
\tilde{{\bf v}}^\lambda_{\alpha,\mu-i+1}&=&R_i{}^jw_\alpha\otimes
v_j^\lambda\,,
\end{eqnarray}
which still satisfy eq.(\ref{eqc}) and which are pseudo-orthogonal
\begin{eqnarray}
(\tilde{{\bf v}}^{\lambda}_I,\,\tilde{{\bf v}}^{\lambda}_J) &\equiv
&\eta^{(2)}_{\alpha\beta}\,\eta^{(\mu)}_{ij}=\eta^{(2\mu)}_{IJ}\,.
\end{eqnarray}
Consider now the matrix
\begin{eqnarray}
T&=&\left((\tilde{v}^{\lambda_1}_{i_1}),\dots
,\,(\tilde{v}^{\lambda_k}_{i_k}),(\tilde{{\bf
v}}^{\lambda_{k+1}}_{I_{k+1}}),\dots,\,(\tilde{{\bf
v}}^{\lambda_{s}}_{I_{s}})\right)\,.
\end{eqnarray}
The matrix $T$ is pseudo-orthogonal
\begin{eqnarray}
T^T\eta T&=&\left(\begin{matrix} \epsilon_1\,\eta^{(\mu_1)} & & & &
& \cr
                             &\ddots & & & & \cr
                              & &\epsilon_k\,\eta^{(\mu_k)}& & & \cr
&&&\eta^{(2\mu_{k+1})}& &\cr
                       &&&&\ddots & \cr
                      &&&& & \eta^{(2\mu_{s})}
\end{matrix}\right)\,,
\end{eqnarray}
and, moreover, $Q_N=T^{-1}Q T$.\par If $Q$ is diagonalizable
$\mu_i=1$, for $i=1,\dots, s$. From eq. (\ref{andone}) we see that
there must exist $q-p$ real eigenvalues $\sigma_i$ with
$\epsilon_i=+1$, while  among the remaining real eigenvalues will be
the same number of $\epsilon_i=+1$ and $\epsilon_i=-1$. From eq.
(\ref{andtwo}) it follows that there can be at most $p$ complex
eigenvalues ($s-k\le p$). From these observations we conclude that
the normal form of a diagonalizable $Q$ can be written, upon a
change of basis, as follows
\begin{eqnarray}
Q_N&=&\left(\begin{matrix}B_1 & & & & & &\cr
                              &\ddots & & & & &\cr
                                & &B_p & & & &\cr
                                   &  & &\sigma_1 & & &\cr
                                      & & & &\ddots & &\cr
                                         & & & & &
                                         \sigma_{q-p}&\cr\end{matrix}\right)\,,\label{normald}
\end{eqnarray}
where each $B_i$ is a $2\times 2$ matrix of the form
\begin{eqnarray}
B_i&=&\left(\begin{matrix}a_i+b_i & c_i\cr -c_i &a_i-b_i
\end{matrix}\right)\,,
\end{eqnarray}
and is meant to be acted on by an $\SO(1,1)$ transformation which
will further reduce it as follows
\begin{eqnarray}
B_i&\longrightarrow &\begin{cases}\left(\begin{matrix}a_i &
\sqrt{c_i^2-b_i^2}\cr -\sqrt{c_i^2-b_i^2} &a_i
\end{matrix}\right) & c_i^2>b_i^2\cr \left(\begin{matrix}a_i +\sqrt{b_i^2-c_i^2}&
0\cr 0 &a_i-\sqrt{b_i^2-c_i^2}
\end{matrix}\right) & b_i^2 >  c_i^2\end{cases}\,.
\end{eqnarray}
In the former case the block will have complex eigenvalues while in
the latter it will have real eigenvalues with opposite signs for
$\epsilon_i$. The normal form (\ref{normald}) can be written as a
generator in the following coset
\begin{eqnarray}
Q_N&\in
&\left(\frac{\frak{sl}(2,\Real)}{\frak{so}(1,1)}\right)^p\times\frak{so}(1,1)^q\,,\label{normspace}
\end{eqnarray}
where the $\frak{so}(1,1)^q$ factors are parameterized by
$a_i,\,\sigma_i$. We shall find that for various symmetric
pseudo-Riemannian spaces $G/H^*$, one can always define a space of
the form (\ref{normspace}), which contains the normal form of a
diagonalizable $Q$. This simplifies considerably the study of
geodesics generated by a diagonalizable $Q$. According to our
previous analysis, a non-diagonalizable $Q$ can be reduced to a
normal form $Q_N$ which is the sum of a matrix $Q_N^{(0)}$ of the
form (\ref{normald}), with degenerate diagonal blocks, and a
nilpotent matrix $Nil$ commuting with $Q_N^{(0)}$
\begin{eqnarray}
Q_N&=&Q_N^{(0)}+Nil\,\,;\,\,\,\,[Q_N^{(0)},\,Nil]=0\,.\label{nondiagonal}
\end{eqnarray}
In what follows we shall give an intrinsic geometrical meaning to
the normal form associated with diagonalizable matrices $Q$,
characterizing it as an element of the tangent space of a suitable
submanifold of the original space, defining a truncation of the
original theory.

\subsection{Group Theory of Kaluza--Klein Reduction}

In the next subsection we present and prove a theorem about the
normal form of a diagonalizable element of the class of
(non-)split coset spaces $\frak{g}/\frak{H}^*$ arising in
Kaluza-Klein reductions involving the time direction. In section 4
we are able to derive explicit expressions for the generating
solution in $D$ dimensions using the normal form. In this
subsection we first introduce the group theoretical ingredients
that are needed to formulate and prove the theorem. In particular,
we need the group theory of Kaluza--Klein reductions. We first
consider the split case. The cases $D>3$ and $D=3$ are considered
separately.

\subsubsection{Split Group $G$}
\paragraph{Dimension $ D>3$.}

Suppose we construct the Euclidean $D$--dimensional theory by
reducing 11-dimensional supergravity first to $D+1$ dimensions on an
Euclidean torus and then by further reducing to $D$ spacelike
dimensions along the time direction. We denote by $G_{D+1}$ and
$H_{D+1}$ the isometry group of the scalar manifold and its maximal
compact subgroup in the $D+1$ dimensional theory, respectively. Let
also ${\bf R}$ denote the $G_{D+1}$-representation of the vector
fields in $D+1$ dimensions, and $R={\rm dim} {\bf R}$. The isometry
group $G$ in $D$ dimensions contains $G_{D+1}\times {\rm SO}(1,1)$,
where the ${\rm SO}(1,1)$ factor acts as a rescaling on the radial
modulus of the timelike internal circle. The theory in $D+1$
dimensions is maximally supersymmetric and therefore both $G_{D+1}$
and $G$ are split groups (i.e.~maximally non-compact real forms of
their complexifications). With respect to the $G_{D+1}\times {\rm
SO}(1,1)$-subgroup, the following branching holds
\begin{eqnarray}
{\bf Adj}(G)&\rightarrow & {\bf Adj}(G_{D+1})_0+{\bf 1}_0+{\bf
R}_{+1}+{\bf R}_{-1}\,,\label{branch>3}
\end{eqnarray}
where the subscript refers to the ${\rm SO}(1,1)$--grading. We shall
denote by $r$ the rank of the coset $G/H^*$\footnote{We define the
\emph{rank} of $G/H^*$ as the maximum number of hermitian,
i.e.~non-compact, Cartan generators in $\frak{g}/\frak{H}^*$. The
rank of $G/H^*$ coincides with the rank of $G/H$. Here we use the
term non-compact to refer to hermitian generators.}, which coincides
with the rank of $G$ (i.e.~the dimension of the Cartan subalgebra)
if $G$ is split. If $\{{\alpha}_i\}$, $i=1,\dots, r$, is a basis of
simple roots of $\frak{g}$ and $\alpha$ a generic positive root, we
describe $\frak{g}$ in terms of a Cartan basis of generators
\begin{eqnarray}
\{t_n\}&=&\{H_{\alpha_i},\,E_\alpha,\,E_{-\alpha}\}\,,\label{cbasis}
\end{eqnarray}
where $H_\alpha=\alpha^i\,H_i$ and $\{H_i\}$ is an orthonormal basis
of Cartan generators. For the sake of simplicity $\alpha$ also
denotes an index running on the corresponding positive roots. The
following conventions are used for the commutation relations (see
appendix \ref{appendix A})
\begin{eqnarray}
[H_\alpha,\,E_\beta]&=&(\alpha\cdot\beta)\,E_\beta\,\,;\,\,\,[E_\alpha,\,E_{-\alpha}]=H_{\alpha}\,.
\end{eqnarray}
A suitable combination $H_0$ of the $H_{\alpha_i}$ generates the
${\rm SO}(1,1)$ complement of $G_{D+1}$ in $G$. The roots $\alpha$
naturally split into the $G_{D+1}$ roots $\beta$ and roots $\gamma$
such that: $\beta(H_0)=0$ and $\gamma(H_0)>0$. Here $\beta(H_0)$ and
$\gamma(H_0)$ indicate the grading of $E_\beta$ and $E_\gamma$ with
respect to $H_0$, respectively. We shall denote by $\beta_\ell$,
$\ell=1,\dots, r_{D+1}=\mbox{rank}(\frak{g}_{D+1})$, the simple
roots of $\frak{g}_{D+1}$. The Cartan subalgebra of $\frak{g}$
correspondingly splits into the direct sum of the Cartan subalgebra
of $\frak{g}_{D+1}$, generated by $H_{\beta_\ell}$, and the
orthogonal one--dimensional space generated by $H_0$. The spaces
${\bf R}_{+1}$ and ${\bf R}_{-1}$ are spanned by the generators
$E_\gamma$ and $E_{-\gamma}$ respectively.
%It is known that ${\bf R}_{+1}$ is
%parametrized by the Peccei-Quinn scalars in $D$--dimensions and can
%be characterized as the maximal abelian subalgebra of $\frak{g}$.
We can consider a basis of generators for $\frak{g}$ which is
orthogonal with respect to the invariant Cartan--Killing metric, and
decompose it as follows
\begin{eqnarray}
\frak{g}&=&\mathcal{K}\oplus \frak{H}^*\,,\nonumber\\
 \mathcal{K}&=&
 \mathcal{K}_{D+1}\oplus \{H_0\}\oplus \mathcal{K}^{(R)}\,,\nonumber\\
\frak{H}^*&=&\{J_\alpha\}=\mathcal{J}_{D+1}\oplus
\mathcal{J}^{(R)}\,,\label{kj>3}
\end{eqnarray}
where in terms of the $\frak{g}$ generators the above spaces have
the following form
\begin{eqnarray}
 \mathcal{K}_{D+1}&=&\{H_{\beta_\ell},\,E_\beta+E_{-\beta}\}\,\,\,;\,\,\,\,\mathcal{K}^{(R)}=\{E_\gamma-E_{-\gamma}\}\,,\nonumber\\
 \mathcal{J}_{D+1}&=&\{E_\beta-E_{-\beta}\}\,\,\,;\,\,\,\,\mathcal{J}^{(R)}=\{E_\gamma+E_{-\gamma}\}\,.
\end{eqnarray}
The Lie algebra $\frak{H}^*$ generates the group $H^*$, its
subalgebra $ \mathcal{J}_{D+1}$ generates the maximal compact
subgroup $H_{D+1}$ of $G_{D+1}$ and $\mathcal{K}_{D+1}$ locally
generates the scalar manifold in $D+1$ dimensions:
$G_{D+1}/H_{D+1}=\exp(\mathcal{K}_{D+1})$. We see that the maximal
compact subgroup $H_c$ of $H^*$ coincides with $H_{D+1}$. Under the
adjoint action of $H_{D+1}$ both spaces $\mathcal{K}^{(R)}$ and
$\mathcal{J}^{(R)}$ transform in the representation ${\bf R}$ of the
$D+1$ dimensional vector fields. We may choose a parametrization of
$G/H^*$ so that it is locally described as
\begin{eqnarray}
\frac{G}{H^*}&=& \exp{(\mathcal{K})}\,.
\end{eqnarray}
The metric on the above space is then the restriction of the Cartan
-Killing metric on $\frak{g}$ into $\mathcal{K}$: its entries are
positive on the non-compact generators in
$\mathcal{K}_{D+1}+\{H_0\}$ and negative on the compact generators
in $\mathcal{K}^{(R)}$. We may also choose a \emph{solvable}
parametrization for $G/H^*$ which consists in describing, in a local
coordinate patch\footnote{The solvable parametrization for $G/H^*$,
in contrast to the $G/H$ case in which $H$ is the maximal compact
subgroup of $G$, holds only locally. To understand this issue, one
can think of the simple case of $dS_2={\rm SO}(1,2)/{\rm SO}(1,1)$,
in which the solvable parametrization describes the \emph{stationary
universe} and thus covers only half the hyperboloid
\cite{Hawking:1973uf,Keurentjes:2005jw}.}, the scalar manifold as a
solvable group manifold generated by the Borel subalgebra $Solv$ of
$G$
\begin{eqnarray}
\frac{G}{H^*}&=&\exp(Solv)\,\,\,;\,\,\,\,Solv=\{H_{\alpha_i},\,E_\alpha\}\,.
\end{eqnarray}
This description is convenient since the parameters of $Solv$ can be
directly identified with the dimensionally reduced string
zero-modes. The space $Solv$ is endowed with a  metric
$(\cdot,\,\cdot)$ defined as
\begin{eqnarray}
(H_i,\,H_j)&=&2\,\delta_{ij}\,\,\,;\,\,\,\,(E_{\beta},\,E_{\beta^\prime})=\delta_{\beta\beta^\prime}\,\,\,;\,\,\,\,(E_{\gamma},\,E_{\gamma^\prime})=
-\delta_{\gamma\gamma^\prime}\,,
\end{eqnarray}
which induces the metric on the manifold. The Borel subalgebra of
$G$ decomposes with respect to the Borel subalgebra $Solv_{D+1}$ of
$G_{D+1}$ as follows
\begin{eqnarray}
Solv&=&Solv_{D+1}\oplus \{H_0\} \oplus {\bf R}_{+1}\,.
\end{eqnarray}
In the solvable parametrization the generators of ${\bf R}_{+1}$
are parametrized by the Peccei--Quinn scalars in the
$D$-dimensional theory \cite{Andrianopoli:1996bq,Cremmer:1997ct}.

\paragraph{The $D=3$ Case}
In $4$ dimensions the electric and magnetic charges together span
an irreducible representation ${\bf R}$ of $G_4$. Upon dimensional
reduction on the time direction and dualization of the vector
fields into scalars, the isometry group $G$ of the resulting
moduli space now contains $G_4\times {\rm SL}(2,\Real)$ with
respect to which its adjoint representation branches as follows
\begin{eqnarray}
{\bf Adj G}&\longrightarrow & {\bf (Adj G_4,1)} + {\bf (1,3)}+{\bf
(R,2)}\,.\label{3branch}
\end{eqnarray}
The generator $H_0$ parametrized by the radial modulus of the
internal circle is the Cartan generator of the ${\rm SL}(2,\Real)$
factor. The positive roots $\alpha$ of $G$ now split into the
$G_4$ positive roots $\beta$, the roots $\gamma$  and a new root
$\beta_0$ such that: $\beta(H_0)=0$, $\gamma(H_0)=1$ and
$\beta_0(H_0)=2$. The ${\rm SL}(2,\Real)$ group is then generated
by $H_0,\,E_{\pm \beta_0}$, being $H_0=H_{ \beta_0}$. The
generator $H_0$ induces then a double grading structure on $Solv$
which decomposes as follows
\begin{eqnarray}
Solv&=& \left(Solv_4 \oplus \{H_0\}\right)_0 \oplus \{E_{
\beta_0}\}_{+2} \oplus {\bf R}_{+1}\,.\label{solvsolv4}
\end{eqnarray}
The space ${\bf R}_{+1}$ is generated by $E_\gamma$ and
parameterized by the scalar fields originating from the $D=4$ vector
fields and the corresponding conserved charges are the electric and
magnetic charges. The generator $E_{ \beta_0}$ is associated with
the axion dual to the Kaluza-Klein vector and the corresponding
conserved charge is the Taub-NUT charge. As a consequence of the
double grading structure, ${\bf R}_{+1}$ is no longer an abelian
subalgebra but, together with $E_{ \beta_0}$ close as a Heisenberg
algebra
\begin{eqnarray}
[E_\gamma,\,E_{\gamma^\prime}]&=&\mathbb{C}_{\gamma\gamma^\prime}\,E_{
\beta_0}\,,
\end{eqnarray}
where $\mathbb{C}_{\gamma\gamma^\prime}$ is a symplectic invariant
matrix. The above properties of the $D=3$ theory are general and
hold also in the non-maximal supergravities (for symmetric scalar
manifolds). Let us now consider the cases in which $G$ and $G_4$ are
split. Similarly to what we did for $D>3$, we can define the
following spaces
\begin{eqnarray}
\frak{g}&=&\mathcal{K} \oplus \frak{H}^*\,,\nonumber\\
 \mathcal{K}&=&
 \mathcal{K}_{4} \oplus \{H_0,\,E_{\beta_0} +E_{-\beta_0}\} \oplus \mathcal{K}^{(R)}\,,\nonumber\\
\frak{H}^*&=&\{J_\alpha\}=\mathcal{J}_{4} \oplus
\{E_{\beta_0}-E_{-\beta_0}\} \oplus \mathcal{J}^{(R)}\,,\label{kj=3}
\end{eqnarray}
where, in terms of the $\frak{g}$ generators, the above spaces have
the following form
\begin{eqnarray}
 \mathcal{K}_{4}&=&\{H_{\beta_\ell},\,E_\beta+E_{-\beta}\}\,\,\,;\,\,\,\,\mathcal{K}^{(R)}=\{E_\gamma-E_{-\gamma}\}\,,\nonumber\\
 \mathcal{J}_{4}&=&\{E_\beta-E_{-\beta}\}\,\,\,;\,\,\,\,\mathcal{J}^{(R)}=\{E_\gamma+E_{-\gamma}\}\,.
\end{eqnarray}
We see that in $D=3$ the maximal compact subgroup $H_c$ of $H^*$
can be written as $H_c=H_{D+1}\times{\rm U}(1)$ where the ${\rm
U}(1)$ factor is generated by $E_{\beta_0}-E_{-\beta_0}$, while,
as in $D>3$, $H_{D+1}=\exp(\mathcal{J}_{D+1})$.\par
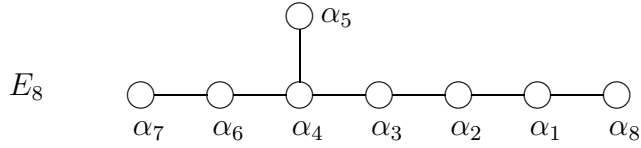
\begin{figure}
\centering
\begin{picture}(100,100)
        \put (-70,35){$E_8$} \put (-20,35){\circle {10}} \put
(-23,20){$\alpha_7$} \put (-15,35){\line (1,0){20}} \put
(10,35){\circle {10}} \put (7,20){$\alpha_6$} \put (15,35){\line
(1,0){20}} \put (40,35){\circle {10}} \put (37,20){$\alpha_4$} \put
(40,65){\circle {10}} \put (48,62.8){$\alpha_5$} \put (40,40){\line
(0,1){20}} \put (45,35){\line (1,0){20}} \put (70,35){\circle {10}}
\put (67,20){$\alpha_{3}$} \put (75,35){\line (1,0){20}} \put
(100,35){\circle {10}} \put (97,20){$\alpha_{2}$} \put
(105,35){\line (1,0){20}} \put (130,35){\circle {10}} \put
(127,20){$\alpha_1$} \put (135,35){\line (1,0){20}} \put
(160,35){\circle {10}} \put (157,20){$\alpha_8$}
\end{picture}
\vskip 1cm \caption{The Dynkin diagrams of $E_{8(8)}$ and the
labeling of simple roots} \label{except2}
\end{figure}

 Let us
consider as an example the Euclidean maximally supersymmetric
theory in $D=3$, in which $G={\rm E}_{8(8)}$, $H^*={\rm
SO}^*(16)$, $G_4={\rm E}_{7(7)}$, ${\bf R}={\bf 56}$ and $H_4={\rm
SU}(8)$. The Dynkin diagram of $\frak{e}_{8(8)}$ is represented in
Fig. \ref{except2}. The simple roots are ordered in such a way
that $\alpha_1,\dots, \alpha_7$ define the $\frak{e}_{7(7)}$
subalgebra.
The decomposition (\ref{3branch}) reads
\begin{eqnarray}
{\bf 248}&\longrightarrow & {\bf (133,1)}+{\bf (1,3)}+{\bf
(56,2)}\,,\label{e8e7}
\end{eqnarray}
 and the root $\beta_0$ in this
representation has the form
\begin{eqnarray}
\beta_0&=&(3,4,5,6,3,4,2,2)\,,
\end{eqnarray}
in the simple root basis $\{\alpha_i\}$. If we introduce the dual
basis of simple weights $\{\lambda^i\}$, $\lambda^i\cdot
\alpha_j=\delta_j^i$, one can verify that $\beta_0=\lambda^8$. The
grading of a generator $E_\alpha$ with respect to $H_0=H_{\beta_0}$,
which is the scalar product $\alpha\cdot \beta_0=\alpha\cdot
\lambda^8$, defines therefore the level of $\alpha$ with respect to
$\alpha_8$. The  decomposition of $\alpha$ into
$\beta,\,\gamma,\,\beta_0$ is nothing else than a \emph{level
decomposition} relative to the root $\alpha_8$.

\subsubsection{Non-split Group $G$}\label{nsg}
In this subsection we  discuss symmetric manifolds with a non-split
isometry group $G$ which is relevant for the case of Kaluza-Klein
reduction of non-maximal supergravity theories \cite{Westra:2005xi}.
We first recall some basic facts. The Lie algebra
$\frak{g}_{\mathbb{C}}$ of the complexification $G_{\mathbb{C}}$ of
$G$ is written in terms of the Lie algebra $\frak{g}$ of $G$ as
$\frak{g}_{\mathbb{C}}=\frak{g}+i\,\frak{g}$. Let $\sigma$ denote
the conjugation with respect to $\frak{g}$:
$\sigma(\frak{g})=\frak{g},\,\sigma(i\,\frak{g})=-i\,\frak{g}$. The
Cartan subalgebra $\frak{h}=\frak{h}[\frak{g}]$ of $\frak{g}$ in
general splits into two orthogonal subspaces:
$i\,\frak{h}_H=i\,\frak{h}_H[\frak{g}]$ consisting of compact (i.e.
having imaginary eigenvalues) generators and
$\frak{h}_K=\frak{h}_K[\frak{g}]$ consisting of non-compact (i.e.
having real eigenvalues) generators. We shall consider the Cartan
subalgebra $\frak{h}$ for which $\frak{h}_K$ has maximal dimension.
In the split case this choice implies $\frak{h}=\frak{h}_K$. In
general $r={\rm dim}(\frak{h}_K)={\rm rank}(G/H)$, $H$, as usual
denoting the maximal compact subalgebra of $G$. The positive roots
of $\frak{g}_{\mathbb{C}}$ split into two spaces:
$\tilde{\Delta}=\tilde{\Delta}[\frak{g}]$ which consists of the
positive roots having a non-vanishing restriction to $\frak{h}_K$,
and $\Delta_0=\Delta_0[\frak{g}]$ consisting of the positive roots
$\alpha$ such that $\alpha(\frak{h}_K)=0$. With each positive root
$\alpha$ we can associate a conjugate one $\alpha^\sigma$ such that
$\sigma(E_{\alpha})=E_{\alpha^\sigma}$. The two roots are related as
follows: $\alpha_{|\frak{h}_K}=\alpha^\sigma_{|\frak{h}_K}$,
$\alpha_{|\frak{h}_H}=-\alpha^\sigma_{|\frak{h}_H}$. One can easily
verify that, for $\alpha\in \Delta_0$, $\alpha^\sigma=-\alpha$. As
in the split case, the scalar manifold $G/H^*$ can be locally
represented as a solvable Lie group $G/H^*=\exp(Solv)$, where the
subalgebra $Solv$ is generated by the non-compact Cartan generators
in $\frak{h}_K$ and by the $\sigma$-invariant combinations of
$E_{\alpha}$ and $E_{\alpha^\sigma}$
\begin{eqnarray}
Solv&=&\frak{h}_K+\{E_{\alpha}+E_{\alpha^\sigma},\,i\,(E_{\alpha}-E_{\alpha^\sigma})\}_{\vert
\alpha\in \tilde{\Delta}}\,.
\end{eqnarray}
If $G$ is the isometry group of the Euclidean theory obtained from
time-reduction of a $D+1$ dimensional supergravity, the space
$\tilde{\Delta}$ also contains the roots $\gamma$ corresponding to
the scalar fields which originate from the vector fields in one
dimension higher. Just for the split case, the space
$\mathfrak{h}_K$ contains the Cartan generator $H_0$ which generates
a rescaling of the radius of the internal time-like dimension. This
generator introduces a grading structure in $Solv$: The shift
generators corresponding to the $\gamma$-roots have grading $+1$
while their hermitian conjugates have grading $-1$. In terms of
$H_0$ it is possible to define a Wick rotation mapping $G/H$ into
$G/H^*$ in precisely the same way as discussed in Appendix
\ref{APPENDIXC} for the split case. The Cartan decomposition of the
algebra $\frak{g}$ defines the algebra $\frak{H}^*$ of $H^*$ and the
space $\mathcal{K}$: $\frak{g}=\frak{H}^*+\mathcal{K}$. The space
$\mathcal{K}$ is spanned by the $\frak{h}_K$ generators, by the
non-compact components of the nilpotent generators in $Solv$
 for $\alpha\neq \gamma$ and the compact components of the shift
 generators corresponding to the $\gamma$ roots, spanning the space $\mathcal{K}^{(R)}$ (here ${\bf R}$ still denotes the $G_{D+1}$--representation
 in which the $D+1$ electric (and magnetic for $D=3$) charges transform). The algebra
 $\frak{H}^*$ consists of the compact Cartan generators in
 $i\,\frak{h}_H$, the compact components of the nilpotent $Solv$-
 generators for $\alpha\neq \gamma$, the non-compact components of the shift
 generators corresponding to the $\gamma$ roots, , spanning the space $\mathcal{J}^{(R)}$, and of the compact
 generators $E_{\alpha}-E_{-\alpha},\,i\,(E_{\alpha}+E_{-\alpha})$
 with $\alpha\in \Delta_0$. Just as for the split case, if we
 replace in $\mathfrak{H}^*$ the subspace $\mathcal{J}^{(R)}$ by
 $\mathcal{K}^{(R)}$, we obtain the algebra $\mathfrak{H}$ of the
 maximal compact subgroup $H$ of $G$. In other words the space $\mathcal{J}^{(R)}$ generates the coset $H^*/H_c$,
 $H_c$ being the maximal compact subgroup of $H^*$.
 \par
  We shall define the \emph{paint} group
 $G_{paint}[G]$ of the group $G$ the maximal subgroup of $H$
 which commutes with $\frak{h}_K$. It is generated by the following  Lie algebra
 $\frak{g}_{paint}$:
 \begin{eqnarray}
\frak{g}_{paint}&=&i\,\frak{h}_H+\{E_{\alpha}-E_{-\alpha},\,i\,(E_{\alpha}+E_{-\alpha})\}_{\vert
\alpha\in {\Delta}_0}\,.
 \end{eqnarray}
$G_{paint}[G]$ is the automorphism group of $Solv$ and was discussed
in \cite{Keurentjes:2002rc,Fre':2005sr,Fre:2006eu}. In the split
case we clearly have $\frak{g}_{paint}=\varnothing$. Let us denote
by $n_+={\rm Card}(\tilde{\Delta})$ and by ${n}_0={\rm
Card}({\Delta}_0)$. Some general relations are
\begin{eqnarray}
{\rm dim}(G_{paint}[G])&=&({\rm rank}(G)-r)+2\,n_0
\,,\nonumber\\{\rm dim}(H)&=&{\rm dim}(H^*)=({\rm
rank}(G)-r)+n_++2\,n_0\,,\nonumber\\{\rm dim}(\frac{G}{H^*})&=&{\rm
dim}(\frac{G}{H})=r+n_+\,.
\end{eqnarray}
\indent Since the space $\mathcal{K}$ contains both compact and
non-compact generators, we may choose a Cartan subalgebra
$\mathfrak{h}$ of $\mathfrak{g}$ for which
$\mathfrak{h}_{\mathcal{K}}=\mathfrak{h}\cap \mathcal{K}$ still
has maximal dimension, but contains compact generators, given by
the intersection
$\mathfrak{h}_{\mathcal{K}}\cap\mathcal{K}^{(R)}$. From general
arguments it follows that ${\rm
dim}\mathfrak{h}_{\mathcal{K}}={\rm dim}\mathfrak{h}_{K}=r$,
though the two spaces are in general inequivalent, since
$\mathfrak{h}_{\mathcal{K}}$ may contain compact generators, while
$\mathfrak{h}_{K}$ by definition is non-compact. For a particular
choice of $\mathfrak{h}$,
$\mathfrak{h}_{\mathcal{K}}\cap\mathcal{K}^{(R)}={\O}$ and
$\mathfrak{h}_{\mathcal{K}}=\mathfrak{h}_{K}$. Thus a choice of
$\mathfrak{h}_{\mathcal{K}}$ is characterized by the number of
compact generators it contains, namely by ${\rm
dim}(\mathfrak{h}_{\mathcal{K}}\cap\mathcal{K}^{(R)})$. The
maximum number of independent compact generators that a space
$\mathfrak{h}_{\mathcal{K}}$ can have is given by the maximum
number of mutually commuting generators in $\mathcal{K}^{(R)}$.
Since $\mathcal{K}^{(R)}$ has the same algebraic properties,
within $\mathfrak{g}$, as $i\,\mathcal{J}^{(R)}$, the maximum
number of commuting generators in $\mathcal{K}^{(R)}$ coincides
with the maximum number of commuting generators in
$\mathcal{J}^{(R)}$. Using the property
$\mathfrak{H}^*/\mathfrak{H}_c=\mathcal{J}^{(R)}$, by definition,
the maximal number of commuting generators in $\mathcal{J}^{(R)}$
is the rank $p$ of the coset $H^*/H_c$. We conclude that
$\mathfrak{h}_{\mathcal{K}}$, for different choices of
$\mathfrak{h}$,  can have at most $p$ independent compact
generators. An important part of our subsequent discussion will be
to characterize these $p$ generators
$\mathcal{J}_k,\,\mathcal{K}_k$, $k=1,\dots, p$, inside
$\mathcal{J}^{(R)}$ and $\mathcal{K}^{(R)}$ respectively. As we
shall show, they define the \emph{normal form} of the
representation ${\bf R}$ under the action of $G_{D+1}$ and are the
non-compact and  compact components respectively of the shift
generators corresponding to $p$ mutually orthogonal
$\gamma$-roots: $\gamma_1,\dots, \gamma_p$. These shift generators
and their hermitian conjugates, close $p$ $\mathfrak{sl}(2,\Real)$
subalgebras together with $p$ Cartan generators $H_k$, in
$\mathfrak{h}_K$. The orthogonal complement of $\{H_k\}$ in
$\mathfrak{h}_K$ generates an $\SO(1,1)^{r-p}$ group which
commutes with the $p$ $\mathfrak{sl}(2,\Real)$ algebras.
\par
 Now we are ready to state the theorem about the generating
geodesic on $G/H^*$ corresponding to a diagonalizable $Q.$

\subsection{A Theorem for Symmetric Spaces}
In analogy with the $\frak{gl}(p+q)/\frak{so}(p,q)$ example, one can
present a general formula for the normal form of a diagonalizable
element $Q$ of a class of  spaces $\frak{g}/\frak{H}^*$ occurring in
the kind of Euclidean Kaluza-Klein supergravities under
consideration. This normal form belongs in general to the following
subspace
\begin{equation}
Q_N \in \Bigl(\frac{\frak{sl}(2,\Real)}{\frak{so}(1,1)}\Bigr)^p
\times \frak{so}(1,1)^{r-p}\,,\label{eqqqqq}
\end{equation}
where the details of this are presented below. But, as in the
$\frak{gl}(p+q)/\frak{so}(p,q)$ example, there is a subspace (of
smaller dimension than the whole space) of elements which are not
'diagonalisable', namely whose
 minimal polynomial of $Q$ has degenerate roots. Then
the above formula should be adjusted with the addition of an extra
nilpotent piece that is constant (the normal form has fixed charge
in this nilpotent subspace), as discussed in Section
\ref{sectionsln}.
%An example of
%the normal form on a non-diagonalizable matrix in
%$\frak{sl}(4,\Real)/\frak{so}(2,2)$ with a $2$ times degenerate
%complex eigenvalue $\lambda\neq \bar{\lambda}$ is
%\begin{eqnarray}
%Q_N&=&Q^{(0)}_N+Nil\,,\nonumber\\Q^{(0)}_N&=&\left(\begin{matrix}A &
%0\cr 0 &
%A\end{matrix}\right)\,\,\,,\,\,\,\,\,Nil=\left(\begin{matrix}{\rm
%Id}_2 & {\rm Id}_2\cr -{\rm Id}_2 & -{\rm
%Id}_2\end{matrix}\right)\,\,\,,\,\,\,\,\,A=\left(\begin{matrix}\lambda_1
%& \lambda_2\cr -\lambda_2 &
%\lambda_1\end{matrix}\right)\,.\label{nondiagonal}
%\end{eqnarray}
In the following, the word diagonalisable will be used in this
generalised sense; the absence of a fixed nilpotent part.\par Let us
anticipate now the content of the general theorem for diagonalizable
$Q$, which will be discussed in detail in the following sections,
giving evidence for it by using the general results of Section
\ref{sectionsln}. We can consider the following general  embeddings
$\frak{g}\subset \frak{gl}({\rm dim}(\frak{g}))$, $\frak{H}^*\subset
\frak{so}(R,{\rm dim}(\frak{g})-R)$ so that we can write:
\begin{eqnarray}
Q&\in & \frac{\frak{g}}{\frak{H}^*}\subset \frac{\frak{gl}({\rm
dim}(\frak{g}))}{\frak{so}(R,{\rm dim}(\frak{g})-R)}\,.
\end{eqnarray}
If $Q$ is diagonalizable, using the results of Section
\ref{sectionsln}, we can write
\begin{equation}
Q_N \in
\left[\left(\frac{\frak{sl}(2,\Real)}{\frak{so}(1,1)}\right)^R
\times \frak{so}(1,1)^{{\rm
dim}(\frak{g})-R}\right]\,\cap\frac{\frak{g}}{\frak{H}^*}=\left(\frac{\frak{sl}(2,\Real)}{\frak{so}(1,1)}\right)^p
\times \frak{so}(1,1)^{r-p}\,,
\end{equation}
where $p$ is defined by the above intersection and is the maximal
number of commuting $\frak{so}(1,1)$ generators in $\frak{H}^*$.
This number will be characterized as the dimension of the normal
form of ${\bf R}$ under the action of $H_c$, maximal compact
subgroup of $H^*$. The same reasoning allows us to conclude that
the normal form of non-diagonalizable matrices Q can be written in
the form (\ref{nondiagonal}), namely as the sum of a generator
$Q_N^{(0)}$ in the space (\ref{eqqqqq}) and a nilpotent generator
$Nil$ commuting with it, though we shall postpone the task of
giving an intrinsic characterization of $Nil$ to a future work.

\subsubsection{The Theorem}\label{THEOREMA}
Consider an Euclidean supergravity arising from a time-like
dimensional reduction, with a pseudo-Riemannian symmetric scalar
manifold of the form $G/H^*$. Let  $Q $ be an element of the space $
\frak{g}/\frak{H}^*$ with $\frak{g}$ a maximal non-compact real
slice of a complex semi-simple lie algebra. Take $p=$
rank$[\frak{H}^*/\frak{H}_c]$ and $r=$ rank$[\frak{g}/\frak{H}]$,
with $\frak{H}_c$ the maximal compact subalgebra of $\frak{H}_*$,
and $\frak{H}$ the maximal compact subalgebra of $\frak{g}$. Then
the normal form of a diagonalisable $Q$ under Adj$ H^*$ is as
follows
\begin{equation}\label{theorem}
\boxed{Q_N \in
\Bigl(\frac{\frak{sl}(2,\Real)}{\frak{so}(1,1)}\Bigr)^p \times
\frak{so}(1,1)^{r-p}\,,}
\end{equation}
where the generators of each of the ${\rm SL}(2,\Real)$ groups are
$H_k,\,\mathcal{K}_k,\,\mathcal{J}_k$, $k=1,\dots, p$,
corresponding to a maximal set of $p$ mutually orthogonal $\gamma$
roots. They define a set of $p$ charges, which in $D=3$ can be
electric and magnetic, associated with the four-dimensional vector
fields.

If $Q$ is not diagonalisable then the above theorem is changed by
the addition of an extra constant nilpotent part, as explained
above. In the next sections we restrict to the diagonalisable cases
since they cover most of the solutions. In the next subsection we
give a general formal proof which holds for both the split and
non-split cases. We shall use general definitions and properties
introduced in subsection \ref{nsg}. It is followed by a constructive
proof, given for the split case only, in which the $H^*$
transformation which turns a generic $Q$ into its normal form $Q_N$
is defined. Although an analogous construction for the non-split
case would follow the same lines, it will not be explicitly given.

\subsubsection{The Proof} \label{PROOF}
\paragraph{Formal proof.} Any diagonalizable element of
$\mathfrak{g}$ can be thought of as an element of a Cartan
subalgebra of $\mathfrak{g}$. This implies that its spectrum
(eigenvalues with their multiplicities) coincides with that of a
suitable element of a given
$\mathfrak{h}=\mathfrak{h}[\mathfrak{g}]$. If we take $Q\in
\mathcal{K}=\mathfrak{g}/\mathfrak{H}^*$, its spectrum coincides
with that of an element $Q_N$ of
$\mathfrak{h}_{\mathcal{K}}=\mathfrak{h}\cap \mathcal{K}$, for a
certain choice of $\mathfrak{h}$. The imaginary and real
eigenvalues of $Q_N$ are associated with the compact and
non-compact elements of $\mathfrak{h}_{\mathcal{K}}$ respectively.
According to the discussion in subsection \ref{nsg}, the right
hand side of (\ref{theorem}), reproduces, for various  choices of
the generator inside each $\frak{sl}(2,R)/\frak{so}(1,1)$
subspaces all possible inequivalent $\mathfrak{h}_{\mathcal{K}}.$
Each coset $\mathfrak{sl}(2,\Real)/\mathfrak{so}(1,1)$ is
generated by one of the $p$ elements of the maximal abelian
subalgebra of $\mathcal{K}^{(R)}$ and by the corresponding $H_k$
generator. Depending on the invariant properties of $Q$, or
equivalently of $Q_N$, its component on each
$\mathfrak{sl}(2,\Real)/\mathfrak{so}(1,1)$ subspace can be
rotated, by means of the corresponding $\SO(1,1)$ transformation,
into the compact or non-compact generator of the coset. Since, as
discussed in subsection \ref{nsg}, there can be at most $p$
compact generators in $\mathfrak{h}_{\mathcal{K}}$, there are
precisely $p$ coset spaces
$\mathfrak{sl}(2,\Real)/\mathfrak{so}(1,1)$ in (\ref{theorem}).
The remaining $\mathfrak{so}(1,1)^{r-p}$ factor represents the
orthogonal complement of the Cartan generators $\{H_k\}$ of
$(\mathfrak{sl}(2,\Real))^p$ within $\mathfrak{h}_{\mathcal{K}}$.

\paragraph{Constructive  proof for the split case.}
Consider the general case in which $G$ is split in a $D$-dimensional
theory. Since we have denoted by $H_c$ the maximal compact subgroup
of $H^*$, and by $\frak{H}_c$ its generating algebra, using eqs.
(\ref{kj>3}), (\ref{kj=3}), we can write
\begin{eqnarray}
\frak{H}^*&=&\frak{H}_c \oplus
\mathcal{J}^{(R)}\,\,\Rightarrow\,\,\,\,\frac{H^*}{H_c}=\exp{\mathcal{J}^{(R)}}\,.\label{hcoset}
\end{eqnarray}

Let $p$ denote the rank of the coset $H^*/H_c$. We notice that
${\bf R}$, besides being a representation of $G_{D+1}$, is also a
representation of $H_c$. By acting on ${\bf R}$ with $H_c$ one can
reduce it to its \emph{normal form} $R_N$. We shall denote by
$H_{cent}=\exp(\frak{H}_{cent})\in H_c$, the centralizer (little
group) of this normal form and by
$G_{cent}=\exp(\frak{g_{cent}})\subset G_{D+1}$, the centralizer
of $R_N$ in $G_{D+1}$. As previously pointed out, the space
$\mathcal{J}^{(R)}$ transforms in the ${\bf R}$ representation
under the adjoint action of $H_c$. This means that the number of
independent entries of the normal form $R_N$ equals the rank $p$
of the coset $H^*/H_c$.  By means of the adjoint action of $H_c$,
$\mathcal{J}^{(R)}$ can be reduced to its normal form
$\mathcal{J}_N^{(R)}$, consisting of $p$ commuting generators
which correspond to $p$ non-compact Cartan generators in
$\frak{H}^*$. By definition of $H_{cent}$ we have
\begin{eqnarray}
\forall J\in \mathcal{J}_N^{(R)} &: &\,\,\,H_{cent}^{-1}J
H_{cent}=J\,.
\end{eqnarray}
With respect to $H_{cent}$ the ${\bf R}$ representation therefore
branches in the following way
\begin{eqnarray}
{\bf R}&\rightarrow& p\times {\bf 1}+{\bf R}_1\,,\label{normal}
\end{eqnarray}
where ${\bf R}_1$ is a reducible representation of $H_{cent}$ of
dimension $r_1$ and the $p$ singlets define the normal form. The
reduction of ${\bf R}$ to $R_N$ is done by fixing the compact
generators in
\begin{equation}
\hat{\mathcal{J}}^{(R_1)}=\frak{H}_c/\frak{H}_{cent}\,.
\end{equation}
Under the adjoint action of $H_{cent}$ the spaces
$\mathcal{K}^{(R)}$ and $\mathcal{J}^{(R)}$ decompose as follows
\begin{equation}
\mathcal{K}^{(R)}=\mathcal{K}^{(R)}_N+\mathcal{K}^{(R_1)}\,,\qquad
\mathcal{J}^{(R)}=\mathcal{J}^{(R)}_N+\mathcal{J}^{(R_1)}\,,
\end{equation}
where the subspaces $\mathcal{K}^{(R_1)},\,\mathcal{J}^{(R_1)}$
transform in the ${\bf R}_1$ representation of $H_{cent.}$. The
$p$-dimensional subspaces $\mathcal{K}^{(R)}_N$ and
$\mathcal{J}^{(R)}_N$ are Abelian and their generators can be
written in the form
\begin{eqnarray}
\mathcal{K}^{(R)}_N&=&\{\mathcal{K}_k\}\equiv\{E_{\gamma_k}-E_{-\gamma_k}\}\,;\,\,\,\mathcal{J}^{(R)}_N=\{\mathcal{J}_k\}\equiv\{E_{\gamma_k}+E_{-\gamma_k}\}\,,
\end{eqnarray}
where $\{\gamma_k\}_{k=1,\dots, p}$ is a maximal set of mutually
orthogonal $\gamma$ roots. The $G_{D+1}$ roots $\beta$, and
$\beta_0$ in $D=3$, then split into roots $\hat{\beta}$ which are
orthogonal to $\gamma_k$ and $r_1$ remaining roots $\tilde{\beta}$
\begin{eqnarray}
\gamma_k\cdot \hat{\beta}&=&0\,\,\,\,\,\,\,\,k=1,\dots, p\,.
\end{eqnarray}
In $D=3$ the root $\beta_0$ is in the $\tilde{\beta}$ group since
$\beta_0\cdot \gamma=1$. The group $G_{cent}$ is then the maximally
non-compact  subgroup of $G_{D+1}$ defined by the roots
$\hat{\beta}$ and
\begin{eqnarray}
\frak{H}_{cent}&=&\{E_{\hat{\beta}}-E_{-\hat{\beta}}\}\,,
\end{eqnarray}
namely $H_{cent}$ is the maximal compact subgroup of $G_{cent}$. %and
%the rank $r_c$ of $G_{cent}/H_{cent}$ is therefore $r-p$.
The generators in $\hat{\mathcal{J}}^{(R_1)}$ are then found to be
\begin{eqnarray}
\hat{\mathcal{J}}^{(R_1)}&=&\{E_{\tilde{\beta}}-E_{-\tilde{\beta}}\}\,.
\end{eqnarray}
As an example consider $D=3$ maximal supergravity. In this case
${\bf R}={\bf 28}_++\overline{{\bf 28}}_-$ as a representation of
$H_c={\rm U}(8),$ and $p=\mbox{rank}(SO^*(16)/{\rm U}(8))=4.$ The
little group in $G_4$ is $G_{cent}={\rm SO}(4,4)$, defined by the
sub-Dynkin diagram $\{\alpha_3,\,\alpha_4,\,\alpha_5,\,\alpha_6\}$
and $H_{cent}={\rm SO}(4)\times{\rm SO}(4)$ is its maximal compact
subgroup. There are eight roots $\gamma$ which are orthogonal to the
$G_{cent}$ roots, namely such that their corresponding charges are
invariant under the action of $G_{cent}$. These eight roots do not
define the normal form yet, since the corresponding generators in
$\mathcal{K}^{(R)}$ are still mapped into one another by a residual
$\SO(2)^4$ group, which therefore has to be fixed. The result are
four roots $\gamma_k$ which define the normal form, which can be
chosen to be
\begin{eqnarray}
\gamma_1&=&\{0, 0, 0, 0, 0, 0, 0, 1\}\,,\nonumber\\
\gamma_2&=&\{2, 2, 2, 2, 1, 1, 0, 1\}\,,\nonumber\\
\gamma_3&=& \{2, 2, 3, 4, 2, 3, 2, 1\}\,,\nonumber\\
\gamma_4&=& \{2, 4, 5, 6, 3, 4, 2, 1\}\,.
\end{eqnarray}
The corresponding generators $E_{\gamma_k}$ define a set of four
conserved quantized charges in $D=4$. The above roots also define
the normal form of a consistent truncation of the maximal theory,
which originates from the $D=4$ STU model and which will be
considered when we study the generating solution of a class of
extreme black holes.\par Let us now go back to the general
discussion. It is useful to define $\hat{G}_{cent}$ as the
subgroup of $G$ obtained by extending $G_{cent}$ by possible ${\rm
O}(1,1)$ factors on whose Cartan generators the $p$ roots
$\gamma_k$ have a trivial value. The rank of
$\hat{G}_{cent}/H_{cent}$ is therefore $r-p$. We can now
reorganize the $\mathcal{K}$ generators in the following subspaces
\begin{eqnarray}
\mathcal{K}&=&\frac{\hat{\frak{g}}_{cent}}{\frak{H}_{cent}}+\{H_{\gamma_k}\}+\hat{\mathcal{K}}^{(R_1)}+\mathcal{K}^{(R)}\,,
\end{eqnarray}
where $\hat{\mathcal{K}}^{(R_1)}$ is the non-compact counterpart of
$\hat{\mathcal{J}}^{(R_1)}$ in $\frak{g}/\frak{H}^*$
\begin{eqnarray}
\hat{\mathcal{K}}^{(R_1)}&=&\{E_{\tilde{\beta}}+E_{-\tilde{\beta}}\}\,.
\end{eqnarray}
Starting from a generic $Q$ in $\mathcal{K}$, the proof now proceed
along the following steps.
\paragraph{Step 1} Through the action of $H_c$ reduce the components
of $Q$ along $\mathcal{K}^{(R)}$ to their normal form in
$\mathcal{K}^{(R)}_N$.
\paragraph{Step 2}
If $Q$ is diagonalisable, there always exists a representative of
the same $H^*$-orbit as $Q$, on which a transformation generated by
$\mathcal{J}^{(R_1)}$ and $\hat{\mathcal{J}}^{(R_1)}$ can set the
components in $\hat{\mathcal{K}}^{(R_1)}$ to zero. As a result we
can find a representative $Q_N$ in the same $H^*$-orbit as the
original $Q$, which lies in the space
\begin{eqnarray}
Q_N&\in&\frac{\hat{\frak{g}}_{cent}}{\frak{H}_{cent}}+\{H_{\gamma_k}\}+\mathcal{K}^{(R)}_N
= \left(\frac{\frak{sl}(2,\Real)}{\frak{so}(1,1)}\right)^p +
\frac{\hat{\frak{g}}_{cent}}{\frak{H}_{cent}}\,,\label{normcent}
\end{eqnarray}
where the $p$ $\frak{sl}(2,\Real)$ algebras are generated by
$H_k\equiv H_{\gamma_k},\,E_{\gamma_k}\pm E_{-\gamma_k}$.
\paragraph{Step 3}
We can still fix $H_{cent}$ to reduce
$\frac{\hat{\frak{g}}_{cent}}{\frak{H}_{cent}}$ into $r-p$ diagonal
entries. We can thus finally write
\begin{eqnarray}
Q_N&\in&\frac{\hat{\frak{g}}_{cent}}{\frak{H}_{cent}}+\{H_{\gamma_k}\}+\mathcal{K}^{(R)}_N=\left(\frac{\frak{sl}(2,\Real)}{\frak{so}(1,1)}\right)^p+
\frak{so}(1,1)^{r-p}\,.
\end{eqnarray}

This concludes the proof of the theorem, see eq.~\eqref{theorem}.
The consequence of this theorem is that the generating geodesic
curve is a solution to the following sigma model
\begin{equation}
\boxed{\d s^2=\sum_{i=1}^p \tfrac{1}{2}(\d\phi^i)^2
-\tfrac{1}{2}\e^{\beta_i\phi^i}(\d\chi^i)^2 +
\sum_{a=1}^{r-p}\tfrac{1}{2}(\d
\Phi^a)^2\,.}\label{generatingsubmanifold}
\end{equation}
This describes the metric on the totally geodesic submanifold
$[\SL(n,\Real)/\SO(1,1)]^p\times \SO(1,1)^{r-p}$ of $G/H^{*}$. The
real numbers $\beta_i$ correspond to the squared length of the roots
$\gamma_i$. \footnote{This statement is true up to an overall
constant that can be traced back to the fact that the form of the
coset-metric is defined up to an overall constant. For a particular
theory, this overall constant gets fixed by supersymmetry.} In the
case of maximal supergravity the cosets are all based on
simply-laced Lie algebras and therefore all $\beta_i$ equal two. The
results for the case of maximal supergravity are summarized in table
\ref{table:maximal supergravities}.

\begin{table}\renewcommand{\arraystretch}{2.2}
\begin{center}
{\scriptsize\begin{tabular}{|c|c|c|c|c|c|}\hline &
{\normalsize $G/H^*$} & {\normalsize $H^*/H_c$} & {\normalsize $p$} & {\normalsize ${\bf R}$} &{\normalsize $\hat{G}_{cent}$} \\
\hline \hline $D=9$ &
$\frac{\GL(2,\Real)}{{\rm SO}(1,1)}$ &${\rm SO}(1,1)$ & 1 & ${\bf 1}$ & ---\\
\hline $D=8$ &
        $\frac{\SL(3,\Real)\times \SL(2,\Real)}{{\rm SO}(2,1)\times{\rm SO}(1,1)}$ & $\frac{{\rm SL}(2,\Real)}{{\rm SO}(2)}\times {\rm SO}(1,1)$ &2& ${\bf 3}$ & ${\rm SO}(1,1)$  \\ \hline
$D=7$ &  $\frac{\SL(5,\Real)}{{\rm SO}(3,2)}$ & $\frac{{\rm
SO}(3,2)}{{\rm SO}(3)\times SO(2)}$ & 2 & ${\bf (3,2)}$
&${\rm SO}(1,1)^2$\\
\hline $D=6$ & $\frac{{\rm O}(5,5)}{{\rm O}(5, \mathbb{C})}$&
$\frac{{\rm O}(5, \mathbb{C})}{{\rm O}(5)}$& 2 & ${\bf 10}$& ${\rm
GL}(2,\Real)\times {\rm SL}(2,\Real) $
 \\ \hline
$D=5$& $\frac{E_{6(+6)}}{U\!Sp(4,4)}$
&$\frac{U\!Sp(4,4)}{U\!Sp(4)\times U\!Sp(4)}$&2&${\bf 16}$& ${\rm GL}(4,\Real)$ \\
\hline
$D=4$ &  $\frac{E_{7(+7)}}{\SU^*(8)}$ &$\frac{\SU^*(8)}{{ U\!Sp}(8)}$ & 3& ${\bf 27}$& ${\rm SO}(4,4)$ \\
\hline $D=3$ & $\frac{{\rm E}_{8(+8)}}{\SO^*(16)}$ &
$\frac{\SO^*(16)}{{\rm U}(8)}$ & 4 & ${\bf 56}$ & ${\rm SO}(4,4)$
\\ \hline
\end{tabular}}
\caption{\it The table displays for each Euclidean maximal
supergravity in $D$ dimensions, the scalar manifold $G/H^*$,
$H^*/H_c$, the number $p$, the representation $R$ of $G_{D+1}$ in
which the vectors (for $D=3$ the electric and magnetic charges) in
$D+1$ dimensions transform and $\hat{G}_{cent}$.}
\label{table:maximal supergravities}
\end{center}
\end{table}

\subsection{Half-maximal Supergravity}\label{HALF}
\label{nmax} In non-maximal supergravity we are dealing with both
split and non-split coset spaces \cite{Westra:2005xi}. The
construction of the normal form of $Q$ given in the previous
sections for the split case can be extended to the case in which
$G=G_D$ is non-split, which typically occur in non-maximal
supergravities. The proof proceeds by following precisely the same
steps as in the split case which we do not repeat here.

All coset spaces in half maximal supergravity are symmetric and are
listed in Table \ref{table: half maximal supergravities} where also
the results for the generating geodesic are given. As in the case of
maximal supergravity the $\beta_i$ are all equal to two and the
numbers $p$ in each dimension is the same as in maximal
supergravity. In fact, if one traces back the 10D origin using
appendix \ref{typeIIred} then one finds that for maximal
supergravity the d.o.f. of the generating submanifold
(\ref{generatingsubmanifold}) lies in the common sector of the 10D
supergravity theories. This explains the fact that we find the same
result for maximal and half-maximal supergravity theories.
\begin{table}\renewcommand{\arraystretch}{2.2}
\begin{center}
{\scriptsize\begin{tabular}{|c|c|c|c|c|c|}\hline
 & {\normalsize $G/H^*$} & {\normalsize $H_c$} & {\normalsize $p$} & {\normalsize ${\bf R}$} & {\normalsize $\hat{G}_{cent}$} \\ \hline
 \hline $D=9$ &
$\SO(1,1)\times\frac{\SO(1,1+n)}{{\rm SO}(1,n)}$ &${\rm SO}(n)$ & 1 & ${\bf n}$ & ${\rm SO}(1,1)\times \SO(n-1)$\\
\hline $D=8$ &
       $\SO(1,1)\times\frac{\SO(2,2+n)}{{\rm SO}(1,1)\times{\rm SO}(1,1+n)}$ & ${\rm SO}(1+n)$
       &2& ${\bf (n+1)+1}$ & ${\rm SO}(1,1)\times{\rm SO}(n)$  \\ \hline
$D=7$ &  $\SO(1,1)\times\frac{\SO(3,3+n)}{{\rm SO}(2,1)\times{\rm
SO}(1,2+n)}$ &  $\SO(2)\times{\rm SO}(2+n)$ & 2 & ${\bf (1,n+2)}+
{\bf (2,1)}$
&${\rm SO}(1,1)\times\SO(1,1+n)$\\
\hline $D=6$ & $\SO(1,1)\times\frac{\SO(4,4+n)}{{\rm
SO}(3,1)\times{\rm SO}(1,3+n)}$ & $\SO(3)\times{\rm SO}(3+n)$& 2 &
${\bf (1,n+3)}+ {\bf (3,1)}$& ${\rm SO}(1,1)\times\SO(2,2+n)$
 \\ \hline
$D=5$& $\SO(1,1)\times\frac{\SO(5,5+n)}{{\rm SO}(4,1)\times{\rm
SO}(1,4+n)}$
&$\SO(4)\times{\rm SO}(4+n)$&2&${\bf (1,n+4)}+ {\bf (4,1)}$& ${\rm SO}(1,1)\times\SO(3,3+n)$ \\
\hline $D=4$ &
$\frac{\SO(2,1)}{\SO(1,1)}\times\frac{\SO(6,6+n)}{{\rm
SO}(5,1)\times{\rm
SO}(1,5+n)}$&$\SO(5)\times{\rm SO}(5+n)$ & 3& ${\bf (1,n+5)}+ {\bf (5,1)}$& ${\rm SO}(4,4+n)$ \\
\hline $D=3$ & $\frac{\SO(8,8+n)}{{\rm SO}(6,2)\times{\rm
SO}(2,6+n)}$ & $\begin{matrix}{\rm SO}(6)\times
\SO(6+n)\times\\\SO(2)^2\end{matrix}$ & 4 & $\begin{matrix}{\bf
(1,n+6)}_++ {\bf (6,1)}_++\\{\bf (1,n+6)}_-+ {\bf
(6,1)}_-\end{matrix}$ & ${\rm SO}(4,4+n)$
\\ \hline
\end{tabular}}
\caption{\it The Table displays for each Euclidean half-maximal
supergravity in $D$--dimensions, the scalar manifold $G/H^*$,
$H_c$, the number $p$, the representation ${\bf R}$ of $G_{D+1}$
in which the vectors (for $D=3$ the electric and magnetic charges)
in $D+1$ dimensions transform and $\hat{G}_{cent}$.} \label{table:
half maximal supergravities}
\end{center}
\end{table}

\subsection{Quarter-maximal Supergravity}\label{QUARTER}
We now discuss the case of quarter-maximal supergravity. These
theories exist in $D\le 6$ dimensions. We consider three cases: the
$D=6\rightarrow D=5, D=5\rightarrow D=4$ and $D=4\rightarrow D=3$
timelike reductions. The results are summarized in Table
\ref{table:N=4supergravities} where the values $\beta_i$ and $p$ can
be found for each case. Below we expand a little on the results of
Table \ref{table:N=4supergravities} starting with the $D=3$
theories, which requires a special care.

\paragraph{$D=3$ theories}
If the three-dimensional theory has a symmetric scalar manifold
$G/H^{*}$, then so has its four-dimensional parent. The latter
manifold $G_4/H_4$ is then a Special K\"ahler manifold, the image
through the c-map of the quaternionic K\"ahler manifold $G/H$. A
feature of these models is that
\begin{eqnarray}
H^*&=& G_4\times {\rm SU}(1,1)\,,\label{hstar}
\end{eqnarray}
With respect to $H^*$ the adjoint representation of $G$ branches as
follows
\begin{eqnarray}
{\bf Adj G}&\longrightarrow & ({\bf Adj G}_4,\,{\bf 1})+({\bf
1},{\bf 3})+({\bf R},{\bf 2})\,,
\end{eqnarray}
which implies that the space $\mathcal{K }=\frak{g}/\frak{H}^*$
defined by the Cartan decomposition of $\frak{g}$, transforms in
the $({\bf R},{\bf 2})$ of $H^*$. A generic element $Q\in
\mathcal{K}$ thus has the form $Q=(Q^{M\,A})$, where $M=1,\dots,
{\rm dim}({\bf R})$ and $A=1,2$.

\par From the general form (\ref{hstar}) of $H^*$ we conclude that
\begin{eqnarray}
p&=&
\mbox{rank}\left(\frac{H^*}{H_c}\right)=\mbox{rank}\left(\frac{G_4}{H_4}\right)+1\,,
\end{eqnarray}
where, as usual, $H_c=H_4\times{\rm U}(1)$. We therefore have that
$r$, defined as the rank of $G/H$, coincides with $p$, i.e.~$p=r$.
Moreover, since the non compact generators in the coset
$\frak{H}^*/\frak{H}_c$ transform under the adjoint action of
$H_c$ in the representation ${\bf R}$, by definition of the rank
of a coset, the number $p$ is precisely the dimension of the
normal form $R_N$ of ${\bf R}$ with respect to the action of
$H_c$. Indeed, through the adjoint action of $H_c$, the generators
in $\frak{H}^*/\frak{H}_c$ can be rotated into the $p-1$
dimensional subspace $\frak{h}_K[\frak{g}_4]$ and the Cartan
subalgebra $\frak{h}_0$ of the ${\rm SU}(1,1)$ factor. These two
spaces together form the non-compact Cartan subalgebra of the
three-dimensional isometry algebra $\frak{g}$, which therefore
defines the normal form $R_N$ of ${\bf R}$:
$\frak{h}_K[\frak{g}]=\frak{h}_K[\frak{g}_4]+\frak{h}_0$. The
group $H_{cent}$, which is the largest subgroup of $H_c$ commuting
with $\frak{h}_K[\frak{g}]$, is also the largest subgroup of $H_4$
commuting with $\frak{h}_K[\frak{g}_4]$. Its completion $G_{cent}$
in $G_4$ coincides with itself and with the paint group of both
$G$ and $G_4$. In other words, for these models, we have
\begin{eqnarray}
\hat{G}_{cent}&=&G_{cent}=H_{cent}=G_{paint}[G]=G_{paint}[G_4]\,.\label{centpaint}
\end{eqnarray}
The group $G_{paint}$ can therefore be characterized as the
centralizer in $G_4$ of the normal form of the representation of the
electric and magnetic charges in four dimensions. The roots in
$\tilde{\Delta}[\frak{g}_4]$, together with $\beta_0$, correspond to
the roots previously denoted by $\tilde{\beta}$ in the split case.
On the other hand $\frak{g}_4$ roots $\hat{\beta}$ have a vanishing
restriction to the $G_4$ non-compact Cartan generators and thus form
the space $\Delta_0[\frak{g}_4]$. Thus for these models we have that
$r=p$ and hence
\begin{eqnarray}
Q_N&\in&\left(\frac{\frak{sl}(2,\Real)}{\frak{so}(1,1)}\right)^p\,.\label{normns}
\end{eqnarray}
The proof of the above statement follows the same lines as the one
given in the previous section. Equation (\ref{normcent}) then
implies (\ref{normns}) in virtue of (\ref{centpaint}).

\paragraph{$D=4$ theories}
We next consider the quarter maximal theories in four dimensions
arising from time reduction of a five-dimensional theory. Again we
have that $r=p$ and therefore we can construct the generating
geodesic as a geodesic in the submanifold
$[\SL(2,\Real/\SO(1,1)]^p$, namely as a solution of the
corresponding consistent truncation. From the algebraic structure
of $Solv$, classified in \cite{alekseevski,Cortes,deWit:1992wf},
we can deduce the form of their sigma-model metric given in table
\ref{table:N=4supergravities}.

\paragraph{$D=5$ theories}
As far as the Euclidean five--dimensional theories originating from
time-reduction of quarter-maximal six--dimensional theories, we
shall restrict as well to those models with a symmetric scalar
manifold. We shall also consider the non-trivial cases in which the
six-dimensional parent theory has a non vanishing number $n_v$ of
vector fields. These models are listed in Table
\ref{table:N=4supergravities}.

The models listed in this table, from top to bottom, are denoted in
the literature by $L^*(q,\,P)$, for certain values of  $q,\,P$:
$L^*(0,P),\,L^*(1,1),\,L^*(2,1),\,L^*(4,1),\,L^*(8,1)$.  The first
model in this table originates from a theory in one dimension higher
with $P$ vector multiplets and  one tensor multiplet besides the
gravitational one. The remaining four models are obtained from a six
dimensional theory with $n_T=q+1$ tensor multiplets and $n_V=2\,q$
vector multiplets. The number of scalar fields in $D=5$ is
$n_V+n_T+1$ while the number of vector fields in $n_V+n_T+2$.  We
can write the metric on the $D=5$ scalar manifold as follows
\begin{eqnarray}
L^*(0,\,P):\,\,\d s^2&=&(\d\varphi_1)^2+(\d\varphi_2)^2-\frac{1}{2}\,e^{\sqrt{2}\,\varphi_1}\,\sum_{m=1}^P \d Y_m^2\,,\nonumber\\
L^*(q,\,1):\,\,\d
s^2&=&(\d\varphi_1)^2+(\d\varphi_2)^2+\frac{1}{2}\,\sum_{m=1}^q\left[e^{-\frac{1}{\sqrt{2}}\,(\varphi_1-\sqrt{3}\,\varphi_2)}\,\d
X_m^2- e^{\sqrt{2}\,\varphi_1}\,\d
Y_m^2\right.\nonumber\\&&\left.-e^{\frac{1}{\sqrt{2}}\,(\varphi_1+\sqrt{3}\,\varphi_2)}\,\d
Z_m^2\,+\dots\right]\,,\nonumber
\end{eqnarray}
where the last expression holds only for the cases $q=1,2,4,8$
considered here and the ellipses indicate interaction terms between
the $X,\,Y$ and $Z$ axions. The scalar fields $Y_m,\,Z_m$ originate
from the $D=6$ vector fields while $X_m$ are the $q$ $D=6$ axions.
In these cases the truncated model is defined by a single axion out
of the $Y_{m}.$ As a result the dilaton $\varphi_2$ decouples from
the remaining scalars and the normal form $Q_N$ belongs to the
following space
\begin{eqnarray}
Q_N&=&\frac{\frak{sl}(2,\Real)}{\frak{so}(1,1)}+\frak{so}(1,1)\,,
\end{eqnarray}
where the $\beta$ parameter for the axion-dilaton system is computed
to be $\sqrt{2}$.
\begin{table}\renewcommand{\arraystretch}{2.3}
\begin{center}
{\scriptsize \begin{tabular}{|c|c|c|c|c|c|c|c|}\hline
   {\normalsize \bf{$G_3/H_3^*$}} & {\normalsize \bf{$G_4/H_4$}} & {\normalsize{\bf $\hat{G}_{cent}=G_{paint}$}} & {\normalsize $p$} & {\normalsize{\bf ${\bf R}$}}  & {\normalsize{\bf $\beta_i$}}\\ \hline
  \hline $\frac{\SU(1,2)}{{\rm S}[\U(1)\times \U(1,1)]}$ &  $---$ & $\SO(2)$
  & 1  &$2\times {\bf 1}$ & 1\\
 \hline $\frac{\SU(2,P+2)}{{\rm S}[\U(1,P+1)\times \U(1,1)]}$ &  $\frac{\U(1,P+1)}{\U(P+1)\times \U(1)}$ & $\U(1)\times \U(P)$
  & 2 &${\bf (P+2)}+\overline{{\bf (P+2)}}$  & $(\sqrt{2},\sqrt{2})$\\
   \hline $\frac{{\rm G}_{2(2)}}{{\rm SL}(2,\Real)\times {\rm SL}(2,\Real)}$ &  $\frac{{\rm SL}(2,\Real)}{{\rm SO}(2)}$
  & $---$ & 2    &${\bf 4}$& $(2/\sqrt{3},2)$\\
    \hline $\frac{{\rm SO}(3,4)}{{\rm SO}(1,2)\times {\rm SO}(2,2)}$ &  $\left(\frac{{\rm SL}(2,\Real)}{{\rm SO}(2)}\right)^2$
  & $---$ & 3   &${\bf (3,1)}+{\bf (1,3)}$&  $(2,2,\sqrt{2})$\\
 \hline $\frac{\SO(4,P+4)}{\SO(2,2)\times \SO(2,P+2)}$ &  $\frac{\SO(2,1)}{\SO(2)}\times \frac{\SO(2,P+2)}{\SO(2)\times \SO(P+2)}$
  & $\SO(P)$ & 4&${\bf (2,P+4)}$& $(2,2,2,2)$\\
 \hline $\frac{{\rm F}_{4(4)}}{{\rm Sp}(6)\times \SO(2,1)}$ &  $\frac{{\rm Sp}(6)}{{\rm U}(3)}$
  & $---$ & 4   &${\bf 14}$& $(2,2,2,2)$\\
\hline $\frac{{\rm E}_{6(2)}}{{\rm SU}(3,3)\times \SU(1,1)}$ &
$\frac{{\rm SU}(3,3)}{{\rm S}[{\rm U}(3)\times {\rm U}(3)]}$
  & $\SO(2)^2$ & 4   &${\bf 20}$& $(2,2,2,2)$\\
\hline $\frac{{\rm E}_{7(-5)}}{{\rm SO}^*(12)\times \SU(1,1)}$ &
$\frac{{\rm SO}^*(12)}{{\rm U}(6)}$
  & $\SO(3)^3$ & 4    &${\bf 32}$& $(2,2,2,2)$\\
 \hline $\frac{{\rm E}_{8(-24)}}{{\rm E}_{7(-25)}\times \SU(1,1)}$ &
$\frac{{\rm E}_{7(-25)}}{{\rm E}_{6(-78)}\times\U(1)}$
  & $\SO(8)$ & 4   &${\bf 56}$ & $(2,2,2,2)$\\
\hline \hline
   {\normalsize \bf{$G_4/H_4^*$}} & {\normalsize \bf{$G_5/H_5$}} & {\normalsize{\bf $\hat{G}_{cent}=G_{paint}$}} & {\normalsize $p$} & {\normalsize{\bf $R$}}  & {\normalsize{\bf $\beta_i$}}\\ \hline\hline $\frac{\SL(2,\Real}{\SO(1,1)}$ &  $---$ & $---$
  & 1   &${\bf 1}$& $\frac{2}{\sqrt{3}}$\\
 \hline $\left(\frac{\SL(2,\Real}{\SO(1,1)}\right)^2$ &  $\SO(1,1)$ & $---$
  & 2  &${\bf 1}_++{\bf 1}_-$ & $(2,\,\sqrt{2})$\\
   \hline $\frac{\SO(2,1)}{\SO(1,1)}\times \frac{\SO(2,P+2)}{\SO(1,1)\times \SO(1,P+1)}$ &  $\SO(1,1)\times \frac{\SO(1,P+1)}{ \SO(P+1)}$
  & $\SO(P)$ & 3   &${\bf (P+2)}+{\bf 1}$& $(2,2,2)$\\
    \hline $\frac{{\rm Sp}(6)}{{\rm GL}(3,\Real)}$ &  $\frac{{\rm SL}(3,\Real)}{{\rm SO}(3)}$
  & $---$ & 3   &${\bf 6}$& $(2,2,2)$\\
 \hline $\frac{\SU(3,3)}{\SL(3,\mathbb{C})\times \SO(1,1)}$ &  $\frac{\SL(3,\mathbb{C})}{\SU(3)}$
  & $\SO(2)^2$ & 3  &${\bf 9}$ & $(2,2,2)$\\
 \hline $\frac{{\rm SO}^*{(12)}}{{\rm SU}^*(6)\times \SO(1,1)}$ &  $\frac{{\rm SU}^*(6)}{{\rm Sp}(6)}$
  & $\SO(3)^3$ & 3  &${\bf 15}$ & $(2,2,2)$\\
\hline $\frac{{\rm E}_{7(-25)}}{{\rm E}_{6(-26)}\times\SO(1,1)}$ &
$\frac{{\rm E}_{6(-26)}}{{\rm F}_{4(-52)}}$
  & $\SO(8)$ & 3   &${\bf 27}$& $(2,2,2)$\\
\hline \hline {\normalsize \bf{$G_5/H_5^*$}} & {\normalsize
\bf{$G_6/H_6$}} & {\normalsize{\bf $\hat{G}_{cent}$}} & {\normalsize
$p$} & {\normalsize{\bf $R$}}  & {\normalsize{\bf $\beta_i$}}\\
\hline
  \hline $\SO(1,1)\times \frac{\SO(1,P+1)}{ \SO(1,P)}$ &  $---$ & $\SO(1,1)\times\SO(P-1)$
  & 1  &${ P}$ & $\sqrt{2}$\\
 \hline $\frac{{\rm SL}(3,\Real)}{{\rm SO}(2,1)}$ &  $\frac{\SO(1,2)}{\SO(2)}$ & $\SO(1,1)$
  & 1   &${\bf 2}$& $\sqrt{2}$\\
   \hline $\frac{\SL(3,\mathbb{C})}{\SU(1,2)}$ &  $\frac{\SO(1,3)}{\SO(3)}$
  & $\SO(1,1)\times\SO(2)$ & 1  &${\bf 4}$ & $\sqrt{2}$\\
    \hline $\frac{{\rm SU}^*(6)}{{\rm Sp}(2,4)}$ &  $\frac{\SO(1,5)}{\SO(5)}$
  & $\SO(1,1)\times\SO(3)^2$ & 1 &${\bf 8}$  & $\sqrt{2}$\\
 \hline $\frac{{\rm E}_{6(-26)}}{{\rm F}_{4(-20)}}$ &  $\frac{\SO(1,9)}{\SO(9)}$
  & $\SO(1,1)\times \SO(7)^+$ & 1  &${\bf 16}$ & $\sqrt{2}$\\
\hline
\end{tabular}}
\caption{\it The symmetric coset spaces in quarter-maximal
supergravity in $D=3,4,5$ obtained from time reduction of $D=4,5,6$
theories. For the last entry, the group $\SO(7)^+$ is the one with
respect to which the ${\bf 8}_c$ of $\SO(8)$ branches in ${\bf
1}+{\bf 7}$ while ${\bf 8}_s\rightarrow{\bf 8}$ and ${\bf
8}_v\rightarrow{\bf 8}$.} \label{table:N=4supergravities}
\end{center}

\end{table}

\section{The Physics I: Einstein Vacuum Solutions }\label{PHYSICSI}

It is natural to consider the uplift of the generating (-1)-brane
solution to a vacuum solution in $D+n$ dimensions. In order to
uplift the solutions from $D>3$ dimensions to $D+n$ dimensions one
uses the Kaluza--Klein Ansatz
\begin{equation}\label{metriek} \d s_{D+n}^2=e^{2\alpha \varphi}\d
s_D^2 + e^{2\beta \varphi}\mathcal{M}_{mn}(\d z^n +A^n) \otimes (\d
z^m+A^m)\,,
\end{equation}
 where
\begin{equation}\label{alpha en beta}
\alpha^2=\frac{n}{2(D+n-2)(D-2)}\,,\qquad
\beta=-\frac{(D-2)\alpha}{n} \,.
\end{equation}
The matrix $\mathcal{M}$ and the scalar $\varphi$ are the moduli
of the $n$-torus and depend on the $D$-dimensional coordinates. In
particular $\mathcal{M}$ is a regular symmetric $n\times n$ matrix
with det$\mathcal{M}=1$ when the torus has Euclidean signature and
det$\mathcal{M}=-1$ when the torus has Lorentzian signature. The
modulus $\varphi$ controls the overall volume and is named the
breathing mode. For a dimensional reduction over a Euclidean torus
the scalars parameterize $\GL(n,\Real)/\SO(n)$ where $\varphi$
belongs to the decoupled $\Real$-part and $\mathcal{M}$ is the
$\SL(n,\Real)/\SO(n)$ part. More precisely $\mathcal{M}=LL^T$
where $L$ is the vielbein matrix of the internal torus and it also
plays the role of the coset representative of
$\SL(n,\Real)/\SO(n)$. For the reduction over the Lorentzian torus
the scalars parameterize $\GL(n,\Real)/\SO(n-1,1)$ and
$\mathcal{M}=L\eta L^T$, where $\eta$ is
$\text{diag}(-1,+1,\ldots,+1)$.

The reduction of pure gravity gives the following $D$-dimensional
Lagrangian
\begin{equation}
\mathcal{L} =\sqrt{-g}\Bigl\{\mathcal{R} -
\tfrac{1}{2}(\partial\varphi)^2 + \tfrac{1}{4}\text{Tr}\partial
\mathcal{M} \partial
\mathcal{M}^{-1}-\tfrac{1}{4}\e^{2(\beta-\alpha)\varphi}\mathcal{M}_{mn}F^m
F^n\Bigr\}\,.
\end{equation}

When $D=3$ the vectors can be dualized to scalars and consequently
there is a symmetry enhancement since the extra scalars combine with
the existing scalars into the coset manifold
$\SL(n+1,\Real)/SO(n-1,2)$\footnote{This means
$\eta=(-1,-1,+1,+1,\ldots,+1)$.}. Note that there is no decoupled
$\Real$ factor in this case. In the next subsections we shall write
down the generating geodesic curves for the three distinct cases
$\SL(n,\Real)/\SO(n)$, $\SL(n,\Real)/\SO(n-1,1)$ and
$\SL(n+1,\Real)/\SO(n-1,2)$. Note that for pure Kaluza--Klein theory
in $D>3$ all geodesics that are related through a
$\SL(n)$-transformation lift up to exactly the same vacuum solution
in $D+n$ dimensions since the $\SL(n)$ corresponds to rigid
coordinate transformations from a $(D+n)$-dimensional point of view.
So, in this sense it is absolutely necessary to understand the
generating geodesic since it classifies higher-dimensional solutions
modulo coordinate transformations. Of course, this is not true for
$D=3$ where $\SL(n+1)$ maps higher-dimensional solutions to each
other that are not necessarily related by coordinate
transformations.

Consider the symmetric coset matrix $\hat{\mathcal{M}}(h)=\eta
\exp{Q_N h}$ with $Q_N$ the normal form of $Q \in
\frak{gl}(n)/\frak{so}(n-1,1)$ (or $\frak{gl}(n)/\frak{so}(n)$) that
generates all other geodesics and $h$ the harmonic function defined
in (\ref{harmonic}). The relation between $\hat{\mathcal{M}}$ and
the moduli $\varphi$ and $\mathcal{M}$ of (\ref{metriek}) is as
follows
\begin{equation}
\hat{\mathcal{M}}=(|\text{det}\hat{\mathcal{M}}|)^{\frac{1}{n}}\mathcal{M}\,,\qquad
|\text{det}{\hat{\mathcal{M}}}|=\exp{\sqrt{2n}\varphi}\,.
\end{equation}

For the uplift of solutions in $D=3$ one has to take into account
the KK vectors since they are dualized to scalars. We only briefly
describe the solutions.

\subsection{Time-dependent Solutions from $\GL(n,\Real)/\SO(n)$}

The generating solution is
\begin{equation}
\hat{\mathcal{M}}(h)=\left(%
\begin{array}{ccc}
  \e^{\lambda_1 h}   &  0         & 0           \\
  0           &  \ddots    & 0           \\
  0           &  0         &\e^{\lambda_n h}    \\
  \end{array}
\right) \,,
\end{equation}
with $h$ given by (\ref{harmonicII}). If we take the $(-1)$-brane
geometry with $k=0$ then the generating solution lifts up to the
Kasner solutions with $\ISO(D-1)$-symmetry \cite{Chemissany:2007fg}
\begin{equation}\label{Kasner}
\d s^2=-\tau^{2p_0}\d \tau^2 + \sum_b \tau^{2p_b}(\d x^b)^2\,,\qquad
b=1,\ldots, D+n-1\,,
\end{equation}
where the power-laws are defined by
\begin{align}
& p_0=(D-2) + \frac{\alpha\sum_i\lambda_i}{\sqrt{2an}}\,,\qquad
p_1=\ldots= p_{D-1} = 1 + \frac{\alpha\sum_i\lambda_i}{\sqrt{2an}}
\,,\\
&p_{D+i-1} = \frac{\sum_i\lambda_i}{2\sqrt{a}}
(\frac{2\beta}{\sqrt{2n}}-\frac{1}{n}) +
\frac{\lambda_i}{2\sqrt{a}}\,.
\end{align}
We defined $a$ in equation (\ref{metricII}) and used that
$||v||^2=\tfrac{1}{2}\sum_i\lambda_i^2$. The numbers $p$ obey the
Kasner constraints
\begin{equation}
 p_0+1    =\sum_{b>0} p_b\,,\,\,\,\,\,(p_0+1)^2=\sum_{b>0} p_b^2\,.
\end{equation}

The higher-dimensional vacuum solutions with $k\pm 1$ are
\begin{equation}
\d s^2= W^{p_0}\Bigl(-\frac{\d t^2}{at^{-2(D-2)}-k} + t^2\d
\Sigma_k^2\Bigr) + \sum_{i=1}^n W^{p_i}(\d z^i)^2\,,
\end{equation}
where the function $W(t)$ is defined as
\begin{equation}
W(t)= \sqrt{a}t^{2-D} + \sqrt{at^{2(D-2)}-k}\,,
\end{equation}
and the various constants $p_0$ and $p_i$ are defined as
\begin{equation}
p_0=
-\frac{\sum_i}{||v||(D-2)}\sqrt{\tfrac{2(D-1)}{(D+n-2)}}\,,\qquad
p_i= -\frac{D-2}{n}p_0 +
\frac{(\sum_j\lambda_j-n\lambda_i)}{n||v||}\sqrt{\tfrac{2(D-1)}{D-2}}\,,
\end{equation}
and the affine velocity is given by $||v||^2 = \tfrac{1}{2}
\sum_i\lambda_i^2$. Note that the $k=-1$ solutions approach flat
Minkowski space in Milne coordinates for $t\rightarrow \infty$,
these solutions are a generalization of the fluxless S-brane
solutions of \cite{Gutperle:2002ai, Townsend:2003fx,Ohta:2003pu,
Chen:2002yq}. For $k=+1$ the solutions do not asymptote to flat
space and they are generalizations of the fluxless solutions
considered in for instance \cite{Lu:1996er}.

\subsection{Time-dependent Solutions from
$\SL(n+1,\Real)/\SO(n+1)$} If we reduce to three dimensions a
symmetry-enhancement of the coset takes place. The dualisation of
the three-dimensional KK vectors generate the coset
$\SL(n+1,\Real)/\SO(n+1)$ instead of the expected
$\GL(n,\Real)/\SO(n)$.  However the generating solution of the
$\SL(n+1,\Real)/\SO(n+1)$-coset has only non-trivial dilatons and is
therefore the same as the generating solution of
$\GL(n,\Real)/\SO(n)$. Nonetheless, there is an important difference
with the time-dependent solutions from $\GL(n,\Real)/\SO(n)$. In
that case a solution-generating transformation $ \in \GL(n,\Real)$
can be interpreted as a coordinate transformation in $D+n$
dimensions and therefore maps the vacuum solution to the same vacuum
solution in different coordinates. In the case of symmetry
enhancement to $\SL(n+1,\Real)$ a solution-generating transformation
is not a coordinate transformation in $D+n$ dimensions. Instead, the
time-dependent vacuum solution  transforms into a "twisted" vacuum
solution. Where the twist indicates off-diagonal terms that cannot
be redefined away. Such twisted solutions with $k=-1$ have received
considerable interest since they can be regular \cite{Wang:2004by,
Jones:2004rg}.

\subsection{Stationary Solutions from $\GL(n,\Real)/\SO(n-1,1)$}
The normal form is given by
\begin{equation}
Q_N=
\left(%
\begin{array}{ccccc}
  \lambda_a & \omega & 0 & \ldots & 0 \\
  -\omega & -\lambda_a & 0 & \ldots & 0 \\
  0 & 0 & 0 & \ldots & 0 \\
  0 & 0 & 0 & \ddots & 0 \\
  0 & 0 & 0 & \ldots & 0 \\
\end{array}%
\right)+
\left(%
\begin{array}{ccccc}
  \lambda_b & 0 & 0 & \ldots & 0 \\
  0 & \lambda_b & 0 & \ldots & 0 \\
  0 & 0 & \lambda_3 & \ldots & 0 \\
  0 & 0 & 0 & \ddots & 0 \\
  0 & 0 & 0 & \ldots & \lambda_{n} \\
\end{array}\right)\,.
\end{equation}
We exponentiate this to $\hat{\mathcal{M}}(h(r))=\eta \e^{Q_N
h(r)}=$
\begin{equation}
 \left(%
\begin{array}{ccccc}
  -\e^{\lambda_b h(r)}f_{+}(r)       &  -\omega\e^{\lambda_b h(r)}\Lambda^{-1}\sinh(\Lambda h(r))  & 0    & \ldots & 0  \\
 -\omega\e^{\lambda_b h(r)}\Lambda^{-1}\sinh(\Lambda h(r))       &  \e^{\lambda_b h(r)}f_{-}(r)    & 0    & \ldots & 0  \\
  0       &  0         &\e^{\lambda_3 h}  & \ldots & 0       \\
  0       &  0         & 0                  & \ddots & \vdots  \\
  0       &  0         & 0                  & \ldots & \e^{\lambda_{n}h}\\
 \end{array}%
\right)\,,
\end{equation}
with
\begin{equation}
f_{\pm}(r)  =\e^{\lambda_b h(r)}\Bigl(\cosh(\Lambda
h(r))\pm\lambda_a\frac{\sinh(\Lambda h(r))}{\Lambda}\Bigr)\,,
\end{equation}
and where we define the $\SO(1,1)$ invariant quantity $\Lambda$ as
\begin{equation}
\Lambda = \sqrt{\lambda_a^2 - \omega^2}\,.
\end{equation}
There exist three distinct cases depending on the character of
$\Lambda$. If $\Lambda$ is real the above expression does not need
rewriting but we can put $\lambda_2$ to zero using a
$\SO(1,1)$-boost and then the generating solution is just the
straight line solution. If $\Lambda=i\tilde{\Lambda}$ with
$\tilde{\Lambda}$ real then the terms with $\cosh(\Lambda h)$ become
$\cos{\tilde{\Lambda}}$ and $\Lambda^{-1}\sinh{\Lambda h}$ become
$\tilde{\Lambda}^{-1}\sin{\tilde{\Lambda h}}$. Finally, if
$\Lambda=0$ then the term $\Lambda^{-1}\sinh{\Lambda h}$ becomes
just $h$ and the term with $\cosh{\Lambda h}$ becomes equal to one.

To discuss the zoo of solutions one should make a classification in
terms of the different signs for $k$, $||v||^2$ and $\Lambda^2$. We
restrict to solutions in spherical coordinates which have $k=+1$.
The other solutions can similarly be found. The solutions with
spherical symmetry have the more interesting properties that they
lift up to vacuum solutions that can be asymptotically flat. These
solutions can be found in appendix \ref{APPENDIXB}.

\section{The Physics II: $D=4,\mathcal{N}=8$  Static Black Holes}\label{PHYSICSII}
Instead of uplifting the generating geodesic to the vacuum in
$D+n$ dimensions as in the previous section, one could also
consider the uplift to $D+1$ dimensions
\cite{Breitenlohner:1987dg, Gaiotto:2007ag}. This is the content
of the coming section. We also generalize the discussion from pure
(Kaluza-Klein) gravity to supergravity.  In particular we describe
the correspondence between $D=3$ instantons and $D=4$ black holes
in maximal supergravity starting with a discussion on the various
dimensional reductions involved. In subsection \ref{bhne} we work
out the generating solution of non-extreme black hole solutions in
$D=4$ maximal supergravity, whose $D=3$ counterparts are generated
by  a diagonalizable $Q$. In subsection \ref{EXTREMAL} we focus on
extreme black holes in $D=4$ instead. For these solutions $Q$ is
nilpotent and therefore our theorem does not apply. However the
space defined on the right hand side of (\ref{theorem}) does
contain nilpotent generators. We shall analyze the black hole
solution generated by a generic combination of these nilpotent
matrices, which, with an abuse of notation, will be denoted by
$Q_N$. The parameters of $Q_N$ coincide with the $D=4$ quantized
charges.  Although our general discussion does not imply that
$Q_N$ is the normal form of a generic nilpotent generator $Q$,
this matrix has a non trivial intersection, for different choices
of its parameters, with all the nilpotent orbits which are
relevant for $D=4$ extreme solutions \cite{Pioline:2006ni}. The
black hole solution generated by $Q_N$ lifts to a known dilatonic
solution of the $\mathcal{N}=2$ STU model (see for instance
\cite{Kallosh:2006ib}). We shall give then in terms of $Q_N$ a
general characterization of the three classes of extreme regular
four-dimensional solutions. Finally, in subsection \ref{gns}, we
will comment on how to generate new solutions starting from this
dilatonic one.

\subsection{Dimensional Reduction}

\paragraph{Time reduction from $D=4$.}
Let us start fixing some general notations about the $D=3$ action
(\ref{SIGMA-ACTION}) as obtained from time reduction of a $D=4$
theory %($8\pi G_N=1$; $i,j=1,2,3$; $\mu,\,\nu=0,1,2,3$)
. The sigma model in $D=3$ is given by \cite{Breitenlohner:1987dg}
\begin{eqnarray}
G_{IJ}\,\d\phi^I\,\d\phi^J&=&4\,(\d U)^2+e^{-4\,U}\omega^2+g_{rs}\,\d\phi^r\,\d\phi^s-2\,e^{-2\,U}\d Z^T\,\mathcal{M}_4\,\d Z\,,\nonumber\\
\omega &=& \d a+Z^T\mathbb{C}\,\d Z\,.\label{metric3}
\end{eqnarray}
where the Ansatz for the $D=4$ space-time metric is
\begin{eqnarray}
\d s^2_4 &=& -e^{2\,U}\,(\d t+A^0_i\,\d x^i)^2+e^{-2\,U}\,g_{ij}\,\d
x^i\,\d x^j\,.\label{met4}
\end{eqnarray}
$g_{ij}$ being the three dimensional metric in the Einstein frame.
We introduced several notations which we now explain. $A^0_i$
denotes the Kaluza-Klein vector in $D=3$ and
$Z=(\zeta^\Lambda,\,\tilde{\zeta}_\Lambda)$ is the symplectic
vector of electric and magnetic potentials, related to the $D=4$
vector fields $A^\Lambda_\mu$ as follows
\begin{eqnarray}
\zeta^\Lambda&=&A^\Lambda_0\,\,\,,\,\,\,\,\,F^0_{ij}=\partial_iA^0_j-\partial_jA^0_i\,,\nonumber\\\partial_iA^\Lambda_j-\partial_jA^\Lambda_i+\zeta^\Lambda\,
F^0_{ij}&=&e\,e^{-2\,U}\,\epsilon_{ijk}\,I^{-1\Lambda\Sigma}\,\left(\partial^k\tilde{\zeta}_\Sigma-R_{\Sigma\Gamma}\,\partial^k\zeta^\Gamma\right)\,.
\end{eqnarray}
Where the matrices $I, R$ are the imaginary and real parts of the
kinetic matrix $\mathcal{N}$ in $D=4$ \cite{Breitenlohner:1987dg}:
$I={\rm Im}(\mathcal{N})<0,\,R={\rm Re}(\mathcal{N})$. The matrix
$\mathcal{M}_4$ is the symplectic matrix in $D=4$ built out of
$I,\,R$ as follows
\begin{eqnarray}
\mathcal{M}_4&=&L_4\,L_4^T=-\left(\begin{matrix}I+R\,I^{-1}\,R &
-R\,I^{-1}\cr -I^{-1}\,R & I^{-1}
\end{matrix}\right)>0\,,\label{MIR}
\end{eqnarray}
$L_4$ being the coset representative of the (homogeneous) scalar
manifold in $D=4$. In terms of the matrix $\mathcal{M}_4$ the sigma
model metric in $D=4$ reads
\begin{eqnarray}
g_{rs}\,d\phi^r\,d\phi^s&=&\frac{1}{2\,c }\,{\rm
Tr}(\mathcal{M}_4^{-1}d\mathcal{M}_4\,\mathcal{M}_4^{-1}d\mathcal{M}_4)\,,
\end{eqnarray}
where $c$ is a constant depending on the $G_4$-representation of
$\mathcal{M}_4$. The matrix $\mathbb{C}$ is the antisymmetric,
symplectic invariant matrix and $a$ is the scalar dual to $A^0_i$
\begin{eqnarray}
F^0_{ij}&=&-e\,e^{-4U}\,\epsilon_{ijk}\,\omega^k\,.
\end{eqnarray}
\paragraph{10D origin} Let us now consider maximal supergravity in
$(3,0)$ dimensions, obtained from a time-reduction of the
four-dimensional theory. In this case $G={\rm E}_{8(8)}$,
$H=\SO(16)$, $H^*=\SO^*(16)$, $G_4={\rm E}_{7(7)}$ and
$H_4=\SU(8)$. Maximal supergravities in any dimension originate
from toroidal reduction of Type II theories. In appendix
\ref{typeIIred} we give the precise group theoretical
characterization of the ten-dimensional origin of the bosonic
fields in $D=3$, namely the correspondence between the three
dimensional scalars arising from the Type II string 0-modes and
the $\frak{ e}_{8(8)}$ positive roots. With respect to the ${\rm
U}(8)$ subgroup of ${\rm SO}^*(16)$, the 56 scalars associated
with $\gamma$ transform in the ${\bf 28}+\overline{{\bf 28}}$.
Upon the action of ${\rm U}(8)$, we can obtain a four dimensional
normal form defined by the following roots $\gamma_i$ (see tables
\ref{alphafield}, \ref{gammafield} for the explicit correspondence
between $\frak{e}_{8(8)}$ roots and dimensionally reduced string
modes)
\begin{eqnarray}
\epsilon_0-\epsilon_4&\leftrightarrow
&A_0^4\,\,,\,\,\,\epsilon_0+\epsilon_4\leftrightarrow
B_{04}\,,\nonumber\\
-\epsilon_5-\epsilon_{10}&\leftrightarrow
&B^5\,\,,\,\,\,\epsilon_5-\epsilon_{10}\leftrightarrow
A_5\,,\nonumber
\end{eqnarray}
where the ten dimensional space-time indices run from $0$ to $9$,
$A_0^4$ and $B_{04}$ are the time-components of the $D=4$ vectors
$A^4_\mu,\,B_{\mu\,4}$, $B^5$ and $A_5$ are the duals of the $D=3$
vectors $A^5_i,\,B_{i\,5}$. %Let us denote by $\Phi_i$ the
%corresponding axions.
The above roots define the coset $[\SL(2,\Real)/\SO(1,1)]^4$. The
Cartan generators of this coset are parametrized by the scalar
fields: $\sigma_0\pm \sigma_4$ and $\sigma_5\pm 2\,\phi_3$. The
sigma model Lagrangian for the above coset reads
\begin{align}
L=&\sum_{k=1}^4 (\partial\varphi_k)^2-e^{-2\,\varphi_k}\,(\partial\chi_k)^2\nonumber\\
=&[\partial (\sigma_0+ \sigma_4)]^2+[\partial (\sigma_0-
\sigma_4)]^2+[\partial (\sigma_5+ 2\,\phi_3)]^2+[\partial (\sigma_5-
2\,\phi_3)]^2+\nonumber\\
& -e^{-2(\sigma_0+ \sigma_4)}\,(\partial
B_{04})^2-e^{-2(\sigma_0- \sigma_4)}\,(\partial
A_{0}^4)^2-e^{2(\sigma_5+ 2\,\phi_3)}\,(\partial
B^5)^2+\nonumber\\&-e^{-2(\sigma_5- 2\,\phi_3)}\,(\partial
A_5)^2\,,\label{Ln}
\end{align}
where, as defined in appendix \ref{typeIIred}, $\sigma_0$ is the
modulus associated with the radius of the time direction $R_0$ and
$\sigma_{m\ge 4}$ are the moduli associated with the radii of the
internal spatial directions $R_m$. In eq. (\ref{Ln}) we have used
the property
\begin{eqnarray}
\varphi_1&=&\sigma_0-
\sigma_4\,\,;\,\,\,\varphi_2=\sigma_0+\sigma_4\,,\nonumber\\\varphi_3&=&-(\sigma_5+2\,\phi_3)\,\,;\,\,\,\varphi_4=\sigma_5-2\,\phi_3\,.\label{phiphi}
\end{eqnarray}
Thus the generating submanifold (\ref{generatingsubmanifold}) is
defined by the sigma model (\ref{Ln}) together with the 4 remaining
decoupled dilatons.

\paragraph{Uplifting to $D=4$} We may think of performing the $D=10\rightarrow
D=3$ reduction through an intermediate step represented by the $D=4$
theory in the Einstein frame. This allows to deduce the relation
between the $D=4$ fields and the quantities in the $D=3$ theory as
originating from the  Type II theories,
\begin{equation}
U=\tfrac{1}{2}\,(\sigma_0-2\,\phi_3)\,,\qquad
\phi_4=\tfrac{1}{2}\,(\sigma_0+2\,\phi_3)\,,
\end{equation} where we denoted the four dimensional
dilaton by $\phi_4.$ The dilaton vector $\vec{h}_4$ in four
dimensions is related to $\vec{h}$ as follows (see appendix
\ref{typeIIred})
\begin{eqnarray}
\vec{h}&=&\vec{h}_4+U\,(\epsilon_0-\epsilon_{10})\,,\nonumber\\
\vec{h}_4&=&\sum_{m=4}^9\sigma_m\,\epsilon_m+\phi_4\,(\epsilon_0+\epsilon_{10})\,.
\end{eqnarray}
We learn then how to deduce the black hole warp factor $U$ from a
solution to the theory described by the $\sigma$-model metric
(\ref{Ln}), by using (\ref{phiphi})
\begin{eqnarray}
U&=&\frac{1}{4}\,\sum_{k=1}^4\varphi_k\,.\label{Uphi}
\end{eqnarray}

\subsection{The generating non-extreme $D=4,\mathcal{N}=8$ black hole
solution}\label{bhne} Having presented the 4D (and 10D) origin of
the generating submanifold in $D=3$ we can uplift the geodesics on
the generating submanifold to black hole solutions. These black
holes are generating in the sense of the hidden $E_{8(8)}$
symmetry on the black hole moduli space in $D=4$. In order to make
contact with the black hole literature we present the instanton
solutions in three dimensions in a different frame from the one
presented in section 2. If we take the $(-1)$-brane metric
solution of section \ref{geometries} with $D=3$ and
$\epsilon=k=+1$ and define a new coordinate $\tau$ via
\begin{equation}
\tau=-\ln\bigl(\tanh(r/2)\bigr)\tfrac{2}{||v||}\,,
\end{equation}
then we find \cite{Breitenlohner:1987dg}
\begin{equation}
g_{ij}\,\d x^i\,\d
x^j=\e^{4A(\tau)}\,\d\tau^2+\e^{2A(\tau)}\,(\d\theta^2+\sin^2(\theta)\,
\d\varphi^2)\,,\qquad
\e^{A(\tau)}=\frac{||v||/2}{\sinh(||v||\tau/2)}\,.\label{met3}
\end{equation}
We denote the generating submanifold (\ref{generatingsubmanifold})
for geodesics on $E_{8(8)}/\SO^{*}(16)$ as (\ref{Ln})
\begin{equation}
L=-\sum_{k=1}^4
(\partial\varphi_k)^2-e^{-2\,\varphi_k}\,(\partial\chi_k)^2-\sum_{a=1}^4
(\partial\Phi_a)^2\,.
\end{equation}

Let us recall the geodesic curves on $\SL(2,\Real)/\SO(1,1)$. If we
restrict to geodesics that pass through the origin at $\tau=0$ the
charge-matrix is given by
\begin{equation}
Q_k=\left(
\begin{array}{cc}
\lambda_k &  \omega_k  \\
-\omega_k & -\lambda_k  \\
\end{array}%
\right)\,\label{SL2CHARGE}.
\end{equation}
The symmetric coset matrix $\mathcal{M}_k=L\eta L^T$ is given by
\begin{equation}
\mathcal{M}_k=\left(
\begin{array}{cc}
 -\e^{\varphi_k}+\e^{-\varphi_k}\chi_k^2 & \e^{-\varphi_k}\chi_k \\
 \e^{-\varphi_k}\chi_k & \e^{-\varphi_k} \\
\end{array}%
\right)\,.\label{SL2SOLVABLE}
\end{equation}
We define
\begin{equation}
\Lambda_k^2=\lambda_k^2-\omega_k^2\,,\qquad
\Lambda_k\equiv|\Lambda_k| \,,
\end{equation}
such that Tr$Q_k^2=2\Lambda_k^2$. The solutions for the geodesic
curves are presented in Table \ref{table:geodesics}.
\begin{table}[ht!]
\begin{center}
{ \begin{tabular}{|c|c|c|}\hline Sgn$\,\Lambda_k^2$
&$\e^{-\varphi_k}$ & $\chi_k$ \\ \hline
$>0$ & $\cosh(\Lambda_k\tau)-\tfrac{\lambda_k}{\Lambda_k}\sinh(\Lambda_k\tau)$ & $-\omega_k(\Lambda_k\coth(\Lambda_k\tau)-\lambda_k)^{-1}$\\
\hline $<0$ &
$\cos(\Lambda_k\tau)-\tfrac{\lambda_k}{\Lambda_k}\sin(\Lambda_k\tau)
$  & $-\omega_k(\Lambda_k\cot(\Lambda_k\tau)-\lambda_k)^{-1}$ \\
\hline $=0$\,,\quad $\omega_k=\mp\lambda_k$ &$ -\lambda_k\tau +1$  &
$\pm \frac{\lambda_k\tau}{1-\lambda_k\tau}$\\ \hline
\end{tabular}}
\caption{\it The geodesic curves on $\SL(2\Real)/\SO(1,1)$.}
\label{table:geodesics}
\end{center}
\end{table}

Using formula (\ref{Uphi}) we can easily uplift to a black hole in
$D=4$. The \emph{extreme} black hole solution is given by
\begin{align}
& \e^{4U}=\Pi_{k=1}^4 \frac{1}{1-\lambda_k\tau} \,,\qquad \e^{2A(\tau)}=\tau^{-2}\,,\label{extremeI} \\
& \e^{-\varphi_k}=1-\lambda_i\tau\,,\qquad \Phi_a=0\,,\\
& \chi_k=\mp
\frac{1}{\tfrac{1}{\lambda_k\tau}-1}\label{extremeIII} \,.
\end{align}

Similarly we can construct the non-extreme solutions. If we avoid
naked singularities and periodic singularities we restrict to
non-extreme solutions with all $\Lambda_i^2>0$. The solution is
\begin{align}
&\e^{4U}=\Pi_{k=1}^4 [\cosh(\Lambda_k\tau) -
\tfrac{\lambda_k}{\Lambda_k} \sinh(\Lambda_k\tau)]^{-1}\,,\\
& \e^{2A(\tau)}=\frac{\tfrac{1}{4}\sum_i\Lambda_i^2+
\tfrac{1}{4}\sum_{a=1}^4\varpi_a^2}{\sinh^2\Bigl(\tau\sqrt{\tfrac{1}{4}\sum_i\Lambda_i^2
+ \tfrac{1}{4}\sum_{a=1}^4\varpi_a^2}\Bigr)}\,, \\
&\e^{-\varphi_k}=\cosh(\Lambda_k\tau)-\tfrac{\lambda_k}{\Lambda_k}\sinh(\Lambda_k\tau)\,,\,\,\qquad \Phi_a=\varpi_a\tau\,,\\
&\chi_k=-\omega_k(\Lambda_k\coth(\Lambda_k\tau)-\lambda_k)^{-1}\,.
\end{align}

Acting with $E_{8(8)}$ on the above solutions gives the most general
single centered static black hole solution. The geodesic velocity of
such a general solution is given by
\begin{eqnarray}
||v||^2=\tfrac{1}{2}\text{Tr}Q^2&=&4(\dot{U})^2+e^{-4\,U}\,(\omega_\tau)^2+g_{st}\,\dot{\phi}^s\,\dot{\phi}^t-2\,e^{-2\,U}\,
\dot{Z}^T\,\mathcal{M}_4\,\dot{Z}\,,
\end{eqnarray}
where the dot denotes the derivative with respect to $\tau$. In
order to relate the 3D charges with the 4D charges we compute the
integrals of motion along a generic geodesic
\begin{eqnarray}
e^{-2\,U}\,
\mathcal{M}_4\,\dot{Z}+e^{-4\,U}\,\mathbb{C}Z\,\omega_\tau
&=&\mathbb{C}\,{\bf Q}\,,\nonumber\\
\dot{U}+e^{-4\,U}\,a\,\omega_\tau-\frac{1}{2}\,e^{-2\,U}\,Z^T\,\mathcal{M}_4\,\dot{Z}&=&m\,,\nonumber\\
e^{-4\,U}\,\omega_\tau&=&n\,,\nonumber\\
\mathcal{M}_4^{-1}\,\dot{\mathcal{M}}_4-c\,e^{-4\,U}\,(Z\,Z^T\mathbb{C}\,\omega_\tau-2\,e^{2\,U}\,Z\,\dot{Z}^T\,\mathcal{M}_4)_{\vert
pr}&=&2 \overline{{\bf Q}}\,,\label{constantsofmotion}
\end{eqnarray}
where $m$ is the ADM mass of the solution, ${\bf
Q}^M=(p^\Lambda,\,q_\Lambda)$ is the symplectic vector of the four
dimensional quantized charges, $n$ is the Taub-NUT charge and
$\overline{{\bf Q}}^M{}_N\in \frak{e}_{7(7)}/\frak{su}(8)$. In the
next subsection we shall examine extreme solutions with vanishing
Taub-NUT charge, namely $n=v=0,$ within the truncated model. In this
case $\tau=-1/r$, with $r$ the usual radial coordinate of a black
hole space-time. The horizon is located at $r=0,\,\tau=-\infty$ and
the radial infinity corresponds to $\tau=0$.\par Recall that the
scalar fields $\phi^I$ originating from higher dimensional theories,
are the parameters of the solvable Lie subalgebra of $G$ defined
through the Iwasawa decomposition
\begin{eqnarray}
\frak{g}&=&\frak{H}^*+Solv\,\,\,;\,\,\,\,\,Solv=\{s_I\}\,,
\end{eqnarray}
so that we can write the coset representative $L$ of $G/H^*$ as
$L=\exp(\phi^I\,s_I)$. Let us denote by $s_0$ the element of $Solv$
parametrized by values $\phi^I_0$ of the scalar fields
$\phi^I(\tau)$ at radial infinity:
$s_0=\phi^I(0)\,s_I=\phi^I_0\,s_I$. The general solution of the
geodesic equations can be written in the form
\begin{eqnarray}
\mathcal{M}&=&L\eta
L^T=e^{s_0}\,\eta\,e^{Q\,\tau}\,e^{s_0^T}\,,\label{MQ}
\end{eqnarray}
where $K\in \frak{g}/\frak{H}^*$. In order to give the parameters
of $Q$ a higher dimensional interpretation (for instance to
identify the electric and magnetic quantized charges) we should
then plug the geodesic solution inside (\ref{constantsofmotion}).

The geodesic is totally defined by the values of the scalar fields
at radial infinity $\phi^I_0$, encoded in the matrix
$\mathcal{M}(0)=e^{s_0}\,\eta\,e^{s_0^T}$ and by the matrix $Q_0$
encoding all the constants of motion in (\ref{constantsofmotion})
\begin{eqnarray}
Q_0&\equiv &
\left(\mathcal{M}^{-1}\frac{d}{d\tau}\mathcal{M}\right)_{\vert
\tau=0}=e^{-s_0^T}\,Q\,e^{s_0^T}\,.
\end{eqnarray}
%The measure along the geodesic can be written in terms of $Q$ as
%follows:
%\begin{eqnarray}
%4\,v^2&=&\frac{1}{4}\,g_{IJ}\,\dot{\phi}^I\,\dot{\phi}^J\propto {\rm
%Tr}(Q_0^2)={\rm Tr}(Q^2)\,.
%\end{eqnarray}
The solution to the geodesic equations, in terms of the scalar
fields, is obtained by solving the following equation
\begin{eqnarray}
e^{\phi^I(\tau)\,s_I}\,\eta\,e^{\phi^I(\tau)\,s_I^T}&=&e^{s_0}\,\eta\,e^{Q\,\tau}\,e^{s_0^T}\,.\label{geoeqs}
\end{eqnarray}
Once $s_0$ is fixed by fixing $Solv$, we can still act on the
geodesic by means of a $G$-transformation in
$e^{s_0}\,H^*\,e^{-s_0}$, isotropy group of the point
$\{\phi^I_0\}$. This allows us to reduce $Q$ to $Q_N$ by virtue of
the previously stated theorem about the normal form of $Q$. If we
decompose $Solv$ with respect to the solvable Lie algebra $Solv_4$
associated with $G_4$, as in (\ref{solvsolv4}), we can make the
dependence of  $L(\tau)$ on the four dimensional fields more
explicit and write
\begin{eqnarray}
L&=&e^{a(\tau)\,E_{\beta_0}}\,e^{\sqrt{2}\,Z^\gamma(\tau)\,s_\gamma}\,e^{\phi^r(\tau)\,s_r}\,e^{U(\tau)\,H_0}\,,\label{L34}
\end{eqnarray}
where $\phi^r$ are the $D=4$ scalar fields parametrizing the
generators $s_r$ of $Solv_4$, $s_\gamma$ are the nilpotent
generators in the space ${R}_+$, corresponding to the $\gamma$ roots
and parametrized by the scalars $Z^\gamma$.

Our discussion so far holds for a generic three dimensional theory
with a homogeneous symmetric scalar manifold. Let us now stick to
the maximal supergravity model where $G$ is a split real form, ${\bf
R}={\bf 56}$ of $\E_{7(7)}$ and $s_\gamma=E_\gamma$.

\subsection{$D=4,\mathcal{N}=8$ Extremal Single Center
Black Holes}\label{EXTREMAL}

Although so far we were mainly concerned with the generating
solution of geodesics with diagonalizable $Q,$ characterized as a
solution of a truncated theory, in this subsection we shall
consider extreme $D=4$ black holes described in $D=3$ within the
same truncation. As we shall see, general properties of this class
of $D=4$ solutions will have a simple mathematical description in
this $D=3$ framework. Let us then focus on regular extreme
solutions in $D=4,$ generated by a $Q=Q_{N}$ in the truncation.
The regularity condition implies the existence of a horizon with
non-vanishing area at which the four dimensional scalar fields
acquire a finite value. From the general form of the four and
three dimensional metrics (\ref{met4}), (\ref{met3}) we deduce the
expression for the horizon area $A_H$ of an extreme solution
\begin{eqnarray}
A_H&=&4\,\pi\,\lim_{\tau\rightarrow
-\infty}\frac{e^{-2\,U}}{\tau^2}\,.
\end{eqnarray}
We see that in order to have a non-vanishing area we should have
$e^{-U}\sim \tau$ at the horizon. Following \cite{Gaiotto:2007ag} we
deduce from eq. (\ref{L34}) that $\mathcal{M}(\tau)$ depends on
$U(\tau)$ through the exponential factor $e^{2\,U\,H_0}$. Since we
are assuming that $U$ is the only source of divergence of
$\mathcal{M}(\tau)$ as $\tau\rightarrow -\infty$, this degree of
divergence depends on the lowest grading of $H_0$ in the adjoint
representation of $\frak{g}$. This grading is $-2$ and corresponds
to the action of $H_0$ on $E_{-\beta_0}$, since $-\beta_0(H_0)=-2$.
Accordingly, the degree of divergence of $\mathcal{M}(\tau)$ in a
regular solution is $\tau^4$, which implies, using the general form
(\ref{MQ}) of the solution to the geodesic equation, that $Q^5=0$.
As we shall prove in subsection \ref{NilpotencyforAttractorBlack},
the truncation (\ref{theorem}) describes matrices $Q$ with a degree
of nilpotency up to $p+1,$ and thus captures the nilpotent orbit
which is relevant for this class of solutions.\par Therefore we
start from the requirement that $Q_N$ be nilpotent. This restricts
$Q_N$ to have the following form
\begin{equation}
Q_N=\sum_{k=1}^4 \sqrt{2}\,Q_k\,n^\pm_{k}\,,\qquad
n^{\pm}_k=H_{\gamma_k}\mp(E_{\gamma_k}-E_{-\gamma_k})\,,\label{QNn}
\end{equation}
where $n^\pm_k$ are nilpotent isometries of the submanifold defining
the normal form (\ref{generatingsubmanifold}). The plus or minus
grading characterizing the nilpotent generators $n^{\pm}_k$ is
referred to the corresponding $\frak{o}(1,1)$ generator
$J_k=E_{\gamma_k}+E_{-\gamma_k}$
\begin{eqnarray}
[J_k,\,n^{\pm}_\ell]&=&\pm
\delta_{k\ell}\,n^{\pm}_\ell\,.\label{JNn}
\end{eqnarray}
The parameters $Q_k$ are related to the $\SL(2,\Real)$-charges in
(\ref{SL2CHARGE}) via $|Q_k|=|\lambda_k|=|\omega_k|$. We shall
choose $Q_k>0$. Their identification as quantized electric or
magnetic depends on the $D=4$ symplectic frame we started from (this
shall be discussed below). We also restrict ourselves to the fields
$\varphi_k,\,\chi_k$ defined by the solvable parametrization of the
submanifold $[\SL(2,\Real)/\SO(1,1)]^4$ defined in
(\ref{SL2SOLVABLE}). The reason for not considering the dilatonic
fields parametrizing the ${\rm SO}(1,1)^4$ factors is that, having
chosen $Q_N$ of the form (\ref{QNn}), these fields would commute
with it and thus be constant along the geodesic. Physically the
axions $\chi_k$, $k=1,\dots, 4$, are identified with the
electric-magnetic potentials of the four-dimensional parent theory.
For the sake of simplicity we start from the origin at radial
infinity, namely we choose $s_0=0$, which would also correspond to
choosing the electric and magnetic potentials $\chi_k$ to vanish for
$r\rightarrow \infty$.

In terms of the harmonic function $H_k=1-\sqrt{2}\,Q_k\,\tau$ the
extreme solution derived above (\ref{extremeI}-\ref{extremeIII})
reads
\begin{eqnarray}
e^{\varphi_k}&=&\frac{1}{H_k}\,\,;\,\,\,
\chi_k=\mp\frac{Q_k}{H_k}\,\tau\,,\label{dilution}
\end{eqnarray}
where the $\mp$ sign in the expression for $\chi_k$ depends on the
choice of $n_k^\pm$ in the definition (\ref{QNn}) of $Q_N$. The
above solution corresponds to a four-charge dilatonic solution. Near
the horizon we have
\begin{eqnarray}
e^{4\,U}=\frac{1}{H_1\,H_2\,H_3\,H_4}&\sim&\frac{1}{(4\,Q_1\,Q_2\,Q_3\,Q_4)}\,\frac{1}{\tau^4}=\frac{1}{(r_H)^4}\,\frac{1}{\tau^4}\,,\label{e4u}
\end{eqnarray}
$r_H$ being the radius of the horizon: $A_H=4\,\pi\,r_H^2$.

%
%Let us now discuss the four dimensional uplift of this general
%solution and its supersymmetry properties.
% If we restrict the four dimensional
%charges to $p^\Lambda,\,q_\Lambda$, with $\Lambda=0,1,2,3$, we can
%verify from our solution that:
%\begin{eqnarray}
%\dot{Z}^M&=&\left(\begin{matrix}\dot{\zeta}^0 \cr {\bf 0}\cr 0\cr
%\dot{\tilde{\zeta}}_i\end{matrix}\right)=\left(\begin{matrix}\frac{q_0}{(H_0)^2}
%\cr {\bf 0}\cr 0\cr
%-\frac{p^i}{(H^i)^2}\end{matrix}\right)=e^{2U}\,\mathbb{C}\mathcal{M}_4\,\left(\begin{matrix}
%0\cr p^i \cr q_0\cr{\bf 0}\end{matrix}\right)\,,\label{zetaq}
%\end{eqnarray}
%where
%\begin{eqnarray}
%e^{2U}\mathcal{M}_4&=&{\rm
%diag}(e^{-2\,\varphi_1},\,e^{2\,\varphi_2},\,e^{2\,\varphi_3},\,e^{2\,\varphi_4},\,e^{2\,\varphi_1},\,e^{-2\,\varphi_2},\,e^{-2\,\varphi_3},\,e^{-2\,\varphi_4})\,.
%\end{eqnarray}
%Equation (\ref{zetaq}) is consistent with the first of
%(\ref{constantsofmotion}).

\par The space $[\SL(2,\Real)/\SO(1,1)]^4$ is a
submanifold of the (para-)quaternionic K\"ahler manifold
\begin{eqnarray}
{\Scr M}_{QK}&=&\frac{\SO(4,4)}{\SO(2,2)\times\SO(2,2)}\subset
\frac{\E_{8(8)}}{\SO^*(16)}\,,
\end{eqnarray}
which originates from the time reduction of the $D=4,
\,\mathcal{N}=2$ STU model characterized by the following scalar
manifold
\begin{eqnarray}
{\Scr M}_{4}^{(STU)}&=&\left(\frac{{\rm SL}(2,\Real)}{{\rm
SO}(2)}\right)^3\,.\label{mstu}
\end{eqnarray}
Thus the generating solution of $D=4$ extreme static black holes
in the maximal theory is also a solution of this quarter-maximal
truncation \cite{Andrianopoli:1997wi}. The embedding of the STU
model inside the maximal theory in $D=4$ can be described as
follows. The central charge matrix ${\bf Z}_{AB}$, $A,B=1,\dots,
8$, of the $D=4,\,\mathcal{N}=8$ theory is a complex antisymmetric
matrix which can be skew-diagonalized using the $\SU(8)$ symmetry
\cite{Ferrara:1980ra}
\begin{eqnarray}
{\bf Z}_{AB}&\stackrel{\SU(8)}{\longrightarrow}&{\bf
Z}_N=\left(\begin{matrix}{\bf Z}_1\,\epsilon &  &&  \bf{0}\cr &{\bf
Z}_2\,\epsilon & &\cr & &{\bf Z}_3\,\epsilon &\cr \bf{0}&&& {\bf
Z}_4\,\epsilon
\end{matrix}\right)\,\,\,,\,\,\,\,\epsilon=\left(\begin{matrix}0 & 1\cr -1 &
0\end{matrix}\right)\,,\label{ZN}
\end{eqnarray}
where ${\bf Z}_k$, $k=1,\dots, 4$, are complex numbers. The normal
form ${\bf Z}_N$ of the central charge matrix is invariant under
the action of $\SU(2)^4\subset \SU(8)$ which is nothing but
$H_{cent}$. Seeing ${\bf Z}_{AB}$ as a function of the scalar
fields and the electric and magnetic charges, the reduction
(\ref{ZN}) can be effected by truncating the $\mathcal{N}=8$ model
to the STU one described by three complex moduli $s,t,u$ and eight
quantized charges in the ${\bf R}_{STU}={\bf (2,2,2)}$ of
$G_4^{(STU)}=\SL(2,\Real)^3$, defined as those charges out of the
56 which are invariant with respect to the action of
$G_{cent}=\SO(4,4)$. The sub-groups of $G_4^{(STU)}$ and
$G_{cent}$ inside $\E_{7(7)}$, being respectively the
\emph{normalizer} and the \emph{centralizer} of ${\bf R}_{STU}$,
commute with one another. Upon reduction to $D=3$, the normal form
$Q_N$ of $Q$ is defined by isometries of the manifold
$[\SL(2)/\SO(1,1)]^4$. Embedding our generating solution in the
STU model allows us to discuss its supersymmetry properties.\par
Let ${\bf a}_1,\,{\bf a}_2,\,{\bf a}_3$ be the
$\frak{e}_{7(7)}\subset\frak{e}_{8(8)}$ positive roots defining
the three $\frak{sl(2,\Real)}$ algebras in $G_4^{(STU)}$. Having
chosen $G_{cent}=\SO(4,4)$ to be identified by the sub-Dynkin
diagram $\Phi_{cent}=(\alpha_3,\dots, \alpha_6)$, the roots ${\bf
a}_i$ are identified as the positive roots orthogonal to
$\Phi_{cent}$
\begin{eqnarray}
{\bf
a}_1&=&\alpha_1+2\,\alpha_2+2\,\alpha_3+2\,\alpha_4+\alpha_5+\alpha_6\,\,\,,\,\,\,\,{\bf
a}_2=\alpha_1\,,\nonumber\\{\bf
a}_3&=&\alpha_1+2\,\alpha_2+3\,\alpha_3+4\,\alpha_4+2\,\alpha_5+3\,\alpha_6+2\,\alpha_7\,.
\end{eqnarray}
The coset representative of the STU model in the solvable gauge has
the form
\begin{eqnarray}
L_{STU}&=&e^{a_i\,E_{{\bf a}_i}}\cdot
e^{\frac{1}{2}\,\tilde{\varphi}_i\,H_{{\bf a}_i}}\in{\Scr
M}_{4}^{(STU)}\,,\label{lstu}
\end{eqnarray}
where, in terms of the six real parameters
$a_i,\,\tilde{\varphi}_i$, the complex scalar fields $s,t,u$ in the
special coordinate frame of ${\Scr M}_{4}^{(STU)}$ read
\begin{eqnarray}
s&=&-a_1-i\,e^{\tilde{\varphi}_1}\,\,,\,\,\,t=-a_2-i\,e^{\tilde{\varphi}_2}\,\,,\,\,\,u=-a_3-i\,e^{\tilde{\varphi}_3}\,.
\end{eqnarray}
 Similarly the eight $\gamma$-roots associated with the
electric and magnetic potentials of the STU model are defined out of
the 56 of the maximal theory as those which are orthogonal to
$\Phi_{cent}$. Written in the Cartan basis $H_0,\,H_{{\bf
a}_1},\,H_{{\bf a}_2},\,H_{{\bf a}_3}$ the eight $\gamma$-roots
associated with the STU model read
\begin{eqnarray}
\gamma^{(1)} &=&\frac{1}{2}\,(1,-1,-1,-1)\,\,,\,\,\,\gamma^{(2)}
=\frac{1}{2}\,(1,1,-1,-1)\,\,,\,\,\,\gamma^{(3)}
=\frac{1}{2}\,(1,-1,1,-1)\,,\nonumber\\
\gamma^{(4)} &=&\frac{1}{2}\,(1,-1,-1,1)\,\,,\,\,\,\gamma^{(5)}
=\frac{1}{2}\,(1,1,1,1)\,\,,\,\,\,\gamma^{(6)}
=\frac{1}{2}\,(1,-1,1,1)\,,\nonumber\\
\gamma^{(7)} &=&\frac{1}{2}\,(1,1,-1,1)\,\,,\,\,\,\gamma^{(8)}
=\frac{1}{2}\,(1,1,1,-1)\,.\label{8gammas}
\end{eqnarray}
Out of the above roots $\gamma^{(n)}$, $n=1,\dots, 8$, we choose a
maximal system of four mutually orthogonal vectors $(\gamma_k)$,
$k=1,\dots, 4$, which defines the normal form. We could choose for
instance
$(\gamma_k)=(\gamma^{(1)},\,\gamma^{(6)},\,\gamma^{(7)},\,\gamma^{(8)})$
or
$(\gamma_k)=(\gamma^{(2)},\,\gamma^{(3)},\,\gamma^{(4)},\,\gamma^{(5)})$.
Let us make the first choice and denote by
$\gamma_k{}^0,\,\gamma_k{}^1,\,\gamma_k{}^2,\,\gamma_k{}^3$ the
components of $\gamma_k$ in the basis $H_0,\,H_{{\bf a}_1},\,H_{{\bf
a}_2},\,H_{{\bf a}_3}$, given in (\ref{8gammas}). From the equation
\begin{eqnarray}
U\,H_0+\frac{1}{2}\,\sum_{i=1}^3\tilde{\varphi}_i\,H_{{\bf
a}_i}&=&\frac{1}{2}\,\sum_{k=1}^4\varphi_k\,H_{\gamma_k}\,,
\end{eqnarray}
we may deduce the relation between $U,\,\tilde{\varphi}_i$ and
$\varphi_k$
\begin{eqnarray}
U&=&\frac{1}{2}\sum_{k=1}^4\gamma_k{}^0\,\varphi_k=\frac{1}{4}\sum_{k=1}^4\varphi_k\,,\nonumber\\
\tilde{\varphi}_1&=&\sum_{k=1}^4\gamma_k{}^1\,\varphi_k=\frac{1}{2}\,(-\varphi_1-\varphi_2+\varphi_3+\varphi_4)\,,\nonumber\\
\tilde{\varphi}_2&=&\sum_{k=1}^4\gamma_k{}^2\,\varphi_k=\frac{1}{2}\,(-\varphi_1+\varphi_2-\varphi_3+\varphi_4)\,,\nonumber\\
\tilde{\varphi}_3&=&\sum_{k=1}^4\gamma_k{}^3\,\varphi_k=\frac{1}{2}\,(-\varphi_1+\varphi_2+\varphi_3-\varphi_4)\,.\label{tp}
\end{eqnarray}
The above relations allow us to write the dilatonic solution
($a_i=0$)  (\ref{dilution}) in terms of the fields $s,t,u$
\begin{eqnarray}
s&=&-i\,\sqrt{\frac{H_1\,H_2}{H_3\,H_4}}\,\,,\,\,\,t=-i\,\sqrt{\frac{H_1\,H_3}{H_2\,H_4}}\,\,,\,\,\,u=-i\,\sqrt{\frac{H_1\,H_3}{H_2\,H_4}}\,\,,\,\,\,\chi_k=\mp\frac{Q_k}{H_k}\,\tau\,.\label{dilution2}
\end{eqnarray}
The above solution clearly exhibits an attractor behavior at the
horizon ($\tau\rightarrow -\infty$) where the scalar fields flow to
the following fixed values
\begin{eqnarray}
s&\rightarrow
&-i\sqrt{\frac{Q_1\,Q_2}{Q_3\,Q_4}}\,\,,\,\,\,t\rightarrow-i\,\sqrt{\frac{Q_1\,Q_3}{Q_2\,Q_4}}
\,\,,\,\,\,u=-i\,\sqrt{\frac{Q_1\,Q_3}{Q_2\,Q_4}}\,.
\end{eqnarray}
Next, we need to identify the parameters $Q_k$ with the quantized
charges ${\bf Q}=(p^\Lambda,\,q_\Lambda)$, $\Lambda=0,\dots, 3$, of
the STU model and $\chi_k$ with the electric-magnetic potentials
$Z=(Z^\Lambda,\,Z_\Lambda)$. This is done by writing the first of
eqs. (\ref{constantsofmotion}) for zero Taub-NUT charge
$\omega_\tau=0$
\begin{eqnarray}
\dot{Z}&=&e^{2U}\,\mathbb{C}\,\mathcal{M}^{STU}_4\,{\bf
Q}\,,\label{zmq}
\end{eqnarray}
where $\mathcal{M}^{STU}_4=L_{STU}\,(L_{STU})^T$ in the
eight-dimensional symplectic representation and its explicit form is
given in the appendix. From this equation we deduce the following
identification
\begin{eqnarray}
Z^0&=&\chi_1\,\,,\,\,\,Z_1=\chi_2\,\,,\,\,\,Z_2=\chi_3\,\,,\,\,\,Z_3=\chi_4\,,\nonumber\\
q_0&=& \mp Q_1\,\,,\,\,\,p^1=\pm Q_2\,\,,\,\,\,p^2=\pm
,Q_3\,\,,\,\,\,p^3=\pm, Q_4\,.\label{QQq}
\end{eqnarray}
\paragraph{BPS and non-BPS solutions} Now we are ready to discuss
the supersymmetry properties of the above dilatonic solutions. To
this end we compute on the solution, at the horizon, the complex
central charge ${\bf Z}$ and matter charges ${\bf Z}_s,\,{\bf
Z}_t,\,{\bf Z}_u$ (we refer the reader to the appendix for a
definition of these charges). When embedding the STU model in the
maximal theory, these charges are naturally identified with the
skew-eigenvalues ${\bf Z}_k$, $k=1,\dots, 4$ of ${\bf Z}_N$. We
start from some general facts about $D=3$ fermionic fields in
quarter maximal theories. As we have seen, general form of $H^*$
is $H^*=\SL(2,\Real)_0\times G_4$. In the $D=3$ theory originating
from the STU model we indeed have $H^*=\SO(2,2)\times
\SO(2,2)=\SL(2,\Real)_{0}\times (\SL(2,\Real))^3$. A fermion in
$D=3$ has the form $\lambda^{M}$, where $M$ runs over the
symplectic ${\bf R}$ representation of $G_4$. Its supersymmetry
variation on the solution reads
\begin{eqnarray}
\delta \lambda^M&=&Q^{M\,A}\,\epsilon_A\,,
\end{eqnarray}
where $A=1,2$, $\epsilon_A$ is the supersymmetry parameter and
$Q^{M\,A}$ is the $H^*$-covariant form of the matrix $Q$ discussed
in section \ref{nmax}. The solution is BPS if there exists at the
horizon ($\tau\rightarrow-\infty$) a Killing spinor, namely a
supersymmetry parameter $\epsilon_A$ for which $\delta
\lambda^M=0$. As discussed in \cite{Pioline:2006ni}, this is the
case if the following factorization occurs: $Q^{M\,A}=C^M\,v^A$.
Indeed this property of the matrix $Q$ ensures that the
supersymmetry variations of $\lambda^M$ vanishes along the
direction $\epsilon_{A}=\epsilon_{AB}\,v^B$, where $\epsilon_{AB}$
is the $\SL(2,\Real)$ invariant tensor. Recall that in the STU
model case ${\bf R}_{STU}={\bf (2,2,2)}$ of $G^{STU}_4$ and thus
we can write $M=(A_1,A_2,A_3)$. Let us consider the various
relevant cases
\begin{itemize}
\item{
{\bf BPS solutions:}
\begin{eqnarray}
Q^{M\,A}&=&Q^{A_1 A_2A_3A}=C^{A_1 A_2A_3}\,v^A\,\,\Rightarrow
\,\,(\mbox{At the horizon})\,\,\,\,{\bf Z}\neq 0\,,\,\,{\bf
Z}_s={\bf Z}_t={\bf
Z}_u=0\,.\nonumber\\&&\label{bpss}\end{eqnarray}}
\item{{\bf non-BPS solutions:}
\begin{eqnarray}
Q^{A_1 A_2A_3A}=C^{A A_2A_3}\,v^{A_1}\,\,&\Rightarrow& \,\,(\mbox{At
the horizon})\,\,\,\,{\bf Z}_s\neq
0\,,\,\,{\bf Z}={\bf Z}_t={\bf Z}_u=0\,,\nonumber\\
Q^{A_1 A_2A_3A}=C^{A A_1A_3}\,v^{A_2}\,\,&\Rightarrow& \,\,(\mbox{At
the horizon})\,\,\,\,{\bf Z}_t\neq
0\,,\,\,{\bf Z}={\bf Z}_s={\bf Z}_u=0\,,\nonumber\\
Q^{A_1 A_2A_3A}=C^{A A_1A_2}\,v^{A_3}\,\,&\Rightarrow& \,\,(\mbox{At
the horizon})\,\,\,\,{\bf Z}_u\neq
0\,,\,\,{\bf Z}={\bf Z}_s={\bf Z}_t=0\,,\nonumber\\
Q^{A_1 A_2A_3A}\,\, \mbox{Not factorized}\,\,&\Rightarrow&
\,\,(\mbox{At the horizon})\,\,\,\,|{\bf Z}|=|{\bf Z}_s|=|{\bf
Z}_t|=|{\bf Z}_u|\,.\label{nonbps}
\end{eqnarray} }
\end{itemize}
This suggests that there could be a connection between the analysis
in (\ref{bpss})-(\ref{nonbps}) and the analysis by Ferrara and Duff
on $q$-bits \cite{Duff:2006rf}, though they do not consider the
three-dimensional theory.

 In all the above cases the entropy  $S_{B-H}$  of the
black hole at the horizon is given by the area law and has the
following expression in terms of the central charges and the
quantized charges \cite{Kallosh:1996uy}
\begin{eqnarray}
S_{B-H}&=&\frac{A_H}{4}=\pi\,(|{\bf Z}|^2+|{\bf Z}_s|^2+|{\bf
Z}_t|^2+|{\bf
Z}_u|^2)\vert_{\mbox{horizon}}=\pi\,\sqrt{|I_4(p,q)|}\,,\label{sbh}
\end{eqnarray}
where $I_4(p,q)$ is the quartic invariant of the ${\bf 56}$ of
$G_4=\E_{7(7)}$. The first three cases in (\ref{nonbps}), where
the factorization occurs, define non-BPS solutions of the
$\mathcal{N}=2$ STU model which are very similar to the BPS
solution in that the role of the central charge and one of the
matter charges are interchanged. In fact, they correspond to BPS
solutions of STU models which are differently embedded in the
parent $\mathcal{N}=8$ model and are characterized by a different
identification of the four $\mathcal{N}=2$ charges ${\bf Z},\,{\bf
Z}_s,\,{\bf Z}_t,\,{\bf Z}_u$ with the $\mathcal{N}=8$ charges
${\bf Z}_1,\,{\bf Z}_2,\,{\bf Z}_3,\,{\bf Z}_4$. These solutions
are thus $1/8$--BPS solutions of the $\mathcal{N}=8$ theory. The
last case in (\ref{nonbps}) define genuine non-BPS solutions of
the $\mathcal{N}=8$ theory.

\par
Let us now discuss the issue of supersymmetry on our simple
dilatonic solution (\ref{dilution2}) and show that all the above
solutions are mapped into one another by a symplectic
transformation on the quantized charges. The first step is to
characterize the $\SL(2,\Real)_0$ group which factorizes
$G_4=(\SL(2,\Real))^3$ in $H^*$ and acts on the index $A$ of
$Q^{M\,A}$. The $\U(1)$ subgroup of $\SL(2,\Real)_0$ correspond to
the K\"ahler transformations on the STU model and is generated by
\begin{eqnarray}
J_{\U(1)}&=&\frac{1}{2}\,(E_{\beta_0}-E_{-\beta_0})+\frac{1}{2}\,\sum_{i=1}^3(E_{{\bf
a}_i}-E_{-{\bf a}_i})\,.
\end{eqnarray}
The remaining two non-compact generators are
\begin{eqnarray}
\tilde{H}_0&=&\frac{1}{2}\,(-J_1+J_2+J_3+J_4)\,\,,\,\,\,\tilde{H}^\prime_0=\frac{1}{2}\,\sum_{i=2}^5(E_{
\gamma^{(i)}}+E_{-\gamma^{(i)}})\,,
\end{eqnarray}
where we recall that $J_k=E_{ \gamma_k}+E_{-\gamma_k}$ and our
choice of the normal form consisted in identifying
$(\gamma_k)=(\gamma^{(1)},\,\gamma^{(6)},\,\gamma^{(7)},\,\gamma^{(8)})$.
We can take $\tilde{H}_0$ as the Cartan generator of
$\frak{sl}(2,\Real)_0$. The Cartan generators of the remaining
$\frak{sl}(2,\Real)^3$ in $\frak{H}^*$ can then be chosen to be
\begin{eqnarray}
\tilde{H}_1&=&\frac{1}{2}\,(J_1-J_2+J_3+J_4)\,,\nonumber\\
\tilde{H}_2&=&\frac{1}{2}\,(J_1+J_2-J_3+J_4)\,,\nonumber\\
\tilde{H}_3&=&\frac{1}{2}\,(J_1+J_2+J_3-J_4)\,.
\end{eqnarray}
Consider first the BPS solution (\ref{bpss}). Modulo an
$\SL(2,\Real)_0$ rotation, we can always take $v^A$ to be a lower
weight vector, namely an eigenvector of $\tilde{H}_0$ with
eigenvalue $- 1/2$. This corresponds to the condition
\begin{eqnarray}
[\tilde{H}_0,\,Q_N]&=&\frac{1}{2}\,Q_N\,.\label{combps}
\end{eqnarray}
From eqs. (\ref{QNn}) and (\ref{JNn}) we see that the only
combination satisfying (\ref{combps}) is
\begin{eqnarray}
Q_N&=&\sqrt{2}\, Q_1\,n_1^-+\sqrt{2}\, Q_2\,n_2^++\sqrt{2}\,
Q_3\,n_3^++\sqrt{2}\, Q_4\,n_4^+\,.
\end{eqnarray}
From eq. (\ref{QQq})  we can read the corresponding quantized
charges of the STU model
\begin{eqnarray}
{\bf Q}^{BPS}&=&(p^\Lambda,\,q_\Lambda)=(0,Q_2,Q_3,Q_4,Q_1,0,0,0)\,.
\end{eqnarray}
In this case we find at the horizon
\begin{eqnarray}
|{\bf Z}|=(4\,Q_1Q_2Q_3Q_4)^{\frac{1}{4}}=(4\,q_0 p^1 p^2
p^3)^{\frac{1}{4}}\,\,,\,\,\,{\bf Z}_s={\bf Z}_t={\bf Z}_u=0\,,
\end{eqnarray}
and thus, from (\ref{sbh}) we find $S_{B-H}=\pi\,|{\bf
Z}|^2=2\,\pi\,\sqrt{q_0 p^1 p^2 p^3}=\pi\,\sqrt{I_4(p,q)}$, where
$I_4(p,q)=4\,q_0 p^1 p^2 p^3>0$ is the quartic invariant of the
${\bf 56}$ of $\E_{7(7)}$, restricted to the chosen normal form
$R_N$.

\par We can make for $Q_N$ a more general choice which does
not correspond to eigenmatrices of the adjoint action of
$\tilde{H}_0$, as in (\ref{combps}), namely take
\begin{eqnarray}
Q_N&=&
\sqrt{2}\,Q_1\,n_1^{-\varepsilon_1}+\sqrt{2}\,Q_2\,n_2^{\varepsilon_2}+\sqrt{2}\,Q_3\,n_3^{\varepsilon_3}+\sqrt{2}\,Q_4\,n_4^{\varepsilon_4}\,,
\end{eqnarray}
where $\varepsilon_k=\pm 1$. The general identification (\ref{QQq})
reads
\begin{eqnarray}
q_0&=& \varepsilon_1\, Q_1\,\,,\,\,\,p^1=\varepsilon_2\,
Q_2\,\,,\,\,\,p^2=\varepsilon_3\, Q_3\,\,,\,\,\,p^3=\varepsilon_4\,
Q_4\,.\label{QQq2}
\end{eqnarray}
 For $\varepsilon_1=\varepsilon_2=\varepsilon_3=\varepsilon_4$
we are back to the BPS solution. For any other choice of
$(\varepsilon_k)$ the solution is non-BPS. In particular, from eqs.
(\ref{QQq}) we see that the corresponding vector of quantized
charges ${\bf Q}$ is related to the BPS one $Q^{BPS}$ by a
symplectic transformation $\mathcal{S}$
\begin{eqnarray}
{\bf Q}&=&\mathcal{S}\,Q^{BPS}\,\,,\,\,\,\mathcal{S}={\rm
diag}(\varepsilon_1,\varepsilon_2,\varepsilon_3,\varepsilon_4,\varepsilon_1,\varepsilon_2,\varepsilon_3,\varepsilon_4)\,.
\end{eqnarray}
Let us consider the relevant cases.
\begin{itemize}\item There are three independent non-BPS solutions
for which
$\varepsilon=\varepsilon_1\,\varepsilon_2\,\varepsilon_3\,\varepsilon_4>0$.
In these cases one can easily verify that the matrix $Q_N$, though
not verifying (\ref{combps}), satisfies
\begin{eqnarray}
[\tilde{H}_i,\,Q_N]&=&\pm\frac{1}{2}\,Q_N\,,\label{combps2}
\end{eqnarray}
for some $i=1,2,3$. It can therefore be written in one of the
factorized forms in (\ref{nonbps}). The corresponding solutions are
characterized at the horizon by ${\bf Z}=0$ and only one non
vanishing matter charge out of ${\bf Z}_s,\,{\bf Z}_t,\,{\bf Z}_u$,
whose norm equals $(4\,Q_1Q_2Q_3Q_4)^{\frac{1}{4}}=(4\,q_0 p^1 p^2
p^3)^{\frac{1}{4}}$. In this case we still have $I_4(p,q)=4\,q_0 p^1
p^2 p^3>0$ and $S_{B-H}=\pi\,\sqrt{I_4(p,q)}$.
\item If
$\varepsilon=\varepsilon_1\,\varepsilon_2\,\varepsilon_3\,\varepsilon_4<0$,
$Q_N$ does not satisfy either (\ref{combps}) or (\ref{combps2}). As
a consequence $Q_N$ does not have a factorized form. Direct
computation shows that, at the horizon, $|{\bf Z}|=|{\bf Z}_s|=|{\bf
Z}_t|=|{\bf
Z}_u|=(\frac{1}{4}\,Q_1Q_2Q_3Q_4)^{\frac{1}{4}}=\frac{1}{2}\,(-4\,q_0
p^1 p^2 p^3)^{\frac{1}{4}}$. In this case $I_4(p,q)=4\,q_0 p^1 p^2
p^3<0$ and $S_{B-H}=\pi\,\sqrt{-I_4(p,q)}=4\,\pi\,|{\bf Z}|^2$.
\end{itemize}
Notice that, in terms of the positive parameters $Q_{k}$,
$k=1,\ldots,4$, the BPS and non--BPS solutions have the same form.
They acquire a different expression once these parameters are
expressed in terms of the quantized charges $p^\Lambda,\,q_\Lambda$.
We can summarize the expression of the dilatonic BPS and non-BPS
solutions in (\ref{e4u}), (\ref{dilution2}) by denoting the complex
$D=4$ scalars $s,t,u$ by $z_1,z_2,z_3$, and writing
\begin{eqnarray}
z_i&=&-i\,\sqrt{\frac{H_1\,
H_{i+1}}{\,H_{j+1}H_{k+1}}}\,,\,\,Z^0=\varepsilon_1\,\frac{q_0}{H_1}\,,\,\,Z^i=-\varepsilon_{i+1}\,\frac{p^i}{H_{i+1}}\,,\,\,e^{4\,U}=\frac{1}{H_1\,H_2\,H_3\,H_4}\label{disolut}\\
H_1&=&1-\sqrt{2}\,\varepsilon_1\,q_0\,\tau\,\,\,,\,\,\,H_{i+1}=1-\sqrt{2}\,\varepsilon_{i+1}\,p^i\,\tau\,,
\end{eqnarray}
where $i,j,k=1,2,3$ and $i\neq j\neq k$. In all the solutions
discussed above $I_4(p,q)=4\,q_0 p^1 p^2 p^3=4\,\varepsilon\, Q_1
Q_2 Q_3 Q_4$, and therefore
$S_{B-H}=\pi\,\sqrt{\varepsilon\,I_4(p,q)}=$ $2\,\pi\,\sqrt{Q_1
Q_2 Q_3 Q_4}$.

\subsection{The Issue of Generating New Solutions and An
Example}\label{gns} Let us consider the issue of generating $D=4$
solutions with generic charges out of the one discussed above. As
we have pointed out earlier, new solutions are generated by acting
with $G/H^*$ on the asymptotic values $\phi^I_0$ of the scalar
fields and with  the stability group $e^{s_0}\,H^*\,e^{-s_0}$ of
$\phi^I_0$ on the tangent space element $Q_N$. Let us consider the
latter action at fixed $\phi^I_0$, say the origin, whose stability
group is therefore just $H^*$. The action of $H^*$ on $Q_N$,
according to our previous analysis, is sufficient to generate the
most general element $Q\in \frak{g}/\frak{H}^*$. In particular,
the action of $H_c=H_4\times \U(1)=\U(8)$ is enough to generate a
solution depending on all the 56 electric-magnetic charges. If
${\Scr O}$ is a global $H^*$ transformation, it will map a
geodesic $\phi^I(\tau)$ defined by $\phi^I_0=\phi^I(0)=0$ and
charge matrix $Q$, into a different geodesic $\phi^{\prime
I}(\tau)$, with $\phi^{\prime I}_0=0$ and matrix $Q^\prime={\Scr
O}^{-T}\,Q\,{\Scr O}^{T}$. Indeed we can start from the general
action of a global $G$ transformation ${\Scr O}$ on $L(\phi^I)$
\begin{eqnarray}
{\Scr O}\,L(\phi^I)&=&L(\phi^{\prime I})\,h\,,\label{ol}
\end{eqnarray}
where $h\in H^*$ is a local matrix depending  on ${\Scr O}$ and
$\phi^I$. Using the $H^*$-invariance property of $\mathcal{M}$ and
the fact that ${\Scr O}$ is in $H^*$, we can act on both sides of
eq. (\ref{MQ}) by ${\Scr O}$ from the left and ${\Scr O}^T$ from the
right, to find
\begin{eqnarray}
\mathcal{M}(\phi^{\prime I})&=&{\Scr O}\mathcal{M}(\phi^{ I})\,{\Scr
O}^T={\Scr O}\eta\,e^{Q\,\tau}\,{\Scr
O}^{-1}=\eta\,e^{Q^\prime\,\tau}\,.
\end{eqnarray}
This clearly applies to a generic $H^*$ transformation.
 The $\U(1)$ factor in $H_c$, generated by
$E_{\beta_0}-E_{-\beta_0}$, will however generate also a NUT charge.
If we are interested in constructing the most general $D=4$ black
hole depending on all the 56 charges at fixed asymptotic values of
the scalar fields and vanishing NUT charge, we would need to
associate the $\U(1)$ action with a suitable $H^*$ boost, generated
by $E_{\gamma}+E_{-\gamma}$, to keep the NUT charge zero. This
combined transformation was not present in the $D=4$ theory and thus
will generate genuinely new $D=4$ black hole solutions belonging to
different $G_4$ orbits. For instance it could create a non trivial
overall phase for the skew--eigenvalues ${\bf Z}_k$ of ${\bf
Z}_{AB}$, which is a $H_4=\SU(8)$--invariant and which is fixed in
the dilatonic solution discussed in the previous section. This
solution indeed is characterized by four invariant parameters,
represented by the moduli $|{\bf Z}_k|$ computed at spatial
infinity. We can generate a 56--charge black hole in two steps: act
by means of a generic transformation in $\SO(2)^3\in G_4^{STU}$,
followed by the combined $\U(1)\times \mbox{(boost)}$
transformation, to generate the 8-charge general solution of the STU
model; Act on this solution by a 48-parameter transformation in
$\SU(8)/[\SO(2)^3\times \SO(4)^2]$ ($\SO(4)^2$ being $H_{cent}$), to
generate the remaining charges.\par We shall consider here, as an
example, the action of ${\Scr O}\in \SO(2)^3\subset G_4^{STU}$ on
the generating solution described in the previous section. The
transformation ${\Scr O}$ can be written as follows
\begin{eqnarray}
{\Scr O}&=&e^{\sum_{i=1}^3\alpha_i\,(E_{{\bf a}_i}-E_{-{\bf
a}_i})}=\bigotimes_{i=1}^3\left(\begin{matrix}\cos(\alpha_i)&-\sin(\alpha_i)\cr
\sin(\alpha_i)&\cos(\alpha_i)\end{matrix}\right)\,.\label{ooo}
\end{eqnarray}
To evaluate the action of ${\Scr O}$ on $L(\phi^I)$ let us observe
that
\begin{eqnarray}
{\Scr O}\,E_{\gamma^{(m)}}\,{\Scr O}^{-1}&=&{\Scr
O}^{-1}{}_m{}^n\,E_{\gamma^{(n)}}\,,
\end{eqnarray}
where ${\Scr O}_m{}^n$ is the $\Sp(8,\Real)$ representation of
${\Scr O}$ in the basis (\ref{8gammas}). The action (\ref{ol}) of
${\Scr O}$ on $L$ is then readily computed
\begin{eqnarray}
{\Scr O}\,L(\phi^I)&=&{\Scr
O}\,e^{\sqrt{2}\,Z^n\,E_{\gamma^{(n)}}}\,L_{STU}(\phi^r)\,e^{U\,H_0}=e^{\sqrt{2}\,Z^{\prime
n}\,E_{\gamma^{(n)}}}\,L_{STU}(\phi^{\prime r})\,e^{U\,H_0}\,h\,,
\end{eqnarray}
where $Z^{\prime n}=Z^{ m}\,{\Scr O}^{-1}{}_m{}^n$. If we use the
complex notation for the $D=4$ STU scalars
$(\phi^r)=(s,t,u)=(z_1,\,z_2,\,z_3)$ we can easily write
$(\phi^{\prime r})=(z_i^\prime)$ in terms of $(\phi^{ r})=(z_i)$
\begin{eqnarray}
z_i^\prime &=&-a_i^\prime-i\,e^{\varphi_i^\prime}=
\frac{\cos(\alpha_i)\,z_i-\sin(\alpha_i)}{\sin(\alpha_i)\,z_i+\cos(\alpha_i)}\,.
\end{eqnarray}
On the dilatonic solution (\ref{dilution2}) the above transformation
yields
\begin{eqnarray}
z_i^\prime
&=&\frac{\cos(\alpha_i)\sqrt{\,H_1\,H_{i+1}}-i\,\sin(\alpha_i)\,\sqrt{H_{j+1}H_{k+1}}}{\sin(\alpha_i)\,\,\sqrt{H_1\,H_{i+1}}+i\cos(\alpha_i)\,\sqrt{H_{j+1}H_{k+1}}}\,.
\end{eqnarray}
We see that the effect of this transformation is to generate
non--trivially evolving axions, consistently with the analysis of
\cite{Kallosh:2006ib}. The quantized charges are as usual deduced
from eq. (\ref{zmq}), which is $G_4$ covariant, and thus we can
write
\begin{eqnarray}
\dot{Z}^\prime&=&e^{2U}\,\mathbb{C}\,\mathcal{M}^{STU}_4(\phi^{\prime
r})\,{\bf Q}^\prime\,,\label{zmq2}
\end{eqnarray}
where ${\bf Q}^{\prime n}={\bf Q}^{ m}\,{\Scr O}^{-1}{}_m{}^n$. The
vectors $Z^\prime$ and ${\bf Q}^\prime$ can be deduced from the
explicit symplectic representation of ${\Scr O}$ given in appendix
\ref{APPENDIXE}. Finally the warp factor $U$ is not affected by the
transformation. \par The action of ${\Scr O}$ on the four charge
solution has generated a seven charge solution, which is still
described, at infinity,  by the four invariants $|{\bf Z}_k|$, since
the effect  of ${\Scr O}$ is to transform ${\bf Z}_k$ by phases
without affecting the overall phase which is still
fixed\cite{Cvetic:1995bj,*Cvetic:1995kv}. To generate the overall
phase the composite $\U(1)+\mbox{boost}$ transformation is needed.
It would be interesting to study the relation between the resulting
$D=4$ five parameter solution and the seed solution constructed in
\cite{Gimon:2007mh}. This analysis will be pursued elsewhere.

\subsection{Nilpotency of $Q$ for Attractor Black Holes}\label{NilpotencyforAttractorBlack}

It is shown in \cite{Gunaydin:2005mx} that the supersymmetry
preserving black hole solutions lead to a nilpotent charge matrix
in 3D. Later on reference \cite{Gaiotto:2007ag} (see also
\cite{Li:2008ar}) demonstrated that the general nilpotent charge
matrices (of a specific degree) classify all the extreme black
holes that posses attractor behaviour. In the beginning of section
\ref{EXTREMAL} we repeated this argument which for instance shows
that in $D=4$ a non-vanishing horizon implies the nilpotency
condition $Q^5=0$. %This can easily be generalised to higher
%dimensions where the metric Ansatz in $D$ dimensions is given by
%\begin{equation}
%\d s^2=-\e^{(D-3)\sqrt{\frac{8}{(D-3)(D-2)}}U(\tau)}\d t^2 +
%\e^{-\sqrt{\frac{8}{(D-3)(D-2)}}U(\tau)}\bigl(\e^{2(D-2)A(\tau)}\d\tau^2+\e^{2A(\tau)}\d\Omega_{D-2}^2\bigr)
%\end{equation}
%The variable $\tau$ is the affine parameter of the corresponding
%geodesic motion and for extreme black holes we have
%\begin{equation}
%\e^{2A}\sim \frac{1}{\tau^{\tfrac{2}{D-3}}}\,.
%\end{equation}
%% The relation
%%with the usual radial coordinate of the black hole spacetime is
%%\begin{equation}
%%\tau=-\frac{1}{(D-3)\,r^{D-3}}\,.
%%\end{equation}
%From this one derives that a finite horizon implies that
%$e^{-2U}\sim \tau^P$ when $\tau\rightarrow-\infty$, where
%\begin{equation}
%P=\sqrt{\frac{2(D-2)}{(D-3)}}\,.
%\end{equation}
%In $D>4$ the grading of $H_0$ in the adjoint of $\frak{g}$ equals
%$-1$. Thus the condition for a finite horizon area translates into
%the nilpotency condition
%\begin{equation}
%Q^{P+1}=0\,.
%\end{equation}
%When the degree of nilpotency is lower, the black hole solutions
%have (classically) a vanishing horizon area. When it is higher,
%there is no horizon.

The discussion of the nilpotenty properties of the charge matrices
is especially simple in our approach. The reason is that a nilpotent
matrix of degree $N$ (i.e.~$Q^N=0$ and $Q^{N-1}\neq 0$) preserves
its nilpotency degree $N$ under $G$ transformations. This also
applies to the number of preserved supersymmetry charges. Therefore
it is sufficient to study the possible nilpotency degrees for the
generating charge matrix $Q_N$. As before we stick to the
diagonalisable case.

The nilpotent generating charge matrices must have the following
form
\begin{equation}
Q_N=\sum_{k=1}^p c_k n_k\,,\qquad n_k=\sqrt{\tfrac{2}{\gamma_k^2}}
H_{\gamma_k}-E_{\gamma_k}+E_{-\gamma_k}\,,
\end{equation}
where $c_i$ is any real number and the operators. To derive the
nilpotency degree $N$ in the \emph{adjoint representation of
$\frak{g}$} we need to calculate commutators. For that reason we
reviewed the canonical commutation relations for semi-simple Lie
algebras in the Cartan-Weyl basis in appendix \ref{appendix A}

We first evaluate the operator adj$n_k$ on a generic step operator
$E_{\beta}$ with $\beta\neq\gamma_k$ and later we evaluate it on an
arbitrary Cartan operator. Since root strings in general can have
length $1,\ldots,4$ commutators can generate the following
possibilities ($\Delta$ denotes the root lattice.)
\begin{itemize}
\item String 1: $\beta,\beta+\gamma_k \in \Delta\,,$
\item String 2: $\beta,\beta-\gamma_k,\beta+\gamma_k \in \Delta\,,$
\item String 3: $\beta,\beta-\gamma_k,\beta+\gamma_k,\beta+2\gamma_k \in
\Delta\,.$
\end{itemize}
There exist more possibilities but it is easy to show that with some
root redefinitions (e.g. $\beta'=\beta-\gamma_k$) it is sufficient
to consider the above three strings. For string 1 one readily finds
\begin{equation}
[n,[n,E_{\beta}]]=0\,.
\end{equation}
Similarly for string 2 we have
\begin{equation}
[n,[n,[n,E_{\beta}]]]=0\,.
\end{equation}
Let us consider string 3. This calculation is a bit more lengthy and
it is useful to introduce the following notation
\begin{equation}
(\text{adj}n_k)^l=x_lE_{\beta} +y_lE_{\beta-\gamma}
+z_lE_{\beta+\gamma} +w_lE_{\beta+2\gamma}\,,
\end{equation}
where the coefficients $x_l, y_l, z_l, w_l$ obey the following
coupled iteration relations
\begin{align}
x_{l+1}&=x_l \sqrt{2/\gamma_k^2}\,\, (\gamma,\beta)-y_lN_{\gamma,\beta-\gamma}+z_lN_{-\gamma,\gamma+\beta}\,,\\
y_{l+1}&=x_lN_{-\gamma,\beta}+y_l\sqrt{2/\gamma_k^2}\,\,(\gamma,\beta-\gamma)\,,\\
z_{l+1}&=-x_lN_{\gamma,\beta}+z_l\sqrt{2/\gamma_k^2}\,\,(\gamma,\gamma+\beta)+w_lN_{-\gamma,\beta+2\gamma}\,,\\
w_{l+1}&=-z_lN_{\gamma,\beta+\gamma}+w_l(\gamma,\beta+2\gamma)\,,
\end{align}
so we are looking for the number $N\in \mathbb{N}$ such that
$x_N=y_N=z_N=w_N=0$. A straightforward computation then gives that
$N=4$. \footnote{We used the following relations: $
(\gamma,\beta)=-\tfrac{1}{2}\gamma^2\,,\qquad
N^2_{\gamma\beta}=2\gamma^2\,,\qquad
N^2_{\gamma,\gamma+\beta}=N^2_{-\gamma,\beta}=\tfrac{3}{2}\gamma^2\,$.}
%\begin{align}
%& x_{1}=a\langle\gamma,\beta\rangle \,,\quad
%y_{1}=N_{-\gamma,\beta}\,,\quad z_{1}=-N_{\gamma,\beta}\,,\quad
%w_{1}=0\,,\\
%& x_{2}=-3\gamma^2\,,\quad
%y_{2}=-2a\gamma^2N_{-\gamma,\beta}\,,\quad z_{2}=0\,,\quad
%w_{2}=N_{\gamma\beta}N_{\gamma,\beta+\gamma}\,,\\
%& x_{3}=\tfrac{9}{2}a\gamma^4\,,\quad
%y_{3}=3\gamma^2N_{-\gamma\beta}\,,\quad
%z_{3}=\tfrac{9}{2}\gamma^2N_{\gamma\beta}\,,\quad
%w_{3}=\tfrac{3}{2}a\gamma^2N_{\gamma\beta}N_{\gamma,\beta+\gamma}\,,\\
%&  x_{4}= y_{4}=z_{4}= w_{4}=0\,.
%\end{align}
Now we evaluate adj$n$ on an arbitrary Cartan operator $H_{\beta}$.
We immediately find  $[n,[n,H_{\beta}]]\sim n$ and thus
adj$n^3(H_\beta)=0$. In sum we have $n^4=0$ and $n^3$ is generically
non-zero.

\subsubsection*{Simply laced algebras and $E_{8(8)}$} Let us use the
above commutation relations in the case of a simply laced algebra.
Then we have that $\gamma_k^2=2$ and that root strings can have at
maximum length two. From the above relations it is then immediately
clear that semi-simple algebras have $n^3=0$.

Now we take $E_{8(8)}$ as an example. There we have to calculate the
degree of nilpotency of the operator
\begin{equation}
\sum_{i=1}^4 c_in_i\,,
\end{equation}
for its adjoint action. Consider for instance the fifth power. From
the previous discussion we see that the following cross-terms might
possible survive the battle (keep in mind that the $n_i$ mutually
commute)
\begin{align}
& n_1n_2n_3n_4^2\,,\qquad \text{+ permutations in the indices}\,,\\
& n_1n_2^2n_3^2 \,,\qquad \quad\text{+ permutations in the
indices}\,.
\end{align}
Both operators can be seen to vanish on an arbitrary step operator
or Cartan operator using the previously derived identities. In case
we consider the fourth power then, given the above, there is one
cross term which does not obviously vanish, namely
\begin{equation}
c_1c_2c_3c_4 n_1n_2n_3n_4\,.
\end{equation}
The adjoint action of this operator with an arbitrary operator from
the Lie algebra can be shown not to vanish in general. We conclude
that for arbitrary $c_i$
\begin{equation} (\sum_{i=1}^4
c_in_i)^5=0\,,\qquad (\sum_{i=1}^4 c_in_i)^4\neq 0\,.
\end{equation}
Clearly if some $c_i$ are zero the story changes. The point is that
the cross-terms should always have at least a $n_i^2$ in order to
vanish. This analysis clearly holds for cases where $p\neq 4$ and we
deduce the general statement: \emph{if there are $p$ non-zero $c_i$
then the nilpotency is of degree $p+1$.}

\subsubsection*{Non-simply laced algebras and $G_{2(2)}$}
For non-simply laced Lie algebras there are no simplifications since
root strings exist up to length four. In the following we consider
the $G_2$ algebra. The $G_2$ algebra appears in the reduction of the
axion-dilaton black hole of $\mathcal{N}=2$ SUGRA to three
dimensions where we have the coset $G_{2(2)}/[\SL(2)\times\SL(2)]$
(see table \ref{table:N=4supergravities}). This model was analyzed
in \cite{Gaiotto:2007ag}. From table \ref{table:N=4supergravities}
we find that
\begin{equation}
Q_N= c_1 n_1 + c_2n_2\,.
\end{equation}
The step operators appearing in $n_1$ and $n_2$ must be mutually
orthogonal \footnote{Take for instance $\gamma_1=\alpha_l+\alpha_s$
and $\gamma_2=\alpha_l+3\alpha_s$, where $\alpha_s$ and $\alpha_l$
denote the short and the long simple root of $G_{2(2)}$.}. Clearly
$Q^7=0$ since all the cross-terms in the product $(n_s + n_l)^7$ at
least contain a $n^4$ and therefore vanish. In fact this clearly
holds for all the other (non-simply laced) algebras: \emph{The
nilpotency degree $N$ obeys $N\leq 3p+1 $}. To know $N$ precisely we
seem to need a case by case study. Let us therefore continue with
$G_{2(2)}$ and the other cases are similar.

For $G_{2(2)}$ we have $Q^5=0$. This can be understood as follows.
The product $(n_1 + n_2)^5$ contains the following terms that are
not obviously zero
\begin{equation}
n_1^3n_2^2\,,\qquad n_2^3n_1^2\,.
\end{equation}
Both terms evaluated on an arbitrary Cartan operator $H_\beta$
clearly vanish. Let us therefore consider step operators $E_\beta$.
If we can argue that an arbitrary $E_{\beta}$ either forms a
$\gamma_1$-string or $\gamma_2$-string with length smaller then four
we are done since in that case $n_1^3$ resp $n_2^3$ vanish on
$E_{\beta}$. This is not too hard to derive for the root-lattice of
$G_2$ \footnote{The reason is simple: $G_{2}$ has only 12 roots. If
there exist two different strings of length four, then also their
negative images exist, and we would end up with more then 12
roots.}.

\section{The Physics III: Euclidean Wormholes}\label{PHYSICSIII}

In this section we discuss wormhole solutions of the Euclidean
theories in $D<10$ obtained from reduction over time. Euclidean
wormhole solutions are discussed in the literature for their
possible role in quantum gravity and holography (see
\cite{Maldacena:2004rf, Arkani-Hamed:2007js,
Bergman:2007ss,Bergshoeff:2005zf} for some recent discussions and
other references). In particular one can study wormhole effects in
string theory which motivates the search for wormhole solutions in
supergravity.

Euclidean wormhole solutions generically suffer from singularity
problems. The singularities are not geometrical since the geometry
is always a smooth wormhole as described in section
\ref{geometries}, but the problem resides in the scalar fields.
The singularities can be circumvented as shown in
\cite{Bergshoeff:2004pg, Arkani-Hamed:2007js, Bergman:2007ss}. For
instance Euclideanized $\mathcal{N}=2$ theories that arise from
CY-compactifications \cite{Bergshoeff:2004pg} seem to allow for
regular Euclidean wormholes. Later examples in Euclideanized
maximal supergravity have been found \cite{Arkani-Hamed:2007js,
Bergman:2007ss}. But as discussed in \cite{Bergman:2007ss,
Cremmer:1998em} there is an issue in how to define the Euclidean
theory. If the Euclidean theory is defined through some liberal
analytical continuation then many possibilities exist of which
many have regular wormhole solutions. However, in here, we take a
more conservative point of view and only consider those Euclidean
theories that are obtained through dimensional reduction over time
of some Lorentzian supergravity theory. This has the advantage
that the Euclidean theory has a well-defined supersymmetry
\cite{Cremmer:1998em,Cortes:2003zd}. These are not the Euclidean
theories that have been considered in those references that
constructed the regular solutions and the question of the
existence of regular wormholes in those models is still an open
one.

In our approach we only need to consider the family of minimal
generating geodesic curves and pick out the ones that represent
regular wormholes. We then know that all regular wormholes can be
obtained by acting with the global symmetry group since the action
of the symmetry group does not affect the smoothness of the
solution. This is a good illustration of the usefulness of the
generating geodesic.

Let us discuss the regularity in the elegant approach of
\cite{Arkani-Hamed:2007js}. The scalar fields trace out geodesics on
moduli space and these geodesics are parameterized by the coordinate
$r$ which starts at $r=0$ (in conformal gauge) on the complete left
of the wormhole  and ends on the very right-end $r=+\infty$ of the
wormhole. Let us calculate the length $d$ of such a curve
\begin{equation}
d=\int^{r=+\infty}_{r=0}\sqrt{|G_{ij}\partial_r\phi^i\partial_r\phi^j|}=\int^{r=+\infty}_{r=0}
\sqrt{2|\mathcal{R}_{rr}|}=\pi \sqrt{2\frac{D-1}{D-2}}\,.
\end{equation}

A singularity occurs when the geodesic on the moduli space is
shorter than $d$. Since then the solution is such that several
geodesics are `patched' together to get the solution defined over
the whole wormhole. This patching introduces singularities in the
scalar fields which are problematic. Let us consider an example. For
the axion-dilaton system $\SL(2,\Real)/\SO(1,1)$ the expression for
the dilaton is something like $\e^{\beta\phi}\sim \sin(h)$ with $h$
the harmonic and $\beta$ the radius of the coset
$\SL(2,\Real)/\SO(1,1)$. Clearly, when the $\sin$ function switches
sign a problem occurs and one has to change the solution to
$\e^{\beta\phi}\sim |\sin(h)|$ which is singular when $h=0$.

If we consider the minimal generating geodesic solutions we can
restrict to the submanifold (\ref{generatingsubmanifold}) Since the
decoupled dilatons in the $\SO(1,1)^{r-p}$-part only make the length
smaller, we consider them to be truncated. Then the maximal length
of the geodesic is the sum of the maximal lengths of the geodesic on
the different $\SL(2)$ - pairs \cite{Arkani-Hamed:2007js}
\begin{equation}
d^2 = \sum_{i=1}^p \frac{4\pi^2}{\beta_i^2}\,,
\end{equation}
where the $\beta_i$ are the different radii of the $\SL(2)$-factors.
The condition of regularity then becomes the inequality
\begin{equation}
\sum_{i=1}^p \frac{4\pi^2}{\beta_i^2} > 2\pi^2 \frac{D-1}{D-2}\,.
\end{equation}

In the following we take diagonalisable $Q$ only. The reason
becomes clear when we study a generic non-diagonalisable case,
described in (\ref{nondiagonal}). This solution can be seen as a
set of $p$ axion-dilaton pairs, each with $\beta=\beta_i$, related
to the $\SL(2,\Real)/\SO(1,1)$ factors, and another decoupled
axion-dilaton pair, excited by $Nil$, for which the solution is
regular and fixed such that it has vanishing velocity squared.
Since $Nil$ commutes with the diagonalizable part of $Q$, this
decoupled pair does neither contribute to the wormhole geometry
nor it introduces irregularities. Thus the criterium for
regularity of this solution is the same as for the axion-dilaton
solution with diagonalizable $Q$.
%
% with $\beta=2$ as in the diagonal case. This
%generalizes to arbitrary cosets.

From the tables presented in section \ref{MATH} we can easily
verify this condition in the various theories we considered. We
find that in $D=3,4$ the quarter-maximal, half-maximal and maximal
supergravity theories behave identically in the sense that the
regularity bound can at most be saturated. In $D=5$ they also
behave identically since the maximal length equals $3/4$ in all
cases which is smaller then the lower bound, so there are no
regular solutions (not even saturated ones). From $D=6$ we only
consider maximal and half maximal supergravity and they again have
the same maximal length which is again too small to lead to any
regular wormhole. So we conclude that for all cases we
investigated in $D>4$ there cannot be regular wormholes.  And for
the $D=3,4$ theories regular wormholes exist in the saturation
case which implies that the singularities are pushed towards the
boundaries of the wormhole solutions.

The similarity between the maximal and half-maximal case is easily
understood; the generating geodesic is carried by moduli that have
their 10D origin in the common sector of both type II and type I
theories. Therefore these geodesics describe exactly the same
solutions. The similarity between quarter-maximal supergravity and
maximal supergravity is smaller; the maximal geodesic length is
still the same but the number of $\SL(2)$-factors and their
$\beta_i$ differ. However in those cases that the quarter maximal
theory is obtained from an orientifolded torus compactification one
again notices that the geodesics have an identical 10D origin as the
geodesics in maximal SUGRA. The orientifold action can identify
moduli and therefore decrease the dimension of the moduli space. In
case two axion-dilaton pairs in the generating submanifold are
identified the number $p$ decreases with one and the $\beta$-factor
decreases with a factor $\sqrt{2}$. This gives a 10D origin for some
of generating submanifold with $\beta_i\neq 2$ and $p<4$ in
Euclidean quarter-maximal theories.

The addition of a negative cosmological constant to our models
gives rise to wormholes that asymptote to Euclidean AdS at the two
boundaries \cite{Arkani-Hamed:2007js, Bergshoeff:2005zf,
Gutperle:2002km}. The effect of the cosmological constant is to
relax the regularity bound as noted in \cite{Arkani-Hamed:2007js,
Bergshoeff:2005zf}. However we have not found a way to add a
cosmological constant, consistent with a supergravity embedding,
in such a way that the axion-dilaton pairs are still free scalars.
The only exception we are aware of is the construction of
Euclidean wormholes in $D=5$ maximal gauged supergravity
\cite{Bergshoeff:2005zf}, obtained from the $S^5$-compactification
of Euclidean IIB supergravity. Unfortunately those wormholes
aren't regular either (unless one performs a more liberal Wick
rotation then the one that is used to define Euclidean IIB ).

From our approach it is also straightforward to understand why
liberal Wick rotations allow regular wormholes. Consider for
instance maximal supergravity in $D=3$ and Wick-rotate several
axions such that $H^*$=SO$(8,8)$. In that case $E_{8(8)}$/H* would
contain a [$\text{SL}(2,\Real)/\text{SO}(1,1)]^8$. \footnote{The 64
axions with negative signature are defined by the level
decomposition with respect to $\alpha_7$ and are the RR fields
(level 1). This defines the Wick rotation.}  In this case the
regularity bound is strictly satisfied.

\section{Discussion}\label{CONCLUSION}

In this paper we introduced a powerful technique, formulated as a
theorem in subsection \ref{THEOREMA}, to generate a large class of
new solutions of supergravity theories by acting with global
symmetries on the so-called {\sl minimal generating solution}. The
solution-generating symmetry is not a symmetry of the
corresponding Lagrangian but only arises upon dimensional
reduction of the supergravity solution over its world volume. In
particular the reduced solution is described by a geodesic curve
on the moduli space. In the case of a symmetric moduli space, the
theorem specifies
  the normal form corresponding to the generating solution. The procedure is both valid for split and
non-split isometry groups. To find the new solutions we never need
to solve any differential equation\footnote{This also holds for
the Einstein equation related to the $(-1)$-brane metrics in the
lower dimension. In section (\ref{section2}) we demonstrated that
the Einstein equation can be reduced to a first-order equation
(\ref{EinsteinII}). When performing a change of coordinates via
$r\rightarrow g(r)$ one finds an expression for the metric without
solving the Einstein equation. }.

We applied the theorem to three cases: (i) Einstein vacuum
solutions, (ii) non-extreme (single-centered) black holes in $D=4$,
$\mathcal{N}=8$ supergravity and (iii) Euclidean wormholes in
symmetric supergravity theories for $D\geq 3$. We also discussed
extreme black holes in the $N=8,\,D=4$ theory, corresponding in
$D=3$ to geodesics with a nilpotent $Q.$

Exponentiating nilpotent matrices $Q$ in the truncated theory we
are able to reproduce the known dilatonic extreme black hole
solutions. Embedding these solutions in the STU model allows us to
discuss its supersymmetric properties. We showed that the
factorization property, which discriminates BPS from non-BPS
solutions, can be given a simple group-theoretical property as
explained in section 5.3. In section 5.4 we explicitly performed a
symmetry-generating transformation on the dilatonic STU black hole
to find a 7-charge solution with varying axions. Furthermore, we
illustrated how to generate $D=4$ solutions with
  generic charges from this minimal one, though leave the details of
  the analysis for future work.

Finally, in the case of wormholes we obtained a full understanding
of the number of regular wormholes for a given symmetric coset. In
particular, we found that in the case of Euclidean supergravities
that follow from the dimensional reduction of a higher-dimensional
Minkowskian supergravity there are no regular wormhole solutions. We
are able to find wormhole solutions that at most saturate the
regularity bound.

Many of the existing solution-generating techniques in the
literature exploit the symmetries of the theory which often
correspond to duality transformations. The benefits of such a
solution-generating technique is a classification of the solutions
in terms of duality orbits. Especially in the case of black holes
this allows one to classify solutions with the same entropy since
duality transformations preserve the number of (quantum)
micro-states. The symmetry we have employed in the case of black
holes is not a duality and therefore does not preserve the entropy.
But we hope to have demonstrated that our solution-generating is
useful in many ways. Let us summarize the strong points:
\begin{itemize}
\item Solutions are constructed without solving any differential
equation.
\item In the case of Einstein vacuum solutions it classifies
solutions up to coordinate transformations for $D>3$. We were able
to reconstruct rather involved solutions in an economic manner.
\item In the case of black holes the generating symmetry commutes
with supersymmetry. Thus an investigation of the susy properties of
the generating solution suffices for knowing the susy of all black
hole solutions.
\item In the case of black holes that possess attractor behavior
the solution is characterized by nilpotent matrices. The generating
symmetry preserves the degree of nilpotency. Thus the nilpotency of
the generating solution is sufficient to understand the nilpotency
of all solutions. We illustrated this for the symmetries $E_{8(8)}$
and $G_{2(2)}$.
\item The construction of BPS and non-BPS solutions
(extreme and non-extreme) is treated on the same footing. Thus
from the point of view of finding solutions this technique is
clearly beneficial. We briefly demonstrated how to find new
solutions for non-BPS STU black holes.
\item When considering  instantons, the generating symmetry \emph{is}
a duality and our theorem thus classifies those solutions in terms
of duality orbits.
\item In the case of Euclidean wormhole solutions we were able to
obtain a full understanding of the regularity of \emph{all} the
solutions for a given symmetric space.
\end{itemize}

Our method also has certain limitations. One of them is that we need
symmetric coset spaces for our theorem to be applicable. It would be
interesting to extend these results to homogenous non-symmetric
spaces.

Due to limitations of time and space we left many other interesting
applications of our theorem untouched. We mention a few of them
here. First of all, we did not discuss multi-centered black hole
solutions. Secondly, we did not determine the dimensions of the
orbits corresponding to the nilpotent elements mentioned above. Some
of these dimensions have already been calculated in
\cite{Gunaydin:2005mx,Gunaydin:2007bg}. It seems plausible that
there exists an explicit expression of these dimensions in terms of
the number $p$ occurring in the theorem.

\section*{Acknowledgments}
We would like to thank L. Andrianopoli, R. D'Auria and Olaf Hohm for
interesting and fruitful discussions and Axel Kleinschmidt, Arjan
Keurentjes, Jan Rosseel and Dennis Westra for useful correspondence.
We also thank each others universities for their hospitality.

The work of T.V.R. is supported in part by the FWO - Vlaanderen,
project G.0235.05 and by the Federal Office for Scientific,
Technical and Cultural Affairs through the "Interuniversity
Attraction Poles Programme – Belgian Science Policy" P5/27. T.V.R
also thanks the Fulbright organisation and the Junta de Andaluc\'ia
for financial support. E.B., W.C. and A.P. are supported by the
European Commission FP6 program MRTN-CT-2004-005104. The work of
A.P. is part of the research programme of the ``Stichting voor
Fundamenteel Onderzoek van de Materie'' (FOM).

\appendix

\section{Conventions}\label{appendix A}
\subsection*{GR conventions}
In our conventions the space time metric is mostly-plus.  The Ricci
tensor evaluated for the Ansatz (\ref{Ansatz}) is given by
\begin{align}
&\mathcal{R}_{rr}=(D-1)\Bigl[-\frac{\ddot{g}}{g}+\frac{ \dot{g} \dot{f}}{gf}\Bigr]\,,\\
&\mathcal{R}_{ab}=-\epsilon\Bigl[\frac{\ddot{g}g}{f^2}-\frac{g\dot{g}\dot{f}}{f^3}+(D-2)(\frac{\dot{g}}{f})^2
\Bigr]g^{D-1}_{ab}+\mathcal{R}^{D-1}_{ab} \,,
\end{align}
where a dot denotes differentiation with respect to $r$. The
Einstein equations are
\begin{align}\label{Einstein equations}
(D-1)\Bigl[-\frac{\ddot{g}}{g}+\frac{ \dot{g} \dot{f}}{gf}\Bigr] - \frac{1}{2}||v||^2f^2g^{2-2D}&=0\,,\\
-\epsilon\Bigl[\frac{\ddot{g}g}{f^2} - \frac{g\dot{g}\dot{f}}{f^3} +
(D-2) (\frac{\dot{g}}{f})^2 \Bigr] g^{D-1}_{ab} +
\mathcal{R}^{D-1}_{ab}&=0\,.
\end{align}
\subsection*{Algebra conventions}
Concerning group theory conventions, we used the Cartan-Weyl basis
for calculating the commutation relations of semi-simple Lie
algebras. Let us denote the root lattice by $\Delta$ and its
elements by $\alpha,\beta,\ldots$. The canonical commutation
relations are
\begin{align}
&[H_{\alpha},H_{\beta}]=0\,,\qquad [H_{\alpha},E_{\beta}]=(\alpha,\beta)E_{\beta}\,,\\
&[E_{\alpha},E_{\beta}]=N_{\alpha,\beta}E_{\alpha+\beta} \quad \text{if}\quad\alpha+\beta \in\Delta\,,\\
&[E_{\alpha},E_{-\alpha}]=H_{\alpha}\,,
\end{align}
where the $N_{\alpha,\beta}$ are given by
\begin{equation}
N_{\alpha,\beta}^2=\tfrac{1}{2}n(m+1)(\alpha,\alpha)\,,
\end{equation}
where $n$ is the integer with the property that $\beta+n\alpha
\in\Delta$ but $\beta+(n+1)\alpha$ is not in $\Delta$ and $m$ is the
integer for which $\beta-m\alpha \in\Delta$ and
$\beta-(m+1)\alpha\not\in \Delta$. Recall that the number $l$
\begin{align}
l=2\frac{(\alpha,\beta)}{(\alpha\alpha)}\in \{-3,-2,-1,0,1,2,3\}\,,
\end{align}
informs us about the length of the string of roots $\beta+k\alpha$.
Imagine the string is
$\beta+n\alpha,\beta+(n-1)\alpha,\ldots,\beta-m\alpha$ then
\begin{equation}
l=m-n
\end{equation}
One can derive the following relations for the $N_{\alpha,\beta}$
\begin{equation}
N_{\alpha,\beta}=-N_{\beta\alpha}=-N_{-\alpha,-\beta}=+N_{\beta,-\alpha-\beta}=N_{-\alpha-\beta,\alpha}\,.
\end{equation}

\section{Stationary Vacuum Solutions from $\frac{GL(n,\Real)}{\SO(n-1,1)}$}\label{APPENDIXB}

The different solutions are classified according to the signs of
$\Lambda$ and $||v||^2$. We only discuss solutions which arise from
the uplift of $-1$-branes with rotational symmetry ($k=+1$). We
refrain from giving the solutions with $\Lambda^2<0$ since these
contain an infinity of singularities.
\\
\begin{itemize}
\item  $||v||^2>0,\quad \Lambda^2>0$:

In this case the matrix $K_N$ can be further diagonalised with a
$\SO(1,1)$ boost that deletes $\omega$. The vacuum solution is then
given by
\begin{equation}
\d s^2 = \tanh^{p_0}(\tfrac{D-2}{2}r)f^2(r)\Bigl(\d r^2 + \d
\Omega^2_{D-1} \Bigr) - \tanh^{p_1}(\tfrac{D-2}{2}r)\d t^2
+\sum_{i=2}^n\tanh^{p_i}(\tfrac{D-2}{2}r)\d z_i^2\,,
\end{equation}
where $f(r)$ can be found in table 1. The coefficients $p$ are given
by
\begin{align}
& p_0 = \frac{\sum_i \lambda_i}{(D-2)||v||}\sqrt{\frac{2(D-1)}{D+n-2}} \,,\\
& p_1 = -\frac{D-2}{n}p_0 + \frac{n\lambda_1 -\sum_j\lambda_j}{n||v||} \sqrt{\tfrac{2(D-1)}{D-2}}\,, \\
& p_i = -\frac{D-2}{n}p_0+\frac{n\lambda_i - \sum_j \lambda_j
}{||v||n} \sqrt{\tfrac{2(D-1)}{D-2}} \,,
\end{align}
and the affine velocity is given by
$||v||^2=\tfrac{1}{2}\sum_i\lambda_i^2$.

\item $||v||^2>0,\quad \Lambda=0$:

In this case $\omega=\lambda_a$ and then the metric reads
\begin{align}
\d s^2 &= \tanh^{p_0}(\tfrac{D-2}{2}r)f^2(r)\Bigl(\d r^2 + \d
\Omega^2_{D-1} \Bigr)\\
&  + \tanh^{p_1}(\tfrac{D-2}{2}r)\Bigl(-\tilde{a}(r)\d t^2 +
\tilde{c}(r)\d x^2 + 2\tilde{b}(r)\d x\d t  \Bigr)
+\sum_{i=3}^n\tanh^{p_{i-1}}(\tfrac{D-2}{2}r)(\d z^i)^2\,,\nonumber
\end{align}
where $f(r)$ is defined in table 1 and the functions $\tilde{a}(r),
\tilde{b}(r)$ and $\tilde{c}(r)$ are given by
\begin{align}
& \tilde{a}(r)=1 + \lambda_a ||v||^{-1}\sqrt{2\tfrac{D-2}{D-1}} \ln\tanh (\tfrac{D-2}{2}r) \,,\\
&\tilde{b}(r)=\lambda_a ||v||^{-1}\sqrt{2\tfrac{D-2}{D-1}}\ln\tanh (\tfrac{D-2}{2}r)\,,\\
& \tilde{c}(r)=1 - \lambda_a ||v||^{-1}\sqrt{2\tfrac{D-2}{D-1}}
\ln\tanh(\tfrac{D-2}{2}r) \,.
\end{align}
The numbers $p$ are given by
\begin{align}
& p_0 = \frac{2\lambda_b+\sum_{i=3}^n\lambda_i}{(D-2)||v||}
\sqrt{2\frac{D-1}{D+n-2}}\,,\label{p_0}\\
& p_1 =-\frac{D-2}{n}p_0 +\frac{(n-2)\lambda_b -\sum_{i=3}^n\lambda_i}{n||v||}\sqrt{2\tfrac{D-1}{D-2}} \,,\label{p_1}\\
& p_{i-1}= -\frac{D-2}{n}p_0 +\frac{-2\lambda_b +n\lambda_i
-\sum_{j=3}^n\lambda_j}{n||v||}\sqrt{2\tfrac{D-1}{D-2}}\,,\qquad
 \text{for} \quad i=3,\ldots n\,,\label{p_i}
\end{align}
and the affine velocity squared $||v||^2$ is simply given by
$\lambda_b^2 + \tfrac{1}{2}\sum_{i=3}^n\lambda_i^2$.
\end{itemize}

The solutions that arise from lightlike geodesics can have
$\Lambda<0$ and $\Lambda=0$. The latter is only possible when all
$\lambda_i=0$ for $i>2$ and is given by

\begin{itemize}
\item $||v||^2=0\,,\quad \Lambda=0$:

This solution is the most simple one.
\begin{equation}
\d s^2 = \d r^2 + r^2\d \Omega^2_{D-1} -\tilde{a}(r)\d t^2 +
2\tilde{b}(r)\d t\d x + \tilde{c}(r)\d x^2 + \d z^i\d z_i\,,
\end{equation}
the harmonic function $h(r)$ on $\Real^D$ is given by $h(r) =a
r^{2-D} +b$ with $a$ and $b$ arbitrary constants of integration. The
functions $\tilde{a}(r), \tilde{b}(r)$ and $\tilde{c}(r)$ are given
by
\begin{equation}
\tilde{a}(r) =  1 +\lambda_a h(r)\,,\quad \tilde{b}(r) = \lambda_a
h(r)\,,\quad \tilde{c}(r) = 1- \lambda_a h(r)\,.
\end{equation}
\end{itemize}

\section{The Wick Rotation from $G/H$ to $G/H^*$}\label{APPENDIXC}
In this appendix we introduce a generalized Wick rotation which maps
a geodesic on $G/H$ in a Minkowskian theory, into a geodesic on
$G/H^*$ in its Euclidean version. In order to map a compactification
on a spatial circle into a one on a time-circle, we need to
analytically continue the internal radius: $R_0\rightarrow i\,R_0$.
This transformation can be seen as the action of a
\emph{complexified} ${\rm O}(1,1)$ transformation
\begin{eqnarray}
\mathcal{O}=i^{H_0}\,,
\end{eqnarray}
on the Minkowskian $D=3$ theory in which the scalar fields span
$G/H$. Consider the following action on the generators $\{t_n\}$ of
$G$
\begin{eqnarray}
t_n&\rightarrow
&\mathcal{O}_n{}^m\,[\mathcal{O}\,t_m\,\mathcal{O}^{-1}]\,.\label{transe8}
\end{eqnarray}
The action of $\mathcal{O}$ on the Cartan generators is trivial
$\mathcal{O}_i{}^j=\delta_i^j$, while it has the following action on
the shift generators
\begin{eqnarray}
\mathcal{O}_\alpha{}^\sigma
&=&i^{-\alpha(H_0)}\,\delta_\alpha^\sigma\,.
\end{eqnarray}
We see that a generic shift generator $E_\alpha$ is mapped into
itself by (\ref{transe8})
\begin{eqnarray}
E_\alpha &\rightarrow
&\mathcal{O}_\alpha{}^{\alpha^\prime}\,[i^{H_0}\,E_{\alpha^\prime}\,i^{-H_0}]=\nonumber\\
&=&\mathcal{O}_\alpha{}^{\alpha^\prime}\,i^{\alpha^\prime(H_0)}E_{\alpha^\prime}=E_\alpha\,.\label{nothing}
\end{eqnarray}
Therefore the transformation (\ref{transe8}) maps $\frak{g}$ into
itself. Let us now denote by $\frak{H}$ the algebra of compact
generators of $H$ and by $\tilde{\mathcal{K}}$ the non-compact
generators in $G/H$
\begin{eqnarray}
\frak{H}&=&\{\tilde{J}_\alpha\}=\{E_\alpha-E_{-\alpha}\}\,,\nonumber\\
\tilde{\mathcal{K}}&=&\{H_{\alpha_i},\,\tilde{K}_\alpha\}=\{H_{\alpha_i},\,E_\alpha+E_{-\alpha}\}\,.
\end{eqnarray}
On a generic element of $G$ the above transformation amounts to a
combination of a change of basis for the matrix representation and a
redefinition of the group parameters. Indeed if we write an element
of $G$ as the product of a coset representative
$\tilde{L}\in\exp(\tilde{\mathcal{K}})$ times an element $\tilde{h}$
of $H$ we have
\begin{eqnarray}
g=\tilde{L}(\varphi,\,\phi)\tilde{h}(\xi)=e^{\phi^\alpha\,\tilde{K}_\alpha}\,e^{\varphi^i\,H_i}\,e^{\xi^\alpha\,\tilde{J}_\alpha}&\rightarrow&
\mathcal{O}\,e^{\phi^{\prime\alpha}\,\tilde{K}_\alpha}\,e^{\varphi^{\prime
i}\,H_i}\,e^{\xi^{\prime
\alpha}\,\tilde{J}_\alpha}\,\mathcal{O}^{-1}\,,
\end{eqnarray}
where the redefined parameters are
\begin{eqnarray}
\varphi^{\prime i}&=&\varphi^{
i}\,\,\,;\,\,\,\,\phi^{\prime\alpha}=\phi^\sigma\,\mathcal{O}_\sigma{}^\alpha\,\,\,;\,\,\,\,
\xi^{\prime\alpha}=\xi^\sigma\,\mathcal{O}_\sigma{}^\alpha\,.
\end{eqnarray}
Let us consider the effect of this transformation on the generators
of the coset representative and of the compact factor
\begin{eqnarray}
\phi^{\prime
\alpha}\,[\mathcal{O}\,\tilde{K}_\alpha\,\mathcal{O}^{-1}]&=&\phi^\alpha\,i^{-\alpha(H_0)}[i^{\alpha(H_0)}E_\alpha+i^{-\alpha(H_0)}\,E_{-\alpha}]=\nonumber\\&=&\phi^\alpha\,(E_\alpha+(-1)^{\alpha(H_0)}\,E_{-\alpha})=\phi^\alpha\,K_\alpha\,,\nonumber\\
\xi^{\prime
\alpha}\,[\mathcal{O}\,\tilde{J}_\alpha\,\mathcal{O}^{-1}]&=&\xi^\alpha\,i^{-\alpha(H_0)}[i^{\alpha(H_0)}E_\alpha-i^{-\alpha(H_0)}\,E_{-\alpha}]=\nonumber\\&=&\xi^\alpha\,(E_\alpha-(-1)^{\alpha(H_0)}\,E_{-\alpha})=\xi^\alpha\,J_\alpha\,,
\end{eqnarray}
where $J_\alpha$ and  $K_\alpha$ differ from $\tilde{J}_\alpha$ and
$\tilde{K}_\alpha$ only for $\alpha=\gamma$, for which
$J_\gamma=E_\gamma+E_{-\gamma}$ and $K_\gamma=E_\gamma-E_{-\gamma}$.
$J_\alpha$ are therefore generators of $H^*$ and $K_\alpha$,
together with $H_{\alpha_i}$ are in $\frak{g}/\frak{H}^*$. The Wick
rotation defines therefore a mapping between two different
representations of the same element $g$ of $G$: One as the product
of a coset representative $\tilde{L}$ in $G/H$ and an element
$\tilde{h}$ of $H$ and the other as a product of a coset
representative $L$ in $G/H^*$ times an element $h$ in $H^*$. The
matrix $\tilde{M}(\varphi^i,\phi^\alpha)=\tilde{L}\tilde{L}^T$ which
describes the scalar fields on $G/H$ transforms as follows
\begin{eqnarray}
\tilde{M}(\varphi^i,\phi^\alpha)&\rightarrow&
\mathcal{O}\tilde{M}(\varphi^{\prime
i},\phi^{\prime\alpha})\mathcal{O}^{T}=L\eta
L^T=M(\varphi^i,\phi^\alpha)\,,
\end{eqnarray}
where $\eta=\mathcal{O}\mathcal{O}^{T}$ and $M$ is the matrix
describing the scalars on $G/H^*$.

\section{Toroidal Reduction of Type II Theories}
\label{typeIIred}
 Let us now consider the metric Ansatz for the
reduction of Type II theory (in the ten dimensional string frame) on
a 7-torus with signature $(1,6)$
\begin{eqnarray}
ds^2&=&G_{mn}\,(dz^m+A^m)\,(dz^n+A^n)+e^{4\,\phi_3}\,g_{ij}\,dx^i\,dx^j\,,
\end{eqnarray}
where $m,\,n=0,4,\dots, 9$, $g_{ij}>0$ is the Euclidean three
dimensional metric in the Einstein frame and $\phi_3$ is the
three-dimensional dilaton
\begin{eqnarray}
\phi_3&=&\phi-\frac{1}{4}\,\log(|{\rm det}(G)|)\,.
\end{eqnarray}
Denoting by $s=0,4,\dots, 9$ the internal rigid index, the vielbein
of the torus read
\begin{eqnarray}
E_m^s=e^{\sigma_m}\,\hat E_m^s\,,
\end{eqnarray}
where $\hat E_m^s$ is an $ {\rm SL}(7,\Real)$ matrix which depends
only on the off--diagonal components of the metric and $\sigma_m$
are the moduli of the internal radii. If we consider a diagonal
dimensional reduction $\hat E_m^s=\delta_m^s$, the internal metric
reads: $G_{mn}=R_m^2\,\eta_{mn}=e^{2\,\sigma_m}\,\eta_{mn}$. We
shall however consider the general case $\hat E_m^s\neq\delta_m^s$
of a non-diagonal dimensional reduction. Since $|{\rm det}(\hat
E)| =1$, we can write the three dimensional dilaton in the
following form
\begin{eqnarray}
\phi_3&=&\phi-\frac{1}{2}\,\sum_{m=0,4}^9\sigma_m\,.
\end{eqnarray}
We may locally associate the $D=3$ scalar fields with the ${\rm
E}_{8(8)}$ Cartan generators and positive roots. The latter split
into the 64 roots $b=\{\beta,\,\beta_0\}$, of level $0,\,2$ with
respect to the root $\alpha_8$, $\beta$ denoting the 63
$\frak{e}_{7(7)}$ positive rots, and 56 roots $\gamma$ of level $+1$
relative to $\alpha_8$. As previously mentioned the roots $\gamma$
correspond to the scalar fields originating from the $D=4$ vector
fields and their duals. We can write the $D=3$ bosonic action as
follows
\begin{eqnarray}
S&=&\int
e\,\left[\mathcal{R}-\partial_\mu\vec{h}\cdot\partial^\mu\vec{h}-\tfrac{1}{2}\,\sum_{b}e^{-2\,\vec{b}\cdot\vec{h}
}\,\left(\partial\Phi_{b}+\dots\right)^2+ \tfrac{1}{2}\,\sum_\gamma
e^{-2\,\vec{\gamma}\cdot\vec{h}
}\,\left(\partial\Phi_\gamma+\dots\right)^2 \right], \nonumber\\ \label{l3}
\end{eqnarray}
where the ellipses represent the couplings among the axions which
are encoded in the scalar manifold metric. These include the
couplings of the axions with the off-diagonal components of the
internal  metric. Here however we are only interested in the
axion-dilaton couplings. To represent the roots and the dilaton
vector $\vec{h}$ it is useful to introduce the following orthonormal
basis $\epsilon_m$, $m=0,4,\dots, 9$
\begin{eqnarray}
\epsilon_0&=&\{1,0,0,0,0,0,0,0\};\,\epsilon_4=\{0,1,0,0,0,0,0,0\}\,\dots\,\epsilon_{10}=\{0,0,0,0,0,0,0,1\}\,.\nonumber
\end{eqnarray}
The dilaton vector reads
\begin{eqnarray}
\vec{h}&=&\sum_{m=0,4}^9\sigma_m\,\epsilon_m+2\,\phi_3\,\epsilon_{10}\,.
\end{eqnarray}
The dilaton part of the Lagrangian density has therefore the
following form
\begin{eqnarray}
-\partial_\mu\vec{h}\cdot\partial^\mu\vec{h}&=&-\sum_{m=0,4}^9(\partial_\mu\sigma_m\partial^\mu\sigma_m)-4\,\partial_\mu\phi_3\partial^\mu\phi_3\,.
\end{eqnarray}
The general form of the action allows us to associate a generic
scalar field $\Phi_\alpha$ with the corresponding positive root
$\alpha$.
 This correspondence is useful if we want to determine an ${\rm
 O}(1,1)$ grading of $\Phi_\alpha$. Indeed consider an ${\rm
 O}(1,1)$ shift transformation on the dilatonic fields
 $\sigma_m,\,\phi_3$ such that $\vec{h}$ transforms as follows
 \begin{eqnarray}
\vec{h}&\rightarrow & \vec{h}+\xi\,\lambda\,,\label{o11}
 \end{eqnarray}
$\lambda$ being a constant vector and $\xi$ a constant parameter.
The kinetic term for $\Phi_\alpha$, which reads
\begin{eqnarray}
-\frac{1}{2}\,e^{-2\,\vec{\alpha}\cdot\vec{h}
}\,\partial_\mu\Phi_\alpha\partial^\mu\Phi_\alpha\,,
\end{eqnarray}
is invariant under the ${\rm
 O}(1,1)$  transformation (\ref{o11}) provided $\Phi_\alpha$
 transforms as follows
 \begin{eqnarray}
\Phi_\alpha&\rightarrow &e^{\alpha\cdot \lambda\,\xi}\Phi_\alpha\,.
 \end{eqnarray}
 Therefore the ${\rm
 O}(1,1)$ grading of $\Phi_\alpha$ is readily computed as the
 scalar product $\alpha\cdot \lambda$. Let us consider for
 instance the rescaling of the radius $R_0$ along the time direction: $R_0\rightarrow
 e^\xi\,R_0$, i.e. $\sigma_0\rightarrow \sigma_0+\xi$. The
 corresponding transformation on $\vec{h}$ reads
 \begin{eqnarray}
\vec{h}&\rightarrow &
\vec{h}+\xi(\epsilon_0-\epsilon_{10})=\vec{h}+\xi\,\lambda^8\,,
 \end{eqnarray}
 where $\lambda^8=\beta_0$ is the highest root of $\frak{e}_{8(8)}$.
 The grading of $\Phi_\alpha$ is $\alpha\cdot \beta_0$.
\par
 The explicit  correspondence between positive roots and
 dimensionally reduced ten dimensional fields, which allows us to
 interpret the $D=3$ scalars in terms of string zero
 modes, is given in table \ref{alphafield}. This table also includes a correspondence between
 $\frak{e}_{8(8)}$ weights and general fluxes, seen as non-propagating fields.
 \begin{table}[!htb]
  \caption{Correspondence between positive roots of
$\frak{e}_{8(8)}$ and dimensionally reduced string
zero-modes.}\label{alphafield}
\begin{center}
\begin{tabular}{|c|c|}\hline
  % after \\:\hline or \cline{col1-col2} \cline{col3-col4} ...
   Field/Flux & root/weight \\\hline\hline
  $B_{nm}$ & $\epsilon_n+\epsilon_m$ \\\hline
  $A_{n}{}^{m}$ & $\epsilon_n-\epsilon_m$ \\\hline
  $C_{n_1\dots n_k}$ & $-\frac{1}{2}\sum_{m=0,4}^9\epsilon_m+\epsilon_{n_1}+\dots \epsilon_{n_k}-\frac{1}{2}\epsilon_{10}$ \\\hline
  $C^{n_1\dots n_k}$ dual to $C_{\mu n_1\dots n_k}$ & $\frac{1}{2}\sum_{m=0,4}^9\epsilon_m-\epsilon_{n_1}-\dots \epsilon_{n_k}-\frac{1}{2}\epsilon_{10}$ \\\hline
  $B^m$ dual to $B_{\mu m}$ & $-\epsilon_n-\epsilon_{10}$ \\\hline
  $A_m$ dual to $A^m_{\mu}$ & $\epsilon_n-\epsilon_{10}$ \\\hline
 $F_{\mu_1\dots\mu_\ell \,n_1\dots n_k}$ & $-\frac{1}{2}\sum_{m=0,4}^9\epsilon_m+\epsilon_{n_1}+
 \dots \epsilon_{n_k}-\frac{(3-2\,\ell)}{2}\epsilon_{10}$ \\\hline
  $H_{nmp}$ & $\epsilon_{n}+\epsilon_{m}+\epsilon_{p}$ \\\hline
 $T_{nm}{}^p$ & $\epsilon_{n}+\epsilon_{m}-\epsilon_{p}$ \\\hline
$Q_{n}{}^{mp}$ & $\epsilon_{n}-\epsilon_{m}-\epsilon_{p}$ \\\hline
 $R^{nmp}$ & $-\epsilon_{n}-\epsilon_{m}-\epsilon_{p}$ \\\hline
\end{tabular}
\end{center}
\end{table}
The scalars $A_m{}^n$  in table  \ref{alphafield} denote the
off-diagonal internal metric moduli, $C$ denote the R-R forms, $F$
their field strengths, $B$ the Kalb-Ramond form, $H$ the
corresponding field strength and $T$ denotes an internal torsion. In
this representation $T$--duality along a direction $z^m$ amounts to
the transformation $\epsilon_m\rightarrow -\epsilon_m$. For instance
the Roman's mass parameter $m$ represents the flux of a 9-form
$F_{\mu\nu\rho\, n_1\dots n_7}$ and corresponds to the ${\rm
E}_{8(8)}$ weight
\begin{eqnarray}
m&\leftrightarrow&\frac{1}{2}\sum_{m=0,4}^9\epsilon_m+\frac{3}{2}\,\epsilon_{10}\,.
\end{eqnarray}
Let us now consider the scalars originating from the $D=4$ vector
fields and their duals. They correspond to the 56 roots $\gamma$,
see table \ref{gammafield}.
\begin{table}[!ht]
 \caption{Correspondence between
$\frak{e}_{8(8)}$ $\gamma$ roots and scalars originating from $D=4$
vector fields and their duals.}\label{gammafield}
\begin{center}
\begin{tabular}{|c|c|}\hline
  % after \\:\hline or \cline{col1-col2} \cline{col3-col4} ...
  Scalar & root $\gamma$ \\\hline\hline
 $C_{i_1\dots i_k \,0}$ & $-\frac{1}{2}\sum_{m=0,4}^9\epsilon_m+\epsilon_{0}+\epsilon_{i_1}+\dots \epsilon_{i_k}-\frac{1}{2}\epsilon_{10}$ \\\hline
 $B_{i0}$ & $\epsilon_i+\epsilon_0$ \\\hline
 $A_{0}^i$ & $-\epsilon_i+\epsilon_0$ \\\hline
  $C^{i_1\dots i_k}$ dual to $C_{\mu i_1\dots i_k}$ & $\frac{1}{2}\sum_{m=0,4}^9\epsilon_m-\epsilon_{i_1}-\dots \epsilon_{i_k}-\frac{1}{2}\epsilon_{10}$ \\\hline
  $B^i$ dual to $B_{\mu i}$ & $-\epsilon_i-\epsilon_{10}$ \\\hline
  $A_i$ dual to $A^i_{\mu}$ & $\epsilon_i-\epsilon_{10}$ \\
 \hline
\end{tabular}
\end{center}
\end{table}
 In Type IIA the R-R scalars in  table
\ref{gammafield} are $C_{ij0},\,C_0,\,C^{ij},\,C$, while in Type IIB
they are $C_{ijk0},\,C_{i0},\,C^{i}$ where $i=4,\dots, 9$.

\section{The STU Model}\label{APPENDIXE}
In this appendix we shall review some geometric properties of the
STU model. This is a $D=4,\,N=2$ supergravity coupled to three
vector multiplets. The scalar sector consists of three complex
fields $(z_i)=(z_1,z_2,z_3)=(s,t,u)$ parametrizing the Special
K\"ahler manifold ${\Scr M}_4^{STU}$ in (\ref{mstu}). This manifold
can be described by the following holomorphic prepotential
\begin{eqnarray}
F(s,t,u)=s\,t\,u\,.
\end{eqnarray}
The K\"ahler potential $K$ has the form:
\begin{eqnarray}
K(z,\bar{z})=-\log[-i\,(s-\bar{s})(t-\bar{t})(u-\bar{u})]\,,
\end{eqnarray}
and the metric is
$g_{i\bar{\jmath}}=\partial_i\partial_{\bar{\jmath}}K=V_i^a\,\bar{V}_{\bar{\jmath}}^{\bar
a}\,\delta_{a\bar a}$. The vielbein are computed to be
$V_i^a=i\,\delta_i^a/(z_i-\bar{z}_i)$.\par All the geometric
quantities characterizing the manifold can be expressed in terms of
the holomorphic symplectic section $\Omega(z)$, which, in the
special coordinate frame, have the following form
\begin{eqnarray}
\Omega(z)&=&(X^\Lambda(z),\,F_\Lambda(z))=(1,\,s,\,t,\,u,\,-stu,\,tu,\,su,\,st)\,,
\end{eqnarray}
where $\Lambda=0,\dots, 3$. It is also useful to define the
covariantly holomorphic section ${\bf V}(z,\bar
z)=(L^\Lambda,\,M_\Lambda)=e^{\frac{K}{2}}\,\Omega$. Next we define
the quantity
\begin{eqnarray}
{\bf U}_i&=&D_i{\bf V}=(\partial_i+\frac{1}{2}\,\partial_iK)\,{\bf
V}=(f^\Lambda{}_i,\,h_{\Lambda\,i})\,,
\end{eqnarray}
and introduce the following square matrices
\begin{eqnarray}
f^\Lambda{}_I&=&\left(\begin{matrix}f^\Lambda{}_i\cr
\overline{L}^\Lambda\end{matrix}\right)\,\,,\,\,\,h_{\Lambda\,I}=\left(\begin{matrix}h_{\Lambda\,i}\cr
\overline{M}_\Lambda\end{matrix}\right)\,,
\end{eqnarray}
where $ I=0,\dots, 3$, and define the complex kinetic matrix
$\mathcal{N}=R+i\,I$ of the $D=4$ vector fields, through the
following equation
\begin{eqnarray}
\overline{\mathcal{N}}_{\Lambda\Sigma}&=&h_{\Lambda\,I}\,(f^{-1})^I{}_\Sigma\,.
\end{eqnarray}
If ${\bf Q}=(p^\Lambda,\,q_\Lambda)$ are the quantized charges, the
complex (scalar dependent) central charge ${\bf Z}$ and matter
charges $({\bf Z}_{a})=({\bf Z}_s,{\bf Z}_t,{\bf Z}_u)$ are defined
as follows:
\begin{eqnarray}
{\bf Z}&=&{\bf V}^T\,\mathbb{C}\,{\bf Q}=L^\Lambda
q_\Lambda-M_\Lambda\,p^\Lambda\,,\nonumber\\
 {\bf Z}_{a}&=& V_a^i\,\nabla_i{\bf Z}=  V_a^i\,{\bf U}_i\,\mathbb{C}\,{\bf Q}=V_a^i\,(f^\Lambda{}_i
 \,q_\Lambda-h_{\Lambda\,i}\,p^\Lambda)\,.
\end{eqnarray}
If we choose as non-vanishing charges $q_0,p_1,\,p_2,\,p_3$ we find
the following expressions for the above charges:
\begin{eqnarray}
{\bf Z}&=& e^{\frac{K}{2}}\,\left({q_0} - {p_3}\,{s}\,{t} - (
{p_2}\,{s} + {p_1}\,{t} ) \,{u} \right)\,,\nonumber\\
{\bf Z}_s&=&i\, e^{\frac{K}{2}}\,\left({q_0} - {p_1}\,{t}\,{u} -( {p_3}\,{t} + {p_2}\,{u} ) \,\bar{s} \right)\,,\nonumber\\
{\bf Z}_t&=&i\, e^{\frac{K}{2}}\,\left({q_0} - {p_2}\,{s}\,{u} - ( {p_3}\,{s} + {p_1}\,{u} ) \,\bar{t} \right)\,,\nonumber\\
{\bf Z}_u&=&i\, e^{\frac{K}{2}}\,\left({q_0} - {p_3}\,{s}\,{t} - ( {p_2}\,{s} + {p_1}\,{t}) \,\bar{u} \right)\,,\nonumber\\
\end{eqnarray}
 Writing the complex scalar fields
in real components
\begin{eqnarray}
s&=&z_1=-a_1-i\,e^{\tilde{\varphi}_1}\,\,,\,\,\,t=z_2=-a_2-i\,e^{\tilde{\varphi}_2}\,\,,\,\,\,u=z_3=-a_3-i\,e^{\tilde{\varphi}_3}\,,
\end{eqnarray}
the relevant blocks of  the symmetric symplectic real matrix
$\mathcal{M}^{STU}_4$, defined in terms of $R,\,I$ by eq.
(\ref{MIR}), read:
\begin{eqnarray}
(\mathcal{M}^{STU}_4)_{\Lambda\Sigma}&=&
e^{\tilde{\varphi}_1+\tilde{\varphi}_2+\tilde{\varphi}_3}\times\mbox{{\small
$\left(\begin{matrix}{{A_1}}^2\,{{A_2}}^2\,{{A_3}}^2 &
{a_1}\,{{A_2}}^2\,{{A_3}}^2 & {a_2}\,{{A_1}}^2\,
   {{A_3}}^2 & {a_3}\,{{A_1}}^2\,{{A_2}}^2 \cr {a_1}\,{{A_2}}^2\,{{A_3}}^2 & {{A_2}}^2\,
   {{A_3}}^2 & {a_1}\,{a_2}\,{{A_3}}^2 & {a_1}\,{a_3}\,{{A_2}}^2 \cr {a_2}\,{{A_1}}^2\,
   {{A_3}}^2 & {a_1}\,{a_2}\,{{A_3}}^2 & {{A_1}}^2\,{{A_3}}^2 & {a_2}\,{a_3}\,{{A_1}}^2 \cr
   {a_3}\,{{A_1}}^2\,{{A_2}}^2 & {a_1}\,{a_3}\,{{A_2}}^2 & {a_2}\,{a_3}\,{{A_1}}^2 & {{A_1}}^2\,
   {{A_2}}^2
     \end{matrix}\right)$}}\,,\nonumber\\
(\mathcal{M}^{STU}_4)_{\Lambda}{}^\Sigma&=&
e^{\tilde{\varphi}_1+\tilde{\varphi}_2+\tilde{\varphi}_3}\times\mbox{{\small
$\left(\begin{matrix} -\left( {a_1}\,{a_2}\,{a_3} \right)  &
{a_2}\,{a_3}\,{{A_1}}^2 & {a_1}\,{a_3}\,
   {{A_2}}^2 & {a_1}\,{a_2}\,{{A_3}}^2 \cr -\left( {a_2}\,{a_3} \right)  & {a_1}\,{a_2}\,
   {a_3} & {a_3}\,{{A_2}}^2 & {a_2}\,{{A_3}}^2 \cr -\left( {a_1}\,{a_3} \right)  & {a_3}\,
   {{A_1}}^2 & {a_1}\,{a_2}\,{a_3} & {a_1}\,{{A_3}}^2 \cr -\left( {a_1}\,{a_2} \right)  & {a_2}\,
   {{A_1}}^2 & {a_1}\,{{A_2}}^2 & {a_1}\,{a_2}\,{a_3}
     \end{matrix}\right)$}}\,,\nonumber\\
(\mathcal{M}^{STU}_4)^{\Lambda\Sigma}&=&
e^{\tilde{\varphi}_1+\tilde{\varphi}_2+\tilde{\varphi}_3}\times\mbox{{\small
$\left(\begin{matrix} 1 & -{a_1} & -{a_2} & -{a_3} \cr -{a_1} &
{{A_1}}^2 & {a_1}\,{a_2} & {a_1}\,
   {a_3} \cr -{a_2} & {a_1}\,{a_2} & {{A_2}}^2 & {a_2}\,{a_3} \cr -{a_3} & {a_1}\,{a_3} & {a_2}\,
   {a_3} & {{A_3}}^2
     \end{matrix}\right)$}}\,,\nonumber\\
A_i^2&\equiv &e^{2\,\tilde{\varphi}_i}+a_i^2=|z_i|^2\,.
\end{eqnarray}
The matrix $\mathcal{M}^{STU}_4$
 can also be written as $L_{STU} (L_{STU})^T $, where $L_{STU}$ is
 the coset representative of ${\Scr M}_4^{STU}$, defined in
 (\ref{lstu}), in the symplectic representation defined by the
 adjoint action of $\frak{g}_{4}^{STU}$, subalgebra of
 $\frak{so}(4,4)$, on the generators $E_{\gamma^{(n)}}$, $n=1,\dots,
 8$. Let us give for completeness the explicit form of the  $\frak{g}_4^{(STU)}$ generators in this representation
\begin{eqnarray}
\sum_{i=1}^3\tilde{\varphi}_i H_{{\bf a}_i}&=&\mbox{diag}(  \tilde{\varphi}  _1  +
 \tilde{ \varphi } _2  +   \tilde{\varphi } _3 , -  \tilde{\varphi}  _1  +    \tilde{\varphi}  _2
+   \tilde{\varphi } _3 ,   \tilde{\varphi}  _1  -   \tilde{\varphi}  _2  +   \tilde{\varphi}   _3 ,
\tilde{\varphi } _1  +  \tilde{ \varphi}  _2  -   \tilde{\varphi}  _3 , \nonumber\\&&-
\tilde{\varphi}  _1  -    \tilde{\varphi}  _2  -   \tilde{\varphi}  _3 ,
  \tilde{\varphi}  _1  -   \tilde{\varphi}  _2  -  \tilde{ \varphi} _3 , -  \tilde{\varphi}  _1
+   \tilde{\varphi}  _2  -   \tilde{\varphi}  _3 , -  \tilde{\varphi}  _1  -   \tilde{ \varphi}
 _2  +   \tilde{\varphi}  _3 )\,,\nonumber\\
\sum_{i=1}^3 a_i E_{{\bf a}_i}&=& \left(\begin{matrix} 0 & {a_1} &
{a_2} & {a_3} & 0 & 0 & 0 & 0 \cr 0 & 0 & 0 & 0 & 0 & 0 & {a_3} &
{a_
    2} \cr 0 & 0 & 0 & 0 & 0 & {a_3} & 0 & {a_1} \cr 0 & 0 & 0 & 0 & 0 & {a_2} & {a_
    1} & 0 \cr 0 & 0 & 0 & 0 & 0 & 0 & 0 & 0 \cr 0 & 0 & 0 & 0 &
    -{a_1} & 0 & 0 & 0 \cr 0 & 0 & 0 & 0 & -{a_2} & 0 & 0 & 0 \cr 0 & 0 & 0 & 0 &
    -{a_3} & 0 & 0 & 0 \cr  \end{matrix}\right)\,\,,\,\,\,E_{-{\bf
    a}_i}=E_{{\bf a}_i}^T\,.
\end{eqnarray}
From the above matrices we can deduce the explicit form of ${\Scr
O}_n{}^m$ in (\ref{ooo}):
\begin{eqnarray}
{\Scr O}_\Lambda{}^\Sigma &=&\left(\begin{matrix}{c_1}\,{c_2}\,{c_3}
& {c_2}\,{c_3}\,{s_1} & {c_1}\,{c_3}\,{s_2} & {c_1}\,{c_2}\,
   {s_3} \cr -\left( {c_2}\,{c_3}\,{s_1} \right)  & {c_1}\,{c_2}\,{c_3} & -\left( {c_3}\,{s_1}\,
     {s_2} \right)  & -\left( {c_2}\,{s_1}\,{s_3} \right)  \cr -\left( {c_1}\,{c_3}\,{s_2}
     \right)  & -\left( {c_3}\,{s_1}\,{s_2} \right)  & {c_1}\,{c_2}\,{c_3} & -\left( {c_1}\,
     {s_2}\,{s_3} \right)  \cr -\left( {c_1}\,{c_2}\,{s_3} \right)  & -\left( {c_2}\,{s_1}\,
     {s_3} \right)  & -\left( {c_1}\,{s_2}\,{s_3} \right)  & {c_1}\,{c_2}\,{c_3} \cr
\end{matrix}\right)\,,\nonumber\\
{\Scr O}_{\Lambda\Sigma} &=&\left(\begin{matrix}-\left(
{s_1}\,{s_2}\,{s_3} \right)  & {c_1}\,{s_2}\,{s_3} & {c_2}\,{s_1}\,
   {s_3} & {c_3}\,{s_1}\,{s_2} \cr -\left( {c_1}\,{s_2}\,{s_3} \right)  & -\left( {s_1}\,{s_2}\,
     {s_3} \right)  & {c_1}\,{c_2}\,{s_3} & {c_1}\,{c_3}\,{s_2} \cr -\left( {c_2}\,{s_1}\,
     {s_3} \right)  & {c_1}\,{c_2}\,{s_3} & -\left( {s_1}\,{s_2}\,{s_3} \right)  & {c_2}\,
   {c_3}\,{s_1} \cr -\left( {c_3}\,{s_1}\,{s_2} \right)  & {c_1}\,{c_3}\,{s_2} & {c_2}\,{c_3}\,
   {s_1} & -\left( {s_1}\,{s_2}\,{s_3} \right)  \cr
\end{matrix}\right)\,,\nonumber\\
{\Scr O}^\Lambda{}_\Sigma &=&\left(\begin{matrix}{c_1}\,{c_2}\,{c_3}
& {c_2}\,{c_3}\,{s_1} & {c_1}\,{c_3}\,{s_2} & {c_1}\,{c_2}\,
   {s_3} \cr -\left( {c_2}\,{c_3}\,{s_1} \right)  & {c_1}\,{c_2}\,{c_3} & -\left( {c_3}\,{s_1}\,
     {s_2} \right)  & -\left( {c_2}\,{s_1}\,{s_3} \right)  \cr -\left( {c_1}\,{c_3}\,{s_2}
     \right)  & -\left( {c_3}\,{s_1}\,{s_2} \right)  & {c_1}\,{c_2}\,{c_3} & -\left( {c_1}\,
     {s_2}\,{s_3} \right)  \cr -\left( {c_1}\,{c_2}\,{s_3} \right)  & -\left( {c_2}\,{s_1}\,
     {s_3} \right)  & -\left( {c_1}\,{s_2}\,{s_3} \right)  & {c_1}\,{c_2}\,{c_3} \cr
\end{matrix}\right)\,,\nonumber\\
{\Scr O}^{\Lambda\Sigma} &=&\left(\begin{matrix}{s_1}\,{s_2}\,{s_3}
& -\left( {c_1}\,{s_2}\,{s_3} \right)  & -\left( {c_2}\,{s_1}\,
     {s_3} \right)  & -\left( {c_3}\,{s_1}\,{s_2} \right)  \cr {c_1}\,{s_2}\,{s_3} & {s_1}\,
   {s_2}\,{s_3} & -\left( {c_1}\,{c_2}\,{s_3} \right)  & -\left( {c_1}\,{c_3}\,{s_2} \right)
      \cr {c_2}\,{s_1}\,{s_3} & -\left( {c_1}\,{c_2}\,{s_3} \right)  & {s_1}\,{s_2}\,
   {s_3} & -\left( {c_2}\,{c_3}\,{s_1} \right)  \cr {c_3}\,{s_1}\,{s_2} & -\left( {c_1}\,{c_3}\,
     {s_2} \right)  & -\left( {c_2}\,{c_3}\,{s_1} \right)  & {s_1}\,{s_2}\,{s_3} \cr
\end{matrix}\right)\,,
\end{eqnarray}
where $c_i=\cos(\alpha_i)$ and $s_i=\sin(\alpha_i)$.

\bibliography{geo}
\bibliographystyle{utphysmodb}

\end{document}